\documentclass{ar-1col}
\usepackage[numbers,compress,square]{natbib}
\usepackage{url}
\usepackage[table]{xcolor}
\usepackage{verbatim}  

\usepackage[titletoc]{appendix}
\appendixtitleon    
\appendixtitletocon 
 
\usepackage[final]{pdfpages}

\jname{Xxxx. Xxx. Xxx. Xxx.}
\jvol{AA}
\jyear{2022}
\doi{10.1146/((please add article doi))}

\usepackage{amssymb,amstext,amsbsy,amsmath,amsfonts}
\usepackage{nicefrac}
\usepackage{MnSymbol}
\usepackage{enumitem}

\DeclareMathOperator{\re}{Re}
\DeclareMathOperator{\im}{Im}

\def\beq{\begin{equation}}
\def\eeq{\end{equation}}
\def\bea{\begin{eqnarray}}
\def\eea{\end{eqnarray}}
\def\beqa{\begin{equation}\begin{array}{l}}
\def\eeqa{\end{array}\end{equation}}
\def\eqlab#1{\label{eq:#1}}

\def\eref#1{\ref{eq:#1}}
\def\Eqref#1{Eq.~\ref{eq:#1}}

\def\half{\mbox{$\frac{1}{2}$}}

\def\thalf{\mbox{\small{$\frac{3}{2}$}}}
\def\quarter{\mbox{$\frac{1}{4}$}}
\def\third{\mbox{$\frac{1}{3}$}}
\def\sixth{\mbox{$\frac{1}{6}$}}

\def\barr{\left(\begin{array}{c}}
\def\earr{\end{array}\right)}
\def\bmat{\left(\begin{array}{cc}}
\def\emat{\end{array}\right)}
\def\al{\alpha}
\def\be{\beta}
 
\def\de{\delta} \def\De{\Delta}

 \def\La{{\Lambda}}

\def\si{\sigma}

\def\nn{\nonumber}
\def\dd{\mathrm{d}}

\DeclareMathOperator\arccosh{arccosh}

\def\bp{\boldsymbol{p}}

\def\bq{\boldsymbol{q}}

\def\2PE{2$\upgamma$}
\def\mTPE{{2\upgamma}}

\usepackage{upgreek}
\usepackage[colorlinks,hidelinks,citecolor=blue,linktoc=all,linkcolor=blue,pagecolor=blue, filecolor=blue,urlcolor=blue]{hyperref}
\begin{document}

\markboth{A.~Antognini, F.~Hagelstein and V.~Pascalutsa}{Nucleon structure in and out of muonic hydrogen}

\title{The proton structure in and out of muonic hydrogen}

\author{Aldo Antognini,$^{1,2}$ Franziska Hagelstein,$^{1,3}$ and Vladimir Pascalutsa$^3$
\affil{$^1$Laboratory for Particle Physics, Paul Scherrer Institute, 5232 Villigen-PSI, Switzerland 
}
\affil{$^2$Institute for Particle Physics and Astrophysics, ETH, 8093 Zurich, Switzerland
}
\affil{$^3$Institut f\"ur Kernphysik,
Johannes Gutenberg Universit\"at Mainz, 55099 Mainz, Germany
}
\affil{email: aldo.antognini@psi.ch, hagelste@uni-mainz.de, pascalut@uni-mainz.de}
}

\begin{abstract}
Laser spectroscopy of muonic atoms has been recently used to probe properties of light nuclei with unprecedented precision. We introduce nuclear effects in hydrogen-like atoms, nucleon structure quantities (form factors, structure functions, polarizabilities) and their effects in the Lamb shift and hyperfine splitting (HFS) of muonic hydrogen ($\mu$H). Updated theory predictions for the Lamb shift and HFS in $\mu$H are presented. We review the challenges of the ongoing effort to measure the ground-state HFS in $\mu$H and its impact on our understanding of the nucleon spin structure. To narrow down this search, we present a novel theory prediction obtained by scaling the measured HFS in hydrogen leveraging radiative corrections. We also summarize recent developments in the spectroscopy of simple atomic and molecular systems and emphasize how they allow for precise determinations of fundamental constants, bound-state QED tests and New Physics searches.


\end{abstract}

\maketitle

\tableofcontents

\section{Introduction}\label{intro}

$\mu$H  --- a hydrogen with the electron replaced by a muon --- 
has an enhanced sensitivity to the proton structure and the short-range effects in general.
The enhancement factor, as compared to H, is of order $(m_\mu/m_e)^3\approx 10^7$,
making $\mu$H a neat laboratory for studies of the proton structure. The same applies to 
other muonic atoms, where the neutron structure can be explored, along with the structure of the atomic nucleus as a whole.

\begin{figure}[t]
\centering
    \includegraphics[width=0.67\textwidth, bb= 0 0 550 500]{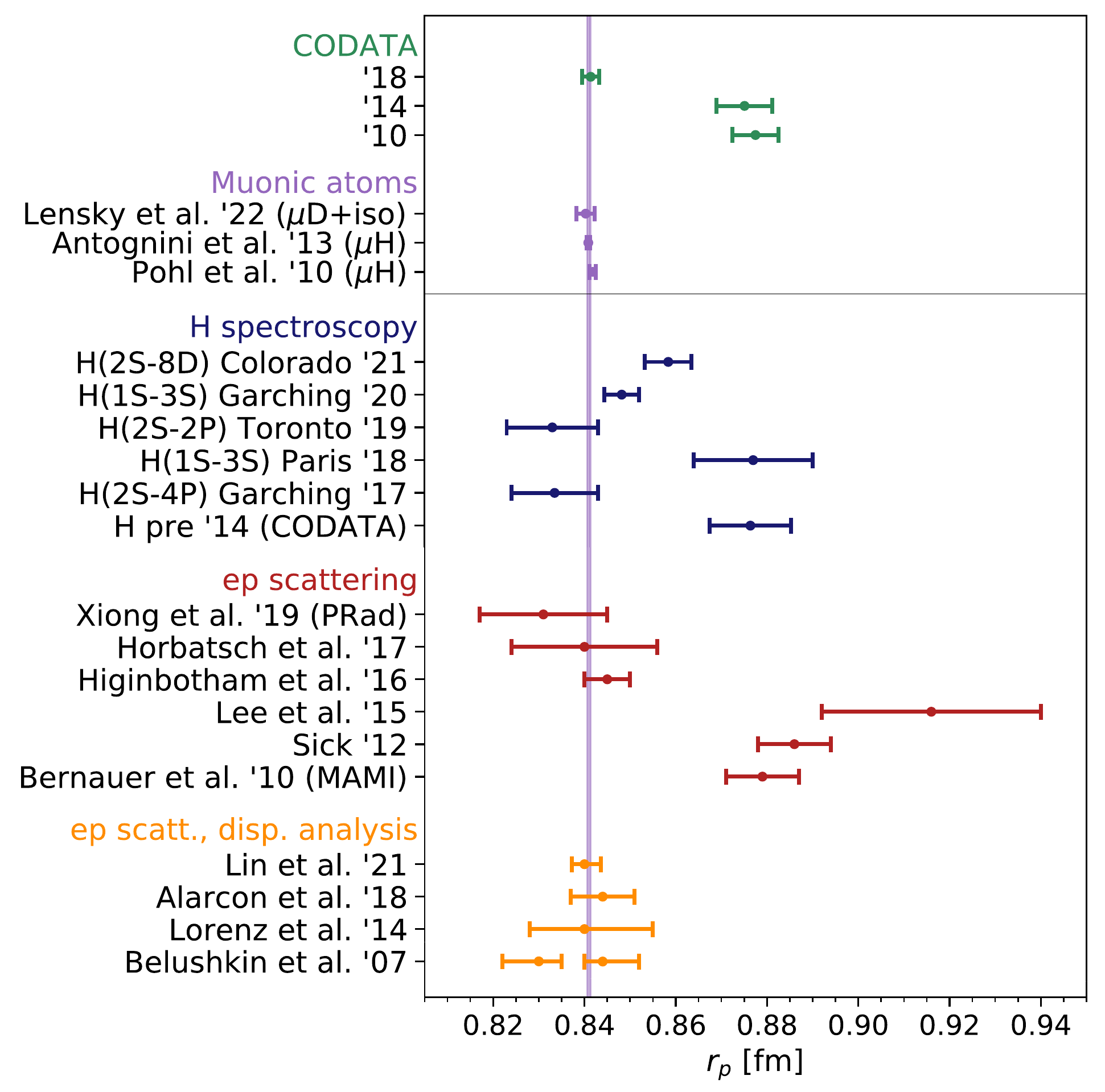}
         \vspace{-0.4cm}
    \caption{
       Selection of recent proton charge radius determinations. 
       For references, see respectively
       (from top to bottom): 
         CODATA \cite{Tiesinga:2021myr,Mohr:2015ccw,Mohr:2012aa}, muonic atoms \cite{Lensky:2022tue,Antognini:1900ns,Pohl:2010zza}, H spectroscopy \cite{Brandt:2021yor,Grinin:2020:Science_H1S3S,Bezginov:2019mdi,Fleurbaey:2018fih,Beyer:2017:Science_2S4P,Mohr:2015ccw}, $ep$ scattering \cite{Xiong:2019umf,Horbatsch:2016ilr,Higinbotham:2015rja,Lee:2015jqa,Sick:2012zz,Bernauer:2010wm}, dispersive analysis of $ep$ scattering \cite{ Lin:2021xrc,Alarcon:2018zbz,Lorenz:2014yda,Belushkin:2006qa}. The vertical band is aligned with the $\mu$H '13 value \cite{Antognini:1900ns}. 
       \label{RadiusSummary}}
\end{figure}

\begin{figure}[t]
\centering
       \includegraphics[width=0.67\textwidth]{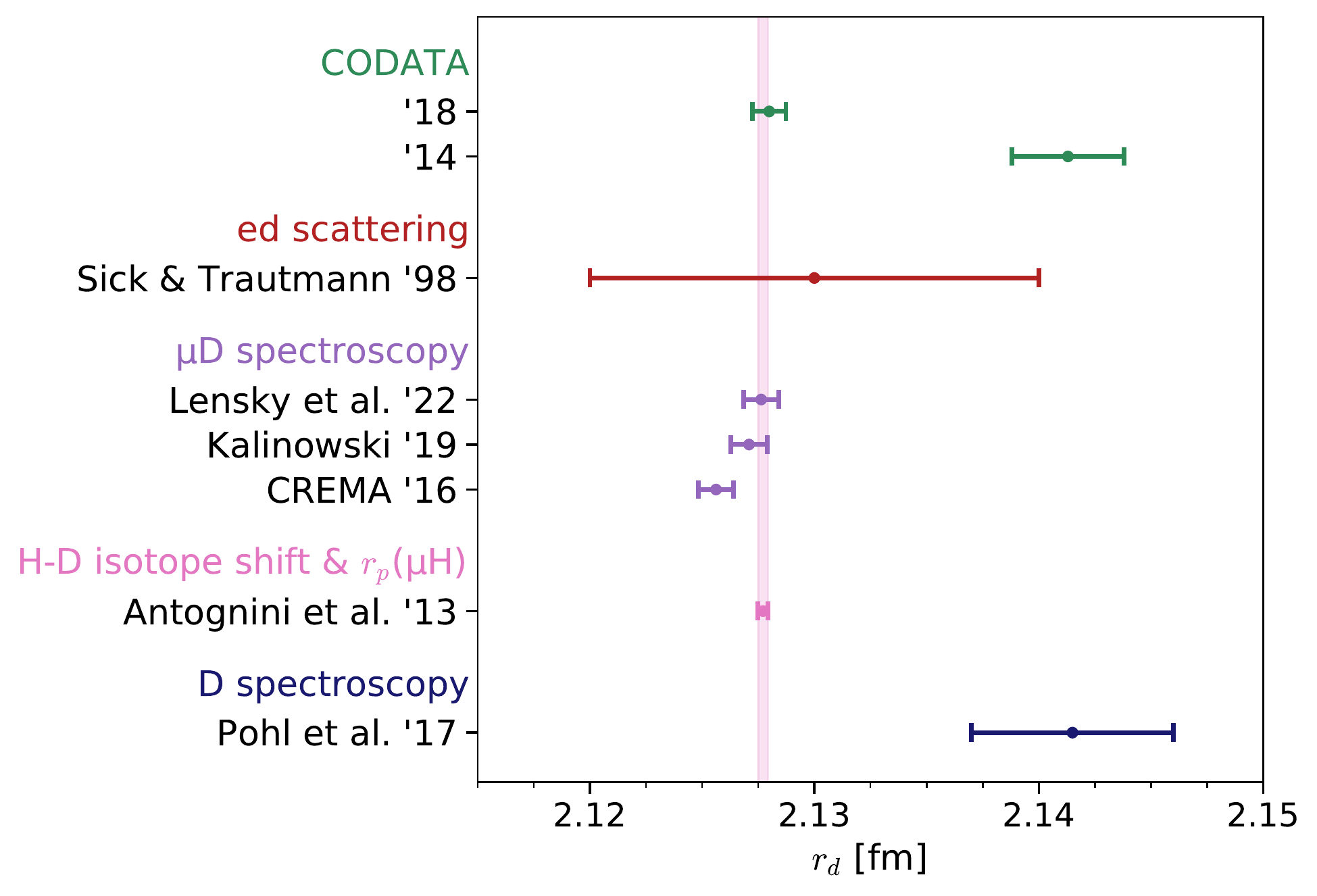}
       \vspace{-0.4cm}
\caption{Deuteron charge radius determinations. For references, see  respectively (from top to bottom):  CODATA \cite{Tiesinga:2021myr,Mohr:2015ccw}, $ed$ scattering \cite{Sick:1998cvq}, $\mu$D spectroscopy \cite{Lensky:2022tue,Kalinowski:2018rmf,CREMA:2016idx},  H-D isotopic shift and $\mu$H  Lamb shift \cite{Antognini:1900ns}, D spectroscopy \cite{Pohl:2016glp}. The vertical band is aligned with the value from the isotopic shift \cite{Antognini:1900ns}.\label{DRadiusSummary}}
\end{figure}

The last decade has witnessed a remarkable breakthrough in the laser spectroscopy of muonic atoms, starting from the long-awaited observation of  the $2S$-$2P$ transition in $\mu$H by the CREMA Collaboration \cite{Pohl:2010zza,Antognini:1900ns}. This transition appeared to be quite far from the predicted value, which made it very difficult to find, and very intriguing when found. 
It inferred a proton charge radius, $r_p$, which was spectacularly ($7\,\sigma$) smaller  than the state-of-the-art value of
that time (see CODATA '10 \cite{Mohr:2012aa} in Fig.~\ref{RadiusSummary}). The CODATA value comprised decades of $r_p$ determinations using the traditional techniques: $ep$ scattering and H spectroscopy. 
This resounding discrepancy, known as the \textit{proton-radius puzzle},  stirred a wealth of activity at the intersection of nuclear, particle, and atomic physics, reaching out to  physics beyond the Standard Model (see Refs.~\citenum{Carlson:2015jba,Karr:2020wgh,Gao:2021sml,Peset:2021iul}, for recent reviews). The subsequent measurement of the $\mu$D Lamb shift \cite{CREMA:2016idx} revealed a similar discrepancy 
for the deuteron charge radius, $r_d$, see Fig.~\ref{DRadiusSummary}. The two discrepancies are, in fact, related by the H-D isotopic shift measurement of the $1S$-$2S$ transition \cite{Jentschura:2011is}, which constrains
the difference, $r_d^2-r_p^2$. They are sometimes commonly referred to as the ``$Z=1$ radius puzzle'', emphasizing that no such discrepancy has been found in 
 muonic helium \cite{Krauth:2021foz}. Using the theory updates of Refs.~\citenum{Lensky:2022tue,Lensky:2022bid,Lensky:2022fif} and \citenum{Kalinowski:2018rmf}, the $r_p$ value obtained from $\mu$D via the isotopic shift  is in agreement with the value extracted directly from $\mu$H on the permille level. 
\begin{marginnote}[]
\entry{H, D}{Hydrogen, deuterium.}
\entry{$\mu$H, $\mu$D}{Muonic hydrogen, muonic deuterium.}
\entry{$ep$, $\mu p$}{Electron-proton, muon-proton scattering.}
\end{marginnote}

Today, more than a decade after the radius-puzzle installment, there is some consensus, adopted also by the CODATA group \cite{Tiesinga:2021myr}, 
that the $\mu$H value is, not only an order-of-magnitude more precise, but also, more accurate. The problem with the previous extractions
may simply lie in unaccounted systematic uncertainties --- a rather boring solution of the puzzle; at least in comparison with
most of the other proposals. 
This view is corroborated by some of the recent (re-)measurements using H. With exception of the 
H($1S$-$3S$) transition measurement by the Paris group~\cite{Fleurbaey:2018fih}, the other four new measurements in H yielded smaller radii than the CODATA '10: three of them in agreement with the muonic results~\cite{Grinin:2020:Science_H1S3S,Bezginov:2019mdi,Beyer:2017:Science_2S4P}, a very recent one 
of the H($2S-8D$)~\cite{Brandt:2021yor} though in some tension, that calls for the need of further experimental determinations.
%
\begin{marginnote}[]
\entry{Proton charge radius}{definition via slope of the electric Sachs form factor, $r_p = \sqrt{ -6 \, G_{Ep}^\prime (0)} $.}
\end{marginnote}

On the side of $ep$ scattering, the recent PRad experiment \cite{Xiong:2019umf}  has found the smaller value of $r_p$, in 
agreement with $\mu$H, confirming  several analysis of scattering data that agree with the muonic result \cite{Horbatsch:2016ilr,Higinbotham:2015rja,Lin:2021xrc,Alarcon:2018zbz,Lorenz:2014yda,Belushkin:2006qa}. The initial-state radiation experiment at MAMI has a larger uncertainty, thus, does not allow to discriminate between the small and large radius scenarios at the moment \cite{Mihovilovic:2019jiz}.

Complementing the picture with these latest results  diffuses the discrepancy quite considerably, see Fig.~\ref{RadiusSummary}. Nonetheless, the jury is still out and a new round of experiments is underway, 
including first measurements from $\mu p $ scattering by MUSE \cite{Strauch:2018ros} and AMBER collaborations \cite{COMPASSAMBERworkinggroup:2019amp}, 
improved $ep$ scattering measurements from PRAD-II \cite{PRad:2020oor} and the PRES Collaboration at MAMI, as well as spectroscopy measurements of H in Rydberg states~\cite{Scheidegger_Merkt_2020}, He$^+$($1S$-$2S$)
\cite{Krauth:2019kei, PhysRevA.79.052505} and simple molecules such as HD$^+$, H$_2^+$ and H$_2$ (see Sec.~\ref{Sec-HD}).

It is also interesting to look beyond the puzzle.
What else can be learned from the muonic atoms, in conjunction
with atomic spectroscopy and scattering experiments? Likewise, how the improved
$r_p$, and other structure information extracted from muonic atoms, will impact
the precision of other experiments, and, more generally, contribute to 
a better understanding of the Standard Model and beyond?

For example, the proton radius from $\mu$H, in combination
with the $1S$-$2S$ transition in H, leads to the most precise determination of the Rydberg constant $R_\infty$. In combination with the H-D isotopic shift, it leads to the most precise determination of the deuteron radius. The latter, combined with the measured Lamb shift in $\mu$D, provides a stringent test for the theory of the deuteron structure, viz., 
the nucleon-nucleon interaction. The proton radius combined with spectroscopy of H, D,  HD$^+$ and other simple systems, can be used to perform precision tests of bound-state QED for few-body systems,
impacting the precision of various fundamental quantities. While at the moment HD$^+$ can barely favor the muonic results \cite{Alighanbari2020}, its potential is enormous.
On the scattering side,
 the precise value of $r_p$ allows for a better determination of the proton electric form factor $G_{Ep}(Q^2)$.
 These are some of the ``ins and outs''
 that will be addressed in this article. 
Obviously, with increasing precision one becomes sensitive to certain scenarios beyond the Standard Model, beyond the ones
proposed as explanation of the puzzle in the first place, e.g., Refs.~\citenum{Pospelov:2017kep,Karshenboim:2014tka,Liu:2018qgl,Carlson:2015poa}.

Another topic of our interest here concerns the ongoing efforts to measure the ground-state hfs in $\mu$H. The CREMA Collaboration is aiming at the measurement
with 1~ppm relative accuracy by means of pulsed laser spectroscopy.
%
In parallel,  two other collaborations, at J-PARC~\cite{Sato:2014uza} and RIKEN-RAL~\cite{Adamczak:2001uvg, Dupays:2003zz, Bakalov:2015xya,Pizzolotto:2021dai, Pizzolotto:2020fue}, are developing measurements of this transition using different techniques. The hfs resonance is two orders of magnitude narrower than the $2S$-$2P$ line width, hence,  difficult to find. We shall examine the prospects for
an accurate prediction of this transition, which will help to guide the upcoming searches, and discuss what can be learned
when this transition is found.
\begin{marginnote}[]
\entry{hfs}{Hyperfine splitting.}
\entry{$\mu$H hfs experiments}{
Ongoing efforts to measure the  $1S$ hfs in $\mu$H with  1 ppm accuracy.}
\end{marginnote}

This paper is organised in the following way. Section \ref{statusSec} provides a brief introduction into the nuclear-structure corrections, with emphasis 
on radiative corrections important for $\mu$H.
Section \ref{Sec3} is devoted to evaluations of the proton-polarizability contribution. Section \ref{Sec4} provides updated theory predictions for the $\mu$H Lamb shift and the hfs in H and $\mu$H.
In Section~\ref{Sec5}, we elaborate more on the central role of HD$^+$ spectroscopy and H to extend the impact of the proton radius measurement to other precision quantities and possible New Physics searches. 
Section \ref{sec:Conclusions} presents a list of future prospects. Note that we are using natural units ($\hbar = c = 1$), unless specified otherwise.

\section{Brief introduction to nuclear effects in hydrogen-like atoms}
\label{statusSec}

Muonic atoms have a distinctly
 small Bohr radius, and therefore a larger sensitivity
to nuclear structure, and short-range effects in general.
While the finite-size contribution is increased by the aforementioned factor of $10^7$, relative to normal atoms, the QED effects contributing to the $2S$-$2P$ splitting increase only by a factor of 50, promoting the finite-size contribution to be the second largest contribution, trumped only by the  one-loop eVP, shown in Fig.\ \ref{DiagramsSec2}(a) and discussed in Sec.~2.1 of the Supplement. The literature accounting for these
effects is very extensive, see, e.g., Refs.~\citenum{Pachucki:1996zza,Eides:2000xc,Borie:2012zz,Indelicato:2012pfa,Karshenboim:2015}.
And nevertheless, the work on accurate calculations of these effects will continue in the foreseeable future, as the ongoing experiments bode new leaps in precision. Important for this program is the progress on the nuclear side, 
since many of the corrections require the input of
nuclear and nucleon form factors and structure functions.
In this section, and Sec.~\ref{Sec3}, we briefly describe how these ingredients are entering the atomic calculations; some numerical results for $\mu$H are discussed in
Sec.~\ref{Sec4}.

\begin{figure}[b]
\includegraphics[width=0.98\textwidth]{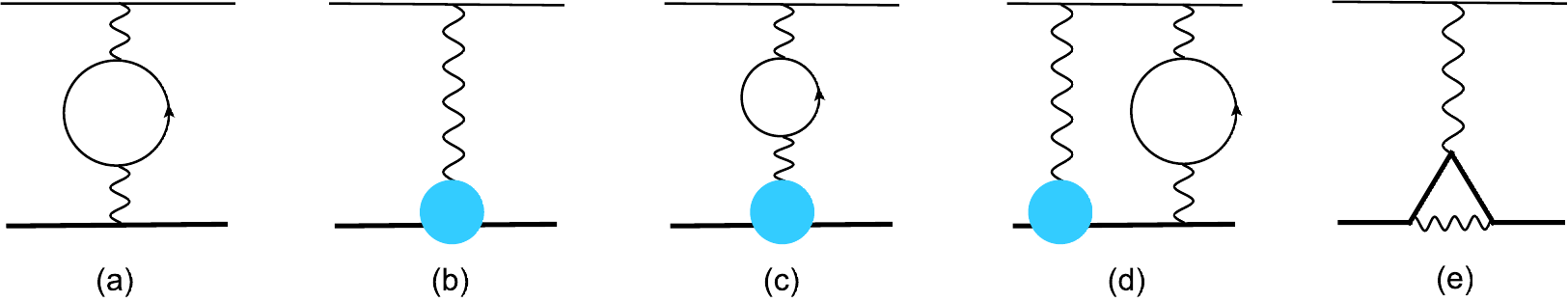}
\caption{Major corrections discussed in the text. The cyan blobs represent the finite-size effects, thin and thick lines the muon and proton, respectively. \label{DiagramsSec2}}
\end{figure}

\begin{marginnote}[]
\entry{Bohr radius}{$a_\mathrm{B}=(Z \alpha m_r)^{-1}$, with reduced mass $m_r = mM/(m+M)$, where $m$ and $M$ are the lepton and nucleus masses.}
\end{marginnote}

The starting point is, of course, the Coulomb problem solved by
using either the Dirac or Schr\"odinger equation \cite{BetheSalpeter}. A short recap of the quantum-mechanical Coulomb problem is given in Sec.~1 of the Supplement. 
It assumes a pointlike nucleus with the electric charge $Ze$, whereas the effects beyond this approximation come as perturbative corrections to the Lamb shift, fine and hyperfine structure. The perturbation series is organized in powers of the fine-structure constant $\al = e^2/4\pi$, and mass ratios. Among the nuclear-structure effects, we distinguish (i) the \textit {finite-size}
effects, which come from the electromagnetic 
distributions in the nucleus [e.g., Fig.\ \ref{DiagramsSec2} (b-d)], and (ii) the \textit {polarizability} effects
[Fig.\ \ref{DiagramsSec3} (a)],
caused by deformations of the distributions within the atom. The former
can be entirely  described by the elastic form factors, such as $G_E(Q^2)$ and $G_M(Q^2)$
in case of a spin-1/2 nucleus (e.g., the proton), whereas  the latter 
require a more complicated input, viz., structure functions. 
The two types of effects are also treated quite differently, as will be seen in the following derivation of the leading and next-to-leading nuclear contributions.

\begin{marginnote}[]
\entry{eVP, $\mu$VP, hVP}{Vacuum polarization due to electrons, muons,  hadrons.}
\end{marginnote}

\subsection{Finite-size effects}
Let us start by discussing the finite-size effects, that can be described through the charge, Friar and Zemach radii, as well as higher moments of the electromagnetic distributions.
\subsubsection{Lamb shift}
The main nuclear effect in the Lamb shift  comes from the nuclear charge distribution $\rho_E(r)$, which, in momentum space,
is described by an electric form factor (eFF) $G_E(Q^2)$, function of the photon virtuality, $Q^2 = \bq^2 - q_0^2$.
The corresponding potential is [see Fig.\ \ref{DiagramsSec2}(b)]:
\beq 
\eqlab{eFFpot}
V_{\mathrm{eFF}}(|\bq|) = -\frac{4\pi Z\al}{\bq^2} \left[G_E(\bq^2) -1\right] 
= \frac{1}{\pi} \int_{t_0}^\infty \frac{ \dd t  }{t}\, \im G_E(t)\, 
\frac{4\pi Z \al}{\bq^2 +t }\,, 
\eeq
neglecting the relativistic effects, such as the dependence on the energy transfer  $q_0$ (retardation)
and recoil corrections, which can be treated within the Breit-potential formalism, 
cf.\ \cite[Ch.\ IX, \S 83]{LandauLifshitz4}. The two forms of the potential in \Eqref{eFFpot} are
related by the once-subtracted dispersion relation:
\beq
G_E (Q^2) = 1 - \frac{Q^2}{\pi} \int_{t_0}^\infty\frac{\dd t}{t} \frac{\im G_E(t) }{t+Q^2 - i0^+},
\eeq 
where the integration is done over the timelike region, where the form factor develops a discontinuity
associated with particle production, with $t_0$ being the lowest threshold. For the proton, for instance, the leading
discontinuity is associated with two-pion production, i.e., $t_0 = 4m_\pi^2 $. 

In principle, the absorptive part of the form factor, $\im G_E(t)$, is known empirically (see, e.g.,
Refs.~\citenum{Lin:2021xrc,Alarcon:2018zbz,Lorenz:2014yda,Belushkin:2006qa}
for the proton). However, here we use the dispersive
representation simply as a convenient analytical tool.   
In coordinate space, where the Coulomb potential is given by $-Z\al/r$, the 
form-factor correction takes the following form,
\beq 
V_{\mathrm{eFF}}(r) = \frac{Z\al}{\pi} \int_{t_0}^\infty \frac{ \dd t  }{t} \im G_E(t)\, 
\frac{1}{r} e^{-r\sqrt{t}}  \,,
\eeq
which, in fact, is the Yukawa potential with the dispersed mass given by $\sqrt{t}$. 
As a result, the 1\textsuperscript{st}-order perturbation-theory contribution to the classic ($2S-2P$) Lamb shift is given by: 
\beq 
\eqlab{LS1PT}
E^{\langle \mathrm{eFF}\rangle}_{2S-2P} \equiv \big< 2S | V_{\mathrm{eFF}} | 2S \big> - \big< 2P | V_{\mathrm{eFF}} | 2P \big> = \frac{(Z\al)^4 m_r^3}{2\pi} \int_{t_0}^\infty \!\!\dd t \, \frac{\im G_E (t)}{(\sqrt{t}+ Z \al m_r)^4}.
\eeq 
Since $Z\al m_r \ll \sqrt{t_0}$, the denominator can be expanded (assuming the atomic Bohr radius is  much larger than the nuclear size), yielding:
\beq
E^{\langle \mathrm{eFF}\rangle}_{2S-2P} 
= \frac{(Z\al)^4 m_r^3}{12} \sum_{k=2}^\infty 
 \frac{(-Z\al m_r)^{k-2}}{(k-2)!} \big< r^k_E\big> =  \frac{(Z\al)^4 m_r^3}{12} \Big( \big< r_E^2\big> - Z\al m_r 
 \big< r_E^3\big>+\ldots\Big) , 
 \eqlab{rmsLS}
\eeq
where $\big< r^k_E\big> $ is the $k$\textsuperscript{th} moment of the charge distribution 
$\rho_E(r)$: 
\beq 
\big< r^k_E\big> = 4\pi \int_0^\infty \dd r \, r^{2+k} \rho_E(r) = \frac{(k+1)!}{\pi} 
\int_{t_0}^\infty \dd t \frac{\im G_E(t)}{t^{1+k/2}}.
\eeq 
The first term in \Eqref{rmsLS} is the famous charge-radius correction, appearing at order $(Z\al)^4$. For a discussion of the self-energy diagram for the bound and free proton, cf.\ Fig.\ \ref{DiagramsSec2}(e), in the context of the proton charge-radius definition, we refer to Sec.~3 of the Supplement.
To compute the next term of order $(Z\al)^5$ correctly, we ought to take this correction to the 2\textsuperscript{nd}-order perturbation theory:
\bea
E^{\langle \mathrm{eFF}\rangle \langle \mathrm{eFF}\rangle}_{2S-2P} &=& \sumint_{n\neq 2} \frac{\big|\big< 2S | V_{\mathrm{eFF}} | nS \big> \big|^2- \big| \big< 2P | V_{\mathrm{eFF}} | nP \big>\big|^2}{E_2 - E_n}\nn\\
&\cong& -   \frac{2(Z\al)^5 m_r^4 }{\pi}\int\limits_0^\infty \!\frac{\dd Q}{Q^4}\, \left[ G_E(Q^2) -1 \right]^2
= \frac{(Z\al)^5 m_r^4}{12}\, \left(
\left<r_E^3\right> - \half \left<r_E^3\right>_{(2)} \right),
\eea
where we only kept terms of order $(Z\al)^5$ and introduced the 3\textsuperscript{rd} Zemach moment \cite{Friar:1978wv}:
\beq \left<r^3_E\right>_{(2)} = \frac{48}{\pi} \int\limits_0^\infty \!\frac{\dd Q}{Q^4}\,
\Big[ G_E^2(Q^2) -1 +\third \left<r_E^2\right> Q^2\Big],
\eeq
 with the corresponding radius called the Friar radius. One sees that the 2\textsuperscript{nd}-order contribution essentially replaces the third radius, appearing in \Eqref{rmsLS},  with the Friar-radius term.  To order $(Z\al)^5$,  the finite-size correction is then given by
\beq 
\eqlab{LSFS}
E^\mathrm{f.s.}_{2S-2P}  = \frac{(Z\al)^4 m_r^3}{12} \big< r_E^2\big> - \frac{(Z\al)^5 m_r^4}{24} \left<r_E^3\right>_{(2)} \,.
\eeq
\begin{marginnote}[]
\entry{The Friar radius}{$r_\mathrm{Friar} = \sqrt[3]{\left<r_E^3\right>_{(2)}}$}
\end{marginnote}
Similarly, to compute the complete order-$(Z\al)^6$ correction, one needs to take this potential to the 3rd-order,
and so forth. The first corrections that affect $P$-levels begin at order $(Z\al)^6$. Thus, up to this order, the entire effect can be deduced from a $\delta(\boldsymbol{r})$-function potential, which provides an easy generalisation for the $nS$-level shift:
\beq 
\eqlab{FSnS}
E^\mathrm{f.s.}_{nS} = \mbox{$\frac{2\pi}{3}$} Z\al \left( \big< r_E^2\big> - \half Z\al m_r \left<r_E^3\right>_{(2)} 
\right)  \phi_{nS}^2 (0) , 
\eeq 
\noindent \begin{marginnote}[]
\entry{Wave function at the origin}{$\phi_{nS}^2 (0) = 1/(\pi a^3_\mathrm{B} n^3)$}
\end{marginnote}
\noindent with $\phi_{nS}^2 (0)$ the wave function at the origin. 
The Friar-radius contribution is obviously playing a more significant role in $\mu$H than in H, and was an initial suspect to resolve the proton-radius puzzle \cite{DeRujula:2010dp}. However, in that scenario,
the Friar radius was so large that the expansion in radii would be invalidated~\cite{Distler:2010zq,Hagelstein:2015aa}. 
The present consensus is that this contribution is at least an order-of-magnitude smaller than what is required for the explanation of the puzzle.
Furthermore, there are relativistic corrections, which can be treated within the Breit-potential formalism, or
alternatively, by considering the two-photon exchange, as will be seen below. Also important are some
radiative
corrections, which come from combining the finite-size and QED effects. In muonic atoms, the eVP
plays an especially prominent role, and produces sizeable radiative corrections to the finite-size effects shown
in Fig.\ \ref{DiagramsSec2} (c) and (d),
considered in Sec.~\ref{Sec:radiativeCorr}.

\subsubsection{Hyperfine splitting}

Assuming a spin-1/2 nucleus, the hfs arises from the magnetisation properties
of the nucleus described by the magnetic form factor (mFF) $G_M(Q^2)$. For the $S$-levels, the corresponding potential is given by (omitting recoil corrections):
\beq
V_{\mathrm{mFF}}^F(|\bq|) = \frac{4\pi Z\al}{3mM} \Big[ F(F+1) -\thalf \Big]\, G_M(\bq^2) = \frac{4 Z\al}{3mM} \Big[ F(F+1) -\thalf \Big]\, \int_{t_0}^\infty \dd t \, 
\frac{\im G_M(t)}{\bq^2 +t }\,, 
\eeq
where $F=0$ or $1$ is the total spin, 
$\kappa_N$ the anomalous magnetic moment of the nucleus; $G_M(0) = 1 + \kappa_N $ is the value of the magnetic moment in units of $Ze/2M$. The corresponding coordinate-space potential is directly proportional to the magnetization density $\rho_M(r)$. Details on the charge and magnetization densities, and the coordinate-space potentials are given in Sec.~2 of the Supplement.

The 1\textsuperscript{st}-order contribution, 
yields the following
hfs interval of the $nS$-level:
\beq 
E_{nS\text{-hfs}}^{\langle \mathrm{mFF} \rangle}  =  \big( 1 - 2Z\al m_r \langle r_M \rangle \big)
\frac{E_\mathrm{F}}{n^3} + O[(Z\al)^6],
\eeq 
where  $E_\mathrm{F}$ is  the Fermi energy,  and 
$\big< r_M \big> = 4\pi \int_0^\infty \dd r \, r^3 \rho_M(r)$ is the linear magnetic radius.
At the 2\textsuperscript{nd} order, the interference with the eFF potential of \Eqref{eFFpot}, gives:
\beq 
E_{nS\text{-hfs}}^{\langle \mathrm{mFF} \rangle \langle \mathrm{eFF} \rangle}
=  Z\al m_r \big( \langle r_M \rangle  - r_\mathrm{Z}   \big) \frac{E_\mathrm{F}}{n^3} + O[(Z\al)^6], 
\eeq 
thus cancelling the linear magnetic radius term from the 1\textsuperscript{st} order, and installing instead the Zemach radius:
\beq 
r_\mathrm{Z} = - \frac{4}{\pi} \int_0^\infty \frac{\dd Q}{Q^2} \left[\frac{G_E(Q^2) G_M(Q^2) }{1+\kappa_N} -1\right].\eqlab{Zemachradius}
\eeq 
The Fermi-energy contribution is not a finite-size effect, as it is already present for 
a pointlike nucleus. The leading finite-size effect in the hfs is therefore of order $(Z\al)^5$, 
\beq 
\eqlab{Zcontr}
E_{nS\text{-hfs}}^{\mathrm{f.s.}} = - (2Z\al m_r/n^3)  E_{\mathrm{F}}\, r_\mathrm{Z} .
\eeq At this order, also the polarizability corrections begin to appear. 
We consider them next.
\begin{marginnote}[]
\entry{The Fermi energy}{$E_\mathrm{F} = \frac{8 (Z\al)^4 m_r^3 (1+\kappa_N) }{3mM} $}
\end{marginnote}

\subsection{Two-photon exchange and polarizability effects}
\label{sec:twogammagen}

\begin{figure}[thb]
\includegraphics[width=0.6\textwidth]{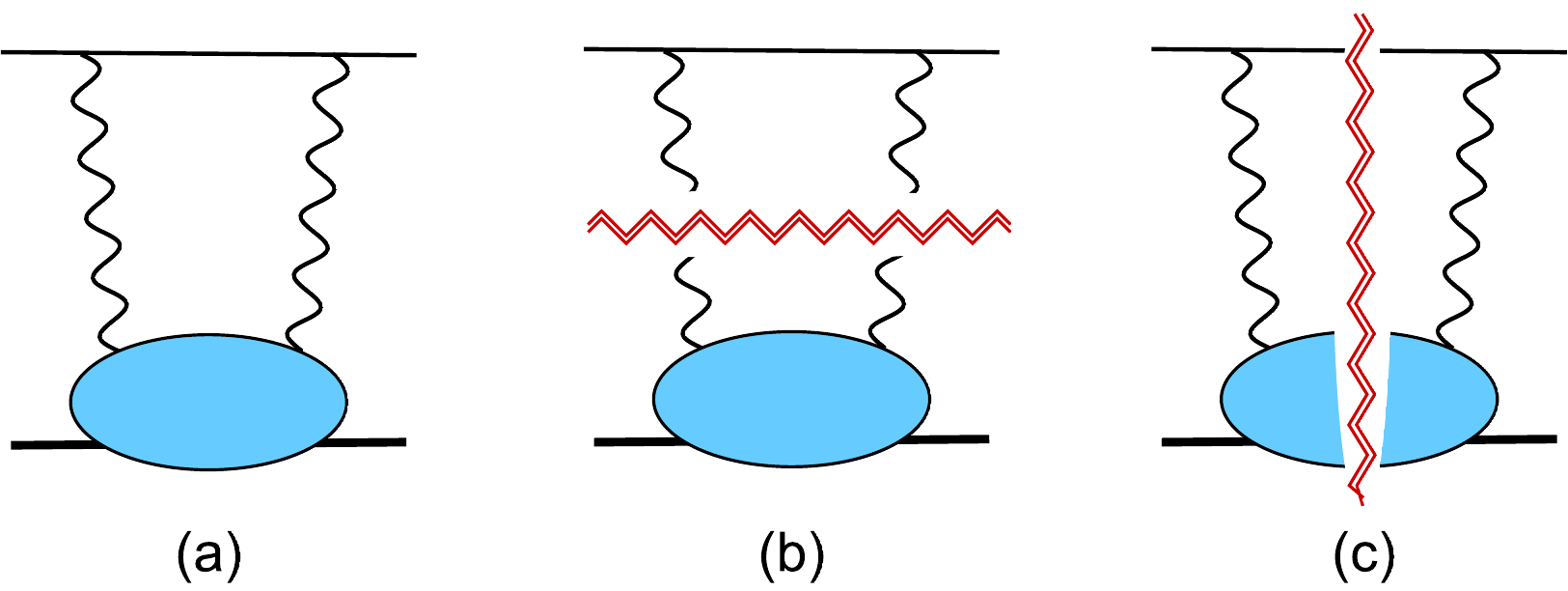}
\caption{The \2PE exchange (a), with the $t$-channel (b) and the $s$-channel (c) cuts. The cyan blobs represent effects from nuclear excitations. \label{DiagramsSec3}}
\end{figure}

Thus far, we considered effects which stem from the one-photon exchange and its iterations, such that
the nucleus stays intact and in its ground state.
There are also effects coming from nuclear excitations, which can only be assessed
through a \2PE exchange, see  Fig.~\ref{DiagramsSec3}(a). 
This description goes beyond the elastic form factors and involves instead
the polarizabilities and inelastic structure functions, as will
be seen in what follows.

The \2PE exchange in Fig.~\ref{DiagramsSec3}(a) introduces, in general, a correction $V_{2\upgamma} (p'-p; p', p)$
which depends on the relative momenta of the initial and final state, $p$ and $p'$, as well as the momentum
transfer $q=p'-p$. These are four-momenta, but the energy effects can safely be neglected, since they are suppressed
by $(Z\al)^2 m_r$. The dependence on $|\bp|=|\bp'|$ is suppressed by $Z\al m_r$ and will, to leading order, be 
neglected too. The dependence on $|\bq| $ is a bit more subtle, because of possible $\log Z\al$ terms.
To see this, let us write a dispersion relation for the graph of Fig.~\ref{DiagramsSec3}(b) in the Mandelstam 
variable $t=q^2= -Q^2$, for fixed $s\simeq (m+M)^2$:
\beq
V_{2\upgamma} (t) = \frac{1}{\pi} \int\limits_0^\infty \!\dd t' \, \frac{\im V_{2 \upgamma}(t') }{ t'- t - i0^+}= V_{2\upgamma} (t=0) + \frac{t}{\pi} \int\limits_0^\infty \!\dd t' \, \frac{\im V_{2 \upgamma}(t') }{ t' ( t'- t - i0^+) } ,
\eqlab{subDR2gamma}
\eeq
where in the second equation we have a once-subtracted relation, thus introducing the forward scattering amplitude, $V_{2\upgamma} (t=0)$. The remainder with non-vanishing momentum transfer is referred to as the off-forward amplitude.

The absorptive part, $\im V_{2 \upgamma}(t)$, corresponds with the discontinuity across the $t$-channel cut, which starts at 0, because
the photons are massless. Because of this, the expansion in $t$ is non-analytic.  When calculating the
level shifts in perturbation theory, this non-analyticity translates into $\log Z\al$ contributions. 

Let us see this for the Lamb shift, where we can make use of the arguments leading to \Eqref{LS1PT} and arrive at:
\begin{subequations}
\bea
E^{\langle 2\upgamma \rangle }_{2S-2P} &=& 
\frac{(Z\al m_r)^3}{8\pi^2} \int\limits_{0}^\infty \!\dd t \, \frac{t\, \im V_{2\upgamma} (t)}{(\sqrt{t}+ Z \al m_r)^4} \\
&=& V_{2\upgamma} (0 )\, \phi_{2S}^2(0) + 
\frac{(Z\al m_r)^3}{8\pi^2} \int\limits_{0}^\infty \!\dd t \, \left[
\frac{t}{(\sqrt{t}+ Z \al m_r)^4} - \frac{1}{t} \right] \im V_{2\upgamma} (t).
\eqlab{2gammasplit}
\eea
\end{subequations}
Let us focus on the leading polarizability effect, coming from
the electric $\al_{E1}$ and magnetic $\be_{M1}$ dipole polarizabilities of the nucleus. 
Knowing how they enter the Compton scattering amplitude, we can obtain their contribution to $\im V_{2\upgamma}(t)$. Note
that our consideration of the \2PE cut in Fig.~\ref{DiagramsSec3}(b) involves only 
the real Compton scattering and static polarizabilities.
The expression is rather lengthy \cite{Hagelstein:2017ldk} 
and we only quote here 
the polarizability potential in the well-known long-range and the singular short-range limit:
\begin{subequations}
\bea
V_{2\upgamma}(r) = \frac{1}{4\pi^2 r} \int\limits_{0}^\infty \dd t  \, \im V_{2\upgamma} (t)\, 
e^{-r\sqrt{t}}  &\stackrel{r\to \infty }{=}& -\frac{\al\, \al_{E1}}{2r^4}  + \frac{\al\, ( 11 \al_{E1} + 5 \beta_{M1})}{4 \pi  m r^5}  + O(1/r^6),\\
&\stackrel{r\to 0 }{=} & \frac{\al m \log(mr)}{2\pi r^3} ( \al_{E1} - \be_{M1}) + O(1/r^3).
\eea
\end{subequations}
Note that the prefactor of $\al$ is coming from the lepton line, whereas the polarizabilities contain $Z^2\al$, hence in total the order is $(Z\al)^2$ as expected.  
The entire potential is negative-definite  (attractive), provided $\al_{E1}> \be_{M1}$, and assuming the electric polarizability is a positive-definite quantity.  
This potential,  however, is
not very useful to compute the contribution to the $S$-states, because of the singular
short-range behavior. 
Anticipating that, we have introduced the subtracted dispersion relation in \Eqref{subDR2gamma},  leading to \Eqref{2gammasplit}. 
The integration over $t$ is now convergent, the short-range contribution regularized, and,
for the leading-$Z\al$ contribution to the $nS$-shift, we obtain:
\beq
E^{\langle 2\upgamma \rangle }_{nS} = \big\{ V_{2\upgamma} (0 )  + 4\al(Z\al) \, \al_{E1} \log [2n (Z\al)^{-1} ]  + O[(Z\al)^3]\big\} \, \phi_{nS}^2(0) .
\eeq 
It remains now to calculate $V_{2\upgamma}(0)$, i.e., the forward  \2PE exchange, as it apparently gives the larger contribution, of order $(Z\al)^5$. 
For this, one can make use of
the  $s$-channel dispersion relations for the forward Compton amplitude, see Fig.~\ref{DiagramsSec3}(c), which
allows one to express everything in terms of integrals of the structure functions, $F_{1,2}(x,Q^2)$, measured in
inclusive electron scattering. Unfortunately, 
one of these dispersion relations requires a subtraction too, which precludes the use of 
a purely data-driven approach. Still, most of the existing calculations are based on the data-driven approach, in conjunction
with some model-building of the subtraction function. Alternatively, one can calculate the entire $(Z\al)^5$ contribution using $\chi$PT
or lattice QCD. More details on evaluations of the
forward \2PE-exchange contribution can found in Sec.~\ref{Sec3}.

Similar consideration can be done for the hfs. There are two important differences: the contribution of order $(Z\al)^6 \log Z\al $
is absent, and the order $(Z\al)^5$ --- the forward \2PE-exchange contribution  ---
can be expressed in terms of the spin structure functions without a subtraction, see Sec.~\ref{Sec3} for details. There is an interesting order-$(Z\al)^6$
contribution coming from the neutral-pion exchange, which couples to the lepton through the
chiral anomaly \cite{Hagelstein:2015lph,Huong:2015naj,Zhou:2015bea,Dorokhov:2017gst}. 
However, it is not very relevant at the
current level of precision. It will become, perhaps, once the $\mu$H $1S$ hfs is measured.

\subsection{Radiative corrections }\label{Sec:radiativeCorr}
The largest corrections in muonic atoms involve the eVP, which
also produces radiative corrections
to the finite-size effects via the mechanisms of Figs.\ \ref{DiagramsSec2} (c) and (d). 
They are respectively referred to as VP1 and VP2. 

\subsubsection{VP2 correction [Fig.~\ref{DiagramsSec2}(d)]}

The diagram 
appears from the interference of the finite-size correction $V_\mathrm{eFF}$, Fig.\ \ref{DiagramsSec2}(b), with the Uehling potential, Fig.\ \ref{DiagramsSec2}(a), at 
the 2\textsuperscript{nd}-order perturbation theory. To avoid lengthy considerations, we cast the finite-size effects into a $\delta(\boldsymbol{r})$-function potential, as remarked above (valid for contributions that only influence the $S$-levels).
The eVP effect then amounts to correcting the wave function at the origin, appearing in \Eqref{FSnS}, as follows:
\beq
\phi^2_{nS} (0) \to \phi^2_{nS} (0)\, \Big[ 1+ \frac{\al}{\pi} \,C_1 (nS) \Big], \eqlab{WFcorrection}
\eeq 
where $C_1 (nS)$ is known analytically for the case of one-loop eVP, see \cite[Eq.~B3]{Karshenboim:2021jsc}. This is the universal eVP correction to any
$\de$-function potential, including the forward \2PE-exchange correction considered in Sec.~3.

\subsubsection{VP1 correction to the Lamb shift [Fig.~\ref{DiagramsSec2}(c)]}

This correction corresponds to the following potential,
\beq 
V_{\mathrm{eFF, \, VP1}}(|\bq|) = -\frac{4\pi Z\al}{\bq^2} \left[G_E(\bq^2) -1\right] \Pi(\bq^2) \,, 
\eeq
where $\Pi(Q^2)$ is the scalar part of the vacuum polarization. Using the dispersive representation, 
the correction to \Eqref{LS1PT} reads:
\bea 
\eqlab{LS1PTrad1}
E^{ \mathrm{VP1}}_{2S-2P}  
&=& \frac{(Z\al)^4 m_r^3}{2\pi} \bigg[\int\limits_{t_0}^\infty \!\!\dd t \, \frac{\im G_E (t)\, \re\Pi(-t) }{(\sqrt{t}+ Z \al m_r)^4} + \int\limits_{4m_e^2}^\infty \!\!\dd t \, \frac{[\re G_E (-t)-1]\, \im\Pi(t) }{(\sqrt{t}+ Z \al m_r)^4}  \bigg] \\
&=&  \frac{(Z\al)^4 m_r^3}{2\pi} \bigg[ \,\int\limits_{4m_e^2}^\infty \!\!\dd t \, \underbrace{[\re G_E (-t)-1]}_{\approx 
\sixth t \big< r_E^2\big>} \, \im\Pi(t) 
\left( \frac{1}{(\sqrt{t}+ Z \al m_r)^4} - \frac{1}{t^2} \right) + \, O\big(\mbox{$\frac{Z\al m_r}{\sqrt{t_0}}$}\big) \bigg]. \nn
\eea 
Within  the indicated approximations 
this correction  affects only the charge-radius contribution: 
\bea
E^{\mathrm{f.s.}}_{2S-2P} & = &  \mbox{$\frac{1}{12}$} (Z\al)^4 m_r^3 \left[ (1+\de^{\mathrm{VP1}}_{2S-2P}) \big< r_E^2\big> - \half Z\al m_r \left<r_E^3\right>_{(2)} 
\right] \nn\\
\de^{\mathrm{VP1}}_{2S-2P} & = & \frac{1}{\pi} \int\limits_{4m_e^2}^\infty {\dd t} \, t \, \left[\frac{1}{(\sqrt{t} +Z \al m_r)^4} -\frac{1}{t^2} \right] \, \im \Pi(t).
\eea
Substituting the one-loop expression for the eVP (i.e, $\im \Pi^{(1)}(t)$ displayed in 
Eq.~14 of the Supplement), we obtain:
\beq
\de^{\mathrm{VP1}(1)}_{2S-2P} = \frac{\al \upkappa }{6 \pi (1-\upkappa^2)^2}  \left[ \upkappa( 4\upkappa^2 - 7) +
   \frac{ 4 \upkappa ^4-10 \upkappa ^2+9 }{\sqrt{1-\upkappa^2}}
    \arccos \upkappa \right], \quad \mbox{with $\upkappa = \frac{Z\al m_r}{2m_e}$}.
\eeq 
For $\mu$H, $\de^{\mathrm{VP1}(1)}_{2S-2P} \simeq 2.155 \times 10^{-3}$. Note that this correction
affects both the $S$- and $P$-levels.

\subsubsection{VP1 correction to hfs [Fig.~\ref{DiagramsSec2}(c)]}
The eVP radiative corrections to the hfs are treated similarly. In this case the potential 
corresponding to the diagram of Fig.\ \ref{DiagramsSec2}(c) is:
\beq 
V_{\mathrm{mFF,\, VP1}}^{\mathrm{hfs }}(|\bq|) = \frac{8\pi Z\al}{3mM}\, G_M(\bq^2) \, \Pi(\bq^2) . 
\eeq
Going through the same steps as in \Eqref{LS1PTrad1}, one finds the following effect on the
ground-state hfs:
\bea 
\eqlab{hfs1PTrad1}
E^{ \mathrm{VP1}}_{1S\text{-hfs}}   
&=&  \frac{8 (Z\al)^4 m_r^3 }{3\pi mM} \bigg[ \int\limits_{4m_e^2}^\infty \!\!\dd t \, \underbrace{\re G_M (-t)}_{\approx 
(1+\kappa_N)} \, \im\Pi(t)
\bigg( \frac{1}{\big(\sqrt{t}+ 2 Z \al m_r\big)^2} - \frac{1}{t} \bigg) \nn\\
&-& 4Z\al m_r \int_{t_0}^\infty \dd t\, \frac{\im G_M(t)}{t^{3/2}} \re \Pi (t)  +\, O\Big(\mbox{$\frac{(Z\al m_r)^2}{t_0}$}\Big) \bigg]. 
\eea 
The first term is not a finite-size correction as it only affects the Fermi-energy term, albeit differently for each $nl$.
For the ground state, its effect is $E_\mathrm{F} (1+ \de^{\mathrm{VP1}}_{1S\text{-hfs}}  )$, where
the one-loop eVP gives:
\bea 
\de^{\mathrm{VP1}}_{1S\text{-hfs}}  &=& \frac{1}{\pi}\int\limits_{4m_e^2}^\infty \!\!\dd t\,  
\bigg[ \frac{1}{\big(\sqrt{t}+ 2 Z \al m_r\big)^2} - \frac{1}{t} \bigg] \, \im\Pi^{(1)}(t)  \nn\\
&=& \frac{\al}{3 \pi \upkappa_1^3} \bigg[ 2\upkappa_1 +\third \upkappa_1^3 + 
\frac{2-\upkappa_1^2+2\upkappa_1^4}{\sqrt{\upkappa_1^2-1}}  \arccosh \upkappa_1  -\pi \bigg], 
\quad \mbox{with $ \upkappa_1 = \frac{Z\alpha m_r}{m_e} $}. 
\eea  
For $\mu$H this amounts to about $2\,\%$ correction to the Fermi energy, or, in absolute terms: $0.37465$ meV. This is a fairly large effect, and one must consider the next term in \Eqref{hfs1PTrad1}, which
eventually leads to a Zemach radius correction, $r_Z (1+
\de^{\mathrm{VP1}}_{\mathrm{Z},1S})$, see~Sec.~5 of the Supplement
for more details.

\subsubsection{Combining VP1 and VP2}
We note that this formalism applies to all the vacuum-polarization contributions, including
hVP, $\mu$VP in H, etc. However, in these cases one can expand in $Z\al$ before the $t$-integration, which simplifies things a lot. For example, 
the VP2 and VP1 corrections become 
equal at leading order, with their combined effect given by:
\beq
\de^{\mathrm{VP1 + VP2}}_{1S\text{-hfs}}  = -\frac{8 Z\al m_r}{\pi}\int\limits_{t_0}^\infty \!\dd t\, \frac{ \im\Pi(t)}{t^{3/2}} 
+ O( Z^2\al^3).\eqlab{muVPinH}
\eeq
Plugging in the one-loop VP with a lepton with mass $m_\ell$, or a charged scalar with mass $m_\pi$, one obtains: $\frac{3}{4}Z \al^2(m_r/m_\ell)$ and $\frac{1}{8 }Z \al^2(m_r/m_\pi)$, respectively.
The latter result can be used to estimate the hVP contribution.

\section{Evaluations of the forward two-photon exchange }\label{Sec3}

It has been long known \cite{Iddings:1965zz,Drell:1966kk} that the forward \2PE exchange, Fig.~\ref{DiagramsSec3}(a), is a convenient way to access the order-$(Z\al)^5$ effects due to inelastic nuclear structure, viz., the polarizability effects. The main ingredient in this calculation is the nuclear Compton scattering amplitude. More specifically, 
the forward VVCS amplitude which, for a spinless or an unpolarized nucleus with spin, is
decomposed into two tensors:
\beq 
T^{\mu\nu}(p, q) = \left(-g^{\mu\nu} +\frac{q^\mu q^\nu}{q^2} \right) T_1(\nu, Q^2)
+ \frac{p^\mu p^\nu}{M^2} T_2(\nu, Q^2),
\eeq 
where $p$ and $q$ are the four-momenta of, respectively, the nucleus, with $p^2=M^2$, and the photon. The scalar amplitudes
$T_1$ and $T_2$ are functions of the photon energy and virtuality, $\nu=p\cdot q/M$, $Q^2=-q^2$. A similar decomposition exists for spin-dependent VVCS, which contributes then to the hfs. For a spin-1/2 nucleus there are two spin amplitudes, $S_1$ and $S_2$. The forward \2PE-exchange contributions to the Lamb shift and hfs have the following generic form:
\begin{marginnote}[]
\entry{VVCS}{Forward doubly-virtual Compton scattering.}
\end{marginnote}
\begin{subequations}
\eqlab{TPEgeneralVVCS}
\bea
E^{\langle 2\upgamma \rangle }_{nS} &=& \phi_{nS}^2(0)\sum_{i=1}^2 \int_{-\infty}^\infty \dd \nu \int_0^\infty  \dd Q^2 \, K_i(\nu, Q^2) \, T_i (\nu,Q^2),\eqlab{TPEgeneralVVCSLS}\\
E^{\langle 2\upgamma \rangle }_{nS\text{-hfs}} &=& \phi_{nS}^2(0) \sum_{i=1}^2 \int_{-\infty}^\infty \dd \nu \int_0^\infty  \dd Q^2 \,\tilde K_i(\nu, Q^2) \, S_i (\nu,Q^2),
\eea
\end{subequations} 
where  $K_i$ and $\tilde K_i$ are some kernels functions. Further notations and formulae can be found in Refs.~\citenum{Hagelstein:2015egb} and \citenum{Pasquini:2018wbl}, as well as Sec.~4 of the Supplement. In particular, it is important to realize that
the Born part of the VVCS amplitudes is expressed in terms of 
the elastic form factors. It yields the finite-size effects,  
considered in the previous section, together with the recoil corrections.  
The non-Born part yields the polarizability contribution. Note that these are not always the same as the \textit{elastic} and \textit{inelastic} \2PE-exchange contributions, which refer to the
contributions of elastic and inelastic parts of the
structure functions.

The VVCS amplitudes can be calculated in $\chi$PT, but the more traditional approach is the data-driven
evaluation using the structure functions. Anticipating the forthcoming discussion, let us remark that the two approaches are presently agreeing on the polarizability
contribution to the Lamb shift of $\mu$H, but disagree for the hfs by several $\si$.
The new experimental data on the proton spin structure from the JLab Spin program
may be very helpful to resolve the latter discrepancy.
\begin{marginnote}[]
\entry{$\chi$PT}{Chiral perturbation theory, an effective-field theory
of low-energy QCD.}
\end{marginnote}

\subsection{Lamb shift in $\mu$H}

The VVCS amplitudes in \Eqref{TPEgeneralVVCS} are not measurable directly, but can be related to the inclusive scattering data by the fundamental principles of unitarity and causality, viz., the optical theorem and dispersion relations \cite{GellMann:1954db,Pascalutsa:2018ced,Pasquini:2018wbl}. Exploiting the $s$-channel cut, see Fig.~\ref{DiagramsSec3}(c), one hopes to express everything in terms of
the nuclear structure functions. 
For the spin-independent amplitudes we have: 
\begin{subequations}
\eqlab{DRs}
\bea
T_1(\nu,Q^2)&=&T_1(0,Q^2) +\frac{32\pi Z^2\al M\nu^2}{Q^4}\int_{0}^1 
\frac{\dd x \,x }{1 - x^2 (\nu/\nu_{\mathrm{el}})^2 - i 0^+}\,F_1 (x, Q^2)\eqlab{T1Subtr},\\
T_2(\nu,Q^2)&=&\frac{16\pi Z^2\alpha M}{Q^2}\int_0^1 \frac{\dd x}{1-x^2(\nu/\nu_\mathrm{el})^2-i0^+}\,F_2(x,Q^2),
\eea
\end{subequations}
where $\nu_\mathrm{el} = Q^2/2M$. 

\renewcommand{\arraystretch}{1.3}
\begin{table}[t]
\caption{Forward \2PE-exchange contributions to the $2S$-shift in $\mu$H, in units of $\upmu$eV.}
\label{results_LS}
\begin{tabular}{l|cc|cc|c}
\hline
\hline
Reference  &$E_{2S}^\mathrm{(subt)}$&$E_{2S}^\mathrm{(inel)}$&$E_{2S}^\mathrm{(pol)}$&$E_{2S}^\mathrm{(el)}$&$E_{2S}^{\langle 2\upgamma \rangle }$\\
\hline
\cellcolor{gray!10}{\textsc{Data-driven dispersive evaluation}}&&&&&\\
\cite{Pachucki:1999zza} Pachucki '99 &$1.9$&$-13.9$&$-12(2)$&$-23.2(1.0)$&$-35.2(2.2)$\\
\cite{Martynenko:2005rc} Martynenko '06 &$2.3$&$-16.1$&$-13.8(2.9)$&&\\
\cite{Carlson:2011zd} Carlson \textit{et al.}~'11\ 
&$5.3(1.9)$&$-12.7(5)$&$-7.4(2.0)$&&\\
\cite{Birse:2012eb} Birse and McGovern '12\ &$4.2(1.0)$&$-12.7(5)$&$-8.5(1.1)$&$-24.7(1.6)$&$-33(2)$\\
\cite{Gorchtein:2013yga} Gorchtein \textit{et al.}'13\ $^{\rm a}$&$-2.3(4.6)$&$-13.0(6)$&$-15.3(4.6)$&$-24.5(1.2)$&$-39.8(4.8)$\\
\cite{Hill:2016bjv}  Hill and Paz '16\ &&&&&$-30(13)$\\
\cite{Tomalak:2018uhr} Tomalak'18&$2.3(1.3)$&&$-10.3(1.4)$&$-18.6(1.6)$&$-29.0(2.1)$\\
\hline
\cellcolor{gray!10}{\textsc{leading-order B$\chi$PT}} &&&&&\\
\cite{Alarcon:2013cba} Alarc\`{o}n \textit{et al.}~'14\ &&&$-9.6^{+1.4}_{-2.9}$&&\\
\cite{Lensky:2017bwi} Lensky \textit{et al.}~'17~$^{\rm b}$&$3.5^{+0.5}_{-1.9}$&$-12.1(1.8)$&$-8.6^{+1.3}_{-5.2}$&&\\
\hline
\cellcolor{gray!10}{\textsc{Lattice QCD}}&&&&&\\
\cite{Fu:2022fgh} Fu \textit{et al.}~'22\ &&&&&$-37.4(4.9)$\\
\hline
\hline
\end{tabular}
\begin{tabnote}

$^{\rm a}$Adjusted values due to a different decomposition into the elastic and polarizability contributions.

$^{\rm b}$Partially includes the $\De(1232)$-isobar contribution.
\end{tabnote}
\end{table}
\renewcommand{\arraystretch}{1}

Unfortunately, the dispersion relation for $T_1$ requires a subtraction, which means not
everything is expressed in terms of the structure functions, here $F_1$ and $F_2$.
The amplitude $T_1(0,Q^2)$, i.e.,  the subtraction function\footnote{The conventional subtraction is done  at $\nu =0$, but, a
subtraction  at $\nu =i Q$ can be used to  diminish the inelastic structure-function contribution and  simplify the calculations \cite{Hagelstein:2020awq}.}
is an additional unknown in this equation.
It is not 
well-constrained by experimental data, and hence, in a
purely data-driven approach its modeling
leaves some room for imagination. 
At the beginning of the proton-radius puzzle, a large subtraction-function contribution was even proposed to resolve the discrepancy \cite{Miller:2012ne}, yielding the
missing 310 $\upmu$eV in the $\mu$H Lamb shift. In all the other existing models, however, this contribution appears to be much smaller, by two orders of magnitude, cf.~$E^\mathrm{(subt)}$ in Table~\ref{results_LS}.  The modest \2PE-exchange contribution was corroborated by $\chi$PT
calculations, where this problem of model-dependence does not arise. These results are also displayed in Table~\ref{results_LS}. Listed in there are the following \2PE-exchange effects in the $\mu$H Lamb shift:
\begin{itemize}
    \item  $E^\mathrm{(subt)}$  the subtraction function,
    \item $E^\mathrm{(inel)}$ the inelastic structure functions,
    \item $E^\mathrm{(pol)} = E^\mathrm{(subt)}+E^\mathrm{(inel)}$, the polarizability contribution,
    \item $E^\mathrm{(el)}$ the elastic structure functions (same as the Friar radius with recoil),
    \item $E^\mathrm{\langle 2\upgamma\rangle}= E^\mathrm{(el)} + E^\mathrm{(pol)}$, the total \2PE exchange.
\end{itemize}
Despite the moderate effect of the subtraction function, 
it does constitute the largest uncertainty of the data-driven evaluations.
Models of the subtraction function for the proton are constrained at  $Q^2=0$  by  the  magnetic polarizability $\beta_{M1}$, and at asymptotically large $Q^2$  by perturbative QCD 
\cite{Birse:2012eb}. There is a new idea \cite{Pauk:2020gjv} of how to further constrain it from
the dilepton electroproduction ($e^-p \rightarrow e^-p\, e^-e^+$), but that would be an 
extremely challenging experiment. There is hope that it can soon be calculated
in lattice QCD
\cite{Can:2020,Hannaford-Gunn:2020pvu,Chambers:2017dov,Fu:2021dja,Can:2022rgi,Fu:2022fgh}.

\begin{table}[b]
\renewcommand{\arraystretch}{1.5}
\caption{Determinations of the proton Zemach radius $r_{\mathrm{Z}p}$, in units of fm.}
\begin{footnotesize}
\label{Table:ZemachRadius}
\centering
\begin{tabular}{c|c|c|c|c|c}
\hline
\hline
\multicolumn{2}{c|}{$ep$ scattering}&\multicolumn{2}{c|}{$\mu$H $2S$ hfs}&\multicolumn{2}{c}{H $1S$ hfs}\\
\hline
Lin \textit{et al.}~\cite{Lin:2021xrc}&
Borah \textit{et al.}~\cite{Borah:2020gte}&
Antognini \textit{et al.}~\cite{Antognini:1900ns}&
B$\chi$PT \cite{Hagelstein:2015lph} &Volotka \textit{et al.}~\cite{Volotka:2004zu}&
B$\chi$PT \cite{Hagelstein:2015lph}\\
$1.054^{+0.003}_{-0.002}$&$1.0227(107)$&$1.082(37)$&$1.041(31)$&$1.045(16)$&$1.012(14)$\\
\hline
\hline
\end{tabular}
\end{footnotesize}
\end{table}
\renewcommand{\arraystretch}{1}

\subsection{Hyperfine splitting in H and $\mu$H}\label{datadrivenTPEHFS}

For the hfs, the \2PE-exchange effects are conventionally split into Zemach-radius, recoil and polarizability contributions \cite{Carlson:2011af}:
\beq
E^{\langle 2\upgamma \rangle }_{nS\text{-hfs}}=\frac{E_\mathrm{F}}{n^3}\left(\Delta_\mathrm{Z}+\Delta_\mathrm{recoil}+\Delta_\mathrm{pol}\right). \eqlab{FS_HFS}
\eeq
All of these effects begin to contribute at order $(Z\al)^5$.
While the elastic  contributions are known to better than $1\,\%$, the absolute uncertainty of the numerically large Zemach-radius contribution is not negligible. 
Still, the largest uncertainty comes from the polarizability contribution. 
In what follows we discuss the Zemach and the polarizability contributions in more detail.

\subsubsection{Zemach radius, correlation with the charge radius}
\label{RzRPCorrelations}

The Zemach-radius contribution,  defined as $\Delta_\mathrm{Z}=-2Z\al m_r r_\mathrm{Z}$, can be evaluated based on empirically known form factors
using \Eqref{Zemachradius}.
For example, the  recent dispersive analysis of the nucleon electromagnetic form factors from the Bonn group \cite{Lin:2021xrc} yields:
\beq
r_{\mathrm{Z}p}=1.054\left(^{+0.003}_{-0.002}\right)_\mathrm{stat}\left(^{+0.000}_{-0.001}\right)_\mathrm{sys}\,\mathrm{fm},\quad \quad
\Delta_\mathrm{Z}(\mu\textrm{H})=-7403^{+21}_{-16}\,\text{ppm}.
\eeq
On the other hand, one can determine this contribution from the experimental hfs, given predictions for the remaining theory contributions.
So far we have the measurements of the $1S$ hfs in H and the $2S$ hfs in $\mu$H. The corresponding extractions of the  
Zemach radius are shown in Table~\ref{Table:ZemachRadius} and compared with the form-factor determinations. Since baryon $\chi$PT (B$\chi$PT) gives a smaller prediction for the polarizability contribution than data-driven evaluations, it also gives a smaller Zemach radius. This discrepancy will be discussed below (cf.~Figure~\ref{fig:HFSpol}).

\begin{figure}[thb]
\centering
 \includegraphics[width=0.7\textwidth]{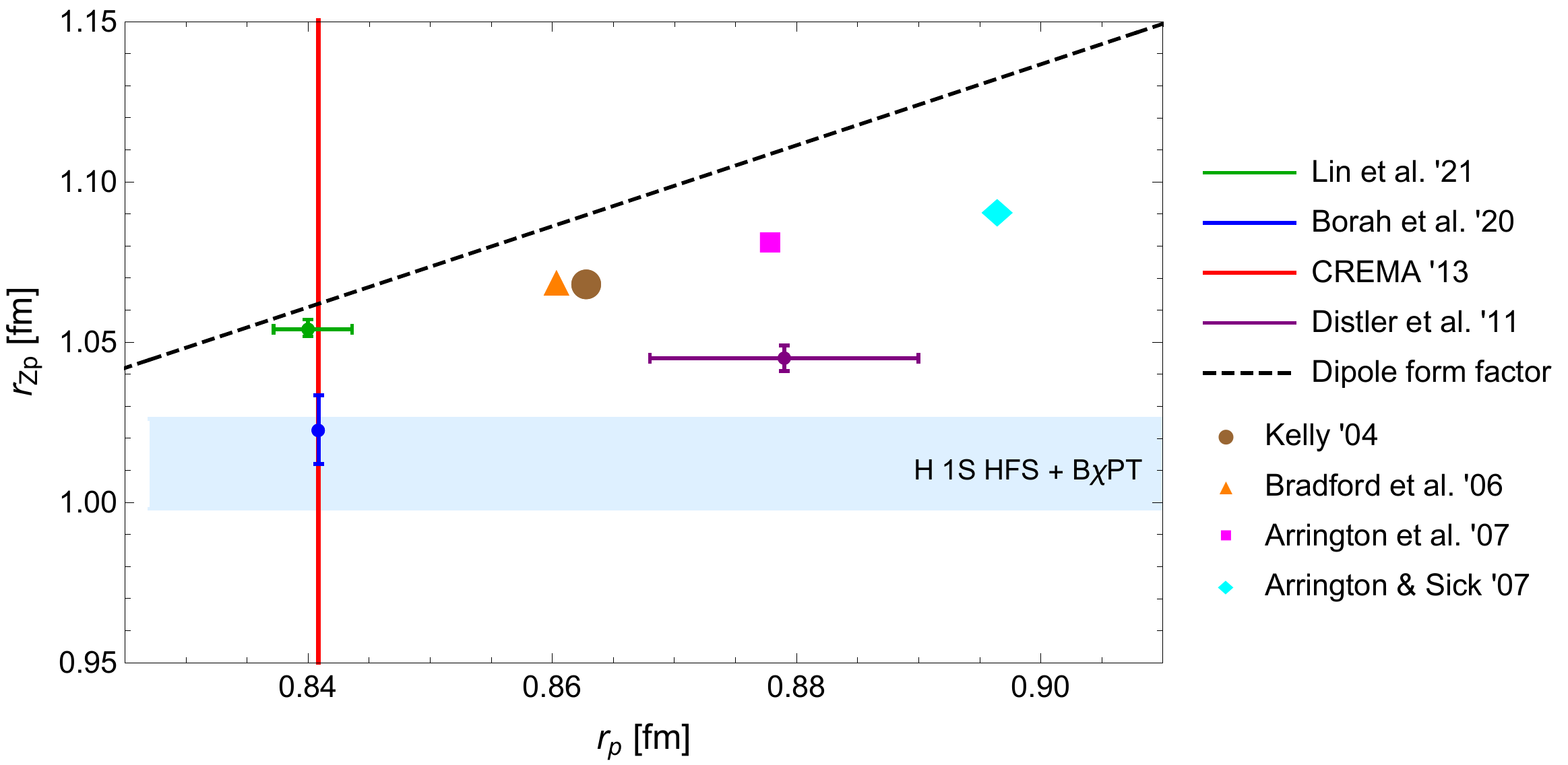}
\caption{Correlation between the Zemach and charge radius of the proton. The shown results are from: Lin et al.~\cite{Lin:2021xrc}, Borah et al.~\cite{Borah:2020gte}, CREMA \cite{Antognini:1900ns}, Distler et al.~\cite{Distler:2010zq}, Kelly \cite{Kelly:2004hm}, Bradford et al.~\cite{Bradford:2006yz}, Arrington et al.~\cite{Arrington:2007ux}, and Arrington \& Sick \cite{Arrington:2006hm}.
        \label{fig:RadiiCorrelation}}
\end{figure}

There is an appreciable linear correlation between the Zemach and charge radius, illustrated in Fig.~\ref{fig:RadiiCorrelation}.
The black dashed line represents the usual dipole approximation, $1/(1+Q^2/\La^2)^2$, for the form factors $G_E$ and $G_M$. This correlation is of course more general, given that the
proton size is set predominantly by one QCD scale, $\La_\mathrm{QCD}$.
Essentially all the empirical parametrizations
of the form factors, shown by data points,  follow this trend too. For comparison, we show our present determination of $r_{\mathrm{Z}p}$ from H (blue band) and $r_{p}$ from $\mu$H (solid red line). The upcoming $1S$ hfs measurement in $\mu$H
is expected to have a big impact on the precise determination of $r_{\mathrm{Z}p}$.

\subsubsection{Polarizability contribution and the spin structure functions}
The polarizability contribution is at least an order of magnitude smaller than the Zemach term, 
but produces a relatively large uncertainty. Here we look at it in more detail. This contribution is usually split into terms, in correspondence with the two spin structure functions, $g_1$ and $g_2$:
\begin{subequations}
\eqlab{POL}
\bea
\Delta_\mathrm{pol}&=&\Delta_1+\Delta_2\equiv \frac{Z\al m}{2\pi (1+\kappa_N) M}\left[\delta_1+\delta_2\right],\\
\delta_1
&=&18\int\limits_0^\infty\frac{\dd Q}{Q} \upkappa_0(Q^2) \, I^\mathrm{(pol)}_1(Q^2)
+16 M^4\int\limits_0^\infty\frac{\dd Q}{Q^3} \int\limits_0^{x_0}\dd x\, \upkappa_1(x,Q^2) \, g_1(x,Q^2),\eqlab{Delta1b}\\
\delta_2&=&96M^2\int\limits_0^\infty\frac{\dd Q}{Q^3}\int\limits_0^{x_0}\dd x\,\upkappa_2(x,Q^2)\, g_2(x,Q^2) , \eqlab{Delta2}
\eea
\end{subequations}
where $x_0$ is the inelastic threshold, which usually is associated with pion production. The kinematical 
functions, $\upkappa_i$, have a particularly simple form for H, since one may neglect the electron mass, 
\beq 
\upkappa_0(Q^2) =1,\quad  \upkappa_1(x,Q^2) = \upgamma( Q^2/x^2 ) \Big[4 +\upgamma(Q^2/x^2 )\Big] - \mbox{$\frac{9}{4}$}  , \quad\upkappa_2(x,Q^2) = \upgamma( Q^2/x^2 ) -\half, \quad  
\eeq
with $\upgamma( t ) \equiv 
\left( 1+ \sqrt{1+\frac{4M^2}{t} } \right)^{-1}$. For the more general form see Eq.~37 of the Supplement. Note that only the recoil corrections to the Zemach term are contained in $\Delta_\mathrm{recoil}$, whereas the polarizability contribution includes the corresponding recoil effects in itself. 

The quantity which stands out in the evaluation of $\Delta_1$ is $I^\mathrm{(pol)}_1(Q^2)$, which is
the polarizability (i.e., non-Born) part of the first moment of $g_1$, 
\beq
I_1^\mathrm{(pol)}(Q^2)=I_1(Q^2)+\quarter F_2^2(Q^2), \quad I_1(Q^2)\equiv \frac{2M^2}{Q^2}\int_0^{x_0}\dd x\, g_1(x,Q^2),
\eqlab{genGDHnonpole}
\eeq
where the Pauli form factor, $F_2(Q^2)$, comes from the non-pole piece of the Born term. 
There is a large cancellation between the two terms 
in $I_1^\mathrm{(pol)}$, which is hard to achieve precisely in empirical evaluations.
In fact, at the real-photon point they cancel exactly, $I_1^\mathrm{(pol)}(0) =0$, 
as a consequence of the GDH sum rule \cite{Drell:1966jv,Gerasimov:1965et}:
$I_1(0)=-\quarter \kappa_N^2$. There is also a sum rule for the slope, $I_1^\mathrm{\prime (pol)}(0)$, 
relating it to the nucleon spin polarizabilities \cite{Pascalutsa:2014zna,Lensky:2017dlc}. However, 
in the data-driven evaluations these relations are only satisfied approximately. In the future, it would be desirable  to develop
the empirical parametrizations of structure functions with built-in constraints from various sum rules.

\begin{table}
\caption{Polarizability contribution to the  hfs of H and $\mu$H, in ppm.}
\label{Table:Summary2PEHFS}
\begin{tabular}{l|c|cc|c|cc}
\hline
\hline
&\multicolumn{3}{c|}{H}&\multicolumn{3}{c}{$\mu$H}\\
\cline{2-7}
Reference &    $\Delta_\mathrm{pol}$& $\Delta_1$& $\Delta_2$&$\Delta_\mathrm{pol}$& $\Delta_1$& $\Delta_2$  \\
\hline
\cellcolor{gray!10}{\textsc{Data-driven disp. eval.}}&&&&&&\\
\cite{Faustov:2006ve} Faustov \textit{et al.}\ '06 &$2.2(8)$&$2.6$&$-0.4$&$470(104)$&$518$&$-48$\\
\cite{Carlson:2008ke,Carlson:2011af} Carlson \textit{et al.}\ '11  &$1.88(64)$&$2.00(63) $&$-0.13(13) $& $351(114)$& $370(112)$& $-19(19)$\\
\cite{Tomalak:2018uhr} Tomalak '18&$1.91(54)$&&&$364(89)$&$429(84)$&$-65(20)$\\
\cite{Zielinski:2017gwp} Zielinski '17&$1.51$&$1.95(95)$&$-0.44$&&&\\
\hline
\cellcolor{gray!10}{\textsc{leading-order B$\chi$PT}} &&&&&&\\
\cite{Hagelstein:2015lph} Hagelstein \textit{et al.}\ '16 & $0.12(55)$&$0.05(52)$&$0.07(17)$&$37(95)$&$29(90)$&$9(29)$\\
\cellcolor{gray!10}{\textsc{$+\,\Delta$(1232) excit.}} &&&&&&\\
\cite{Hagelstein:2018bdi} Hagelstein \textit{et al.}~'18 &$-0.16$&$0.48$&$-0.64$&$-13$&$84$&$-97$\\
\hline
\hline
\end{tabular}
\end{table}

\begin{figure}[b]
       \includegraphics[width=0.7\textwidth]{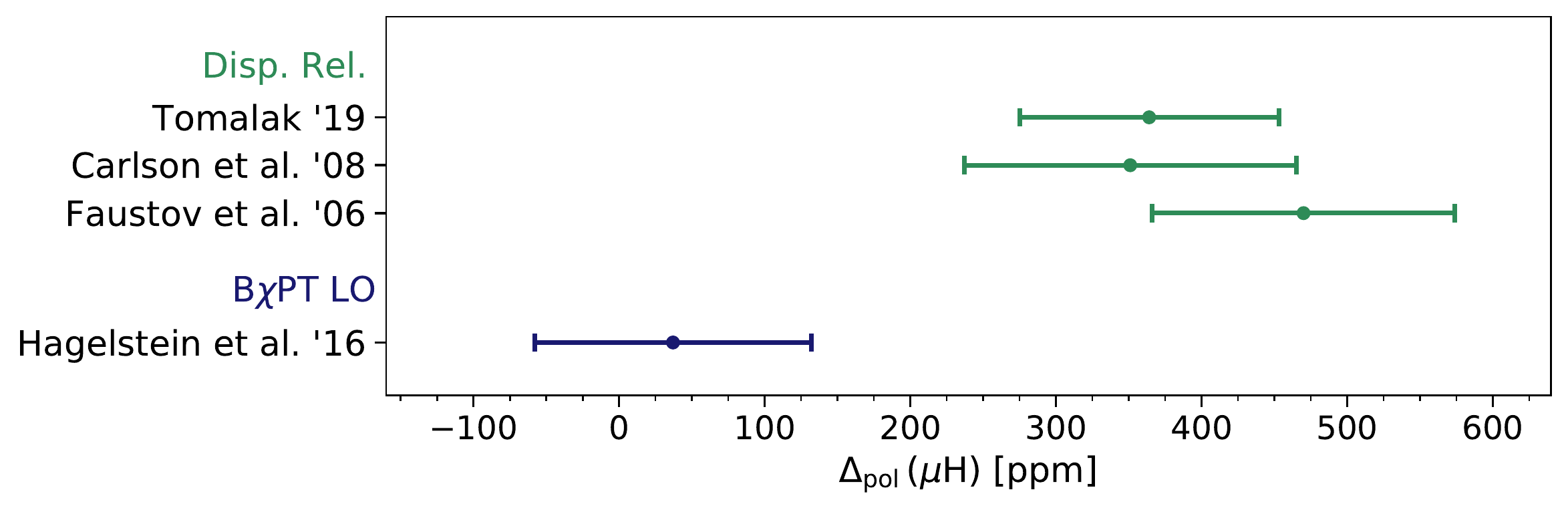}
\caption{The polarizability contribution to the hfs in 
$\mu$H. For the
corresponding values and references, see Table~\ref{Table:Summary2PEHFS}. \label{fig:HFSpol}}
\end{figure}

The present data-driven evaluations also suffer from the poor knowledge of $g_2$.
The data are scarce in the entire kinematic region relevant to $\Delta_2$ \cite{CLAS:2017qga}.  The 
data from the JLab g2p experiment \cite{Slifer:2009ik,Zielinski:2016khh,JeffersonLabHallAg2p:2022qap} may soon improve this situation. Their preliminary data 
have been used by Zielinski \cite{Zielinski:2017gwp} to estimate the effect of $\De_2$ in H, see Table~\ref{Table:Summary2PEHFS}.
In the table, we also show the result of leading-order (LO) B$\chi$PT \cite{Hagelstein:2015lph}, which finds a relatively small polarizability effect. The uncertainty of the LO calculation is estimated as $30\,\%$ [$\simeq (M_\Delta-M_p)/\mathrm{GeV}$] for the contributions from the longitudinal-transverse and helicity-difference cross sections, $\sigma_{LT}$ and $\sigma_{TT}$, respectively, see Eq.~20 of the Supplement for their relation to the spin structure functions. An inclusion of the $\De(1232)$-resonance excitation \cite{Hagelstein:2018bdi} does not change this situation. It increases the effect in the individual $\De_1$ and $\De_2$
contributions, but cancels out from the total $\De_\mathrm{pol}$, as can be seen from comparing the last two rows of the Table. A complete next-to-leading-order B$\chi$PT calculation, as is done for Compton scattering observables \cite{Lensky:2009uv,Lensky:2015awa},  is needed here to elucidate this result and reduce the uncertainty.

Figure \ref{fig:HFSpol} provides a graphic illustration of the present discrepancy between the data-driven evaluations and the LO B$\chi$PT. 
The upcoming $1S$ hfs measurement in $\mu$H will be able to address this discrepancy because, combined with the H, it allows for
a separate assessment of the Zemach and polarizability contributions. More details on this separation are given in the following section.

\section{Theory updates and  future $\mu$H experiments}
\label{Sec4}

\subsection{Lamb shift in $\mu$H}\label{Sec4LS}
The  two CREMA measurements of $2S_{1/2}^{F=1}-2P_{3/2}^{F=2}$ and $2S_{1/2}^{F=0}-2P_{3/2}^{F=1}$ transitions \cite{Antognini:1900ns}  allowed
for a determination  of the $2S$ hfs, discussed further-on, and the Lamb shift:
\beq
  \eqlab{eq:LSmeas_p}
  E_{2P-2S}^\mathrm{exp}(\mu{\rm H}) = 202.3706\,(19)\,_{\rm stat}~(12)\,_{\rm syst} ~{\rm meV}
  ~ = ~ 202.3706\,(23)\,_{\rm total} ~{\rm meV}.
\eeq

On the theory side, the updated summary for the $\mu$H Lamb shift (taking into account the latest results from Refs.~\citenum{Korzinin:2013uia,Karshenboim:2015,Karshenboim:2018iyl,Karshenboim:2021jsc}) is given in \Eqref{LStheory}. The most important improvement comes from the NLO calculation of the hVP \cite{Karshenboim:2021jsc}. The accuracy is still limited by the \2PE exchange, finite-size effects and the hVP.
\begin{textbox}[h]
\section{The Lamb shift of $\mu$H (theory update):}
\beq
  E_{2P-2S}(\mu{\rm H})    =\Big[\underbrace{205.0074}_\text{Uehling}\;\underbrace{+1.0153}_\text{$r_p$ indep.}\; \underbrace{+0.0114(3)}_\text{hVP}\;\underbrace{+0.0006(1)-5.2275(10)\left(\frac{r_p}{\text{fm}}\right)^2}_\text{f.s.\ corr.}\;\underbrace{- E_{2S}^{\langle\mTPE\rangle}}_\text{$2\upgamma$ exchange} \Big]\,  \mbox{meV},\eqlab{LStheory}
\eeq
\end{textbox}

Using the best data-driven evaluation of the \2PE-exchange~\cite{Birse:2012eb}, $\Delta\mathrm{E}_{2S}^{\langle 2\upgamma\rangle}=-33(2)\,\upmu$eV, we obtain:
\beq
r_p(\mu\text{H})=0.84099(12)_\text{sys}(23)_\text{stat}(3)_\text{hVP}(8)_\text{f.s.}(23)_\text{2$\upgamma$}\,\text{fm}=0.84099(36)\,\text{fm}. \eqlab{rpNEW}
\eeq
The uncertainty of the radius is limited in equal parts by the precision of the $2S$-$2P$  measurements and  the prediction of the \2PE-exchange contribution,
with the measurement accuracy  limited by statistics. 
The systematic uncertainty of 300~MHz  is mainly given by the frequency uncertainty  of the laser pulses delivered by the  Raman cell, the last stage of the laser system used to generate the pulses at 6~$\upmu$m. The typical atomic physics systematics such as Stark, collisional and Zeeman shifts  are strongly suppressed in the  tightly-bound $\mu$H atom.

\begin{marginnote}[]
The CREMA setup can be upgraded to improve  the $\mu$H($2S$-$2P$) measurements by a factor 5.
\end{marginnote}
An upgrade of the CREMA-2010 setup~\cite{Pohl:2010zza} holds the potential
of improving the $2S$-$2P$ measurements by at least a  factor of 5,
reachable by increasing the statistics by 25 and reducing 
the systematics by 3.
The statistical improvement could be achieved mainly by having a longer data-taking time
(from 1 week to 5 weeks), and by increasing the laser pulse energy (from 0.2 mJ to 1 mJ), accompanied by slight overall
improvements of the setup, including X-ray detection efficiency, muon beam rate,
multi-pass cavity performance and laser repetition rate. The systematic uncertainty
  could be reduced by using novel optical parametric
down-conversion technologies under  development for the
measurement of the hfs in $\mu$H. This technology, capable of delivering pulses with few mJ energy and a bandwidth  smaller than 100~MHz in the  6~$\upmu$m region, enables increasing both the laser
pulse energy and the frequency control.

\begin{figure}[h]
\hsize38pc
\centering   
\includegraphics[width=1.2\linewidth]{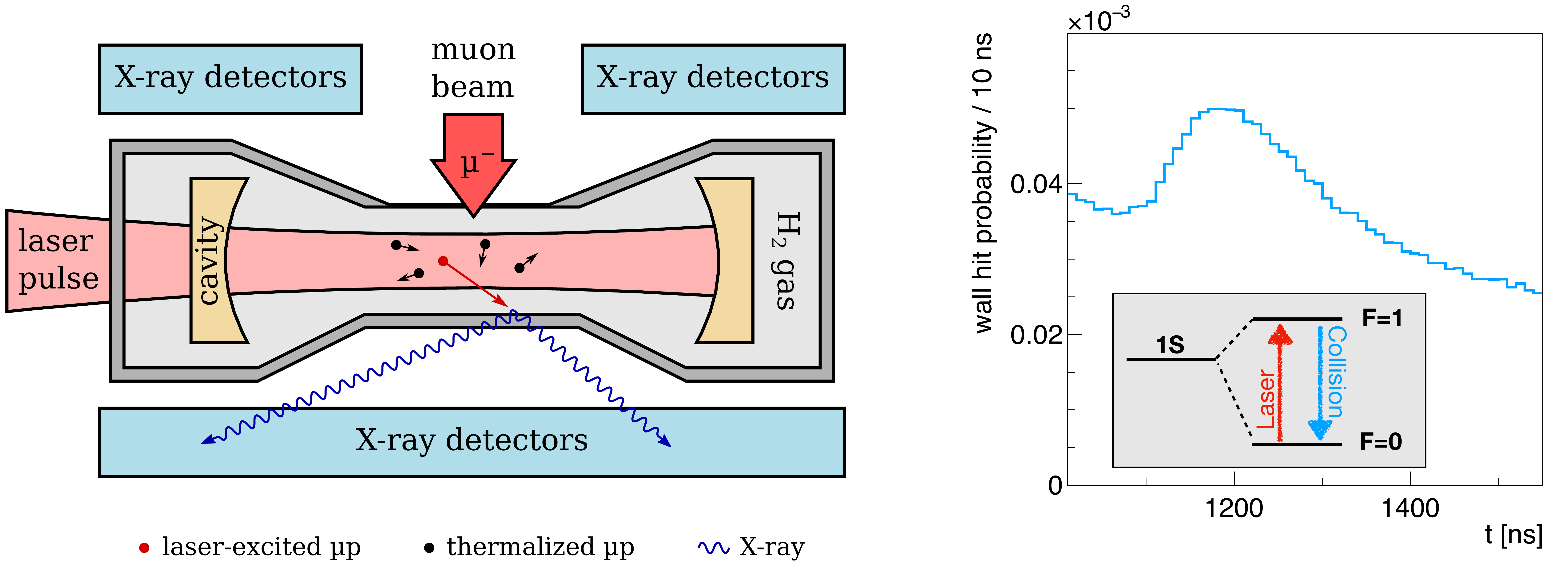}
\caption{Setup, principle and level scheme of the CREMA hfs experiment.
(Left) 
The setup in which the muon beam is stopped in a hydrogen-gas target and the formed $\mu$H atoms are excited by the laser pulse.   A successful excitation of the hfs transition leads  to a $\mu$H atom with extra
  kinetic energy that efficiently
 diffuses to one of the target walls where X-rays are produced. 
(Right) Probability (normalized to the number of entering muons) that a $\mu$H is reaching the target walls versus time at typical target conditions and laser performance. The laser excitation occurs at 1.0~$\upmu$s. The laser induced events are clearly visible.}
        \label{fig:target-region}
\end{figure}

\begin{textbox}[h]\section{Principle of the CREMA hfs experiment}
\begin{footnotesize}

The hfs experiment by the CREMA Collaboration  follows the  sequence
illustrated in Fig.\ \ref{fig:target-region}. A negative muon  of  11~MeV/c momentum passes an entrance detector triggering the laser system and is
stopped in a H$_2$ gas target ($\sim 1$~mm thickness, $0.5$~bar
pressure, 20~K temperature), wherein a $\mu$H atom is formed. While the laser pulse is being generated, the $\mu$H atom is de-exciting to the $F=0$ sublevel (see inset in Fig.\ \ref{fig:target-region}) of the $1S$-state and thermalizing to the H$_2$ gas temperature.
After $1~\upmu$s, the $\mu$H is thermalized and the
generated laser pulse of few-mJ energy at a wavelength of 6.8~$\upmu$m (equivalent to a
frequency of 44~THz and an energy of 0.18~eV) is coupled into a multi-pass
cavity surrounding the muon stopping region.
The multiple reflections occurring in this toroidal cavity allow the
illumination of a disk-shaped volume with a diameter of 15~mm and a
thickness of 0.5~mm with a laser fluence of $O(10)$~J/cm$^2$. 
The on-resonance laser pulse excites the muonic atom from the singlet
$F=0$ to the triplet $F=1$ sublevels.
Within a short time, an inelastic collisions between the $\mu$H atom
and one H$_2$ molecule of the gas target de-excites the $\mu$H atom
from the triplet back to the singlet sublevels.
In this process, the hfs transition energy 
is converted into kinetic energy: on average the $\mu$H atom acquires 0.1~eV kinetic
energy, the rest goes to the H$_2$ molecule.
With this extra kinetic energy, which is much larger than the thermal
energy, the $\mu$H atoms start diffusing in the H$_2$ gas reaching  the target walls $100-400$ ns after laser excitation, as shown by the peak in Fig.~\ref{fig:target-region} (right).
At the gold-coated target walls the muon is transferred from  $\mu$H
to the nucleus, forming muonic gold ($\mu$Au$^*$) in highly
excited states.
The $\mu$Au$^*$ de-excitation produces various X-rays of MeV energy
which are used as signature of a successful laser-induced transition, so that
the hfs resonance can be exposed by counting the number of $\mu$Au
cascade events after laser excitation as a function of the laser frequency.
\end{footnotesize}
\end{textbox}

\subsection{Hyperfine splitting in $\mu$H}\label{Sec4HFS}

The improved $2S-2P$ measurements discussed above will also improve the precision of the $2S$ hfs measurement. 
However, a new level of precision will be reached in the upcoming CREMA measurement of $1S$ hfs \cite{Amaro:2021goz}. The schematics of this experiment
are shown in Fig.~\ref{fig:target-region} explained in the insert.  On the theory side, the updated summary for the hfs in $\mu$H is given in Equation 40.
Compared with a previous compilation by Peset et al.~\citenum{Peset:2021iul}, we have included  hVP \cite{Faustov:1997rc}, weak \cite{Eides:2000xc}, and two-loop eVP corrections in 2\textsuperscript{nd} and 3\textsuperscript{rd}-order perturbation theory \cite{Karshenboim2009}, as well as some higher-order radiative corrections \cite{Brodsky:1966vn}. For the radius-independent term, we are keeping the error estimate from Refs.~\citenum{Peset:2016wjq}, which does  take into account missing higher-order recoil corrections. The radiative corrections to the \2PE exchange are discussed in Sec.~5 of the Supplement. 

\begin{textbox}[htb]
\section{The hyperfine splitting of $\mu$H (theory update):}

\bea
E_{1S\text{-hfs}} &=&  \Big[\underbrace{182.443}_{E_\mathrm{F}}\; \underbrace{+1.350(7)}_\text{QED+weak}\;\underbrace{+0.004}_\text{hVP}
\;\underbrace{-1.30653(17)\left(\frac{r_{\mathrm{Z}p}}{\text{fm}}\right)+ E_\mathrm{F}\,\Big(1.01656(4)\,\Delta_\mathrm{recoil}+1.00402\,\Delta_\mathrm{pol}\Big)}_\text{\2PE incl.\ radiative corr.}\Big]\,\text{meV}, \eqlab{HFStheory1S}\\
E_{2S\text{-hfs}}&=&  \Big[\underbrace{22.8054}_{\frac18 E_\mathrm{F}} \;\underbrace{+0.1524(8)}_\text{QED+weak}\;\underbrace{+0.0006(1)}_\text{hVP}
\;\underbrace{-0.16319(2)\left(\frac{r_{\mathrm{Z}p}}{\text{fm}}\right) +\mbox{$\frac{1}{8}$}E_\mathrm{F}\,\Big(1.01580(4)\,\Delta_\mathrm{recoil}+1.00326\,\Delta_\mathrm{pol}\Big)}_\text{\2PE incl.\ radiative corr.}\Big]\,\text{meV}.\eqlab{HFStheory2S}\nn
\eea
\end{textbox}

Once a high-precision measurement of the $1S$ hfs in $\mu$H is available, it can be used together with H to accurately disentangle
the Zemach and polarizability contributions, $\Delta_{\mathrm{Z}}$ and $\Delta_\mathrm{pol}$, with unprecedented precision. This is possible because the 
eVP corrections to the \2PE exchange differ between H and $\mu$H, cf.\ Eqs.~\eref{HFStheory1S} and \eref{HFSHtheory}.  Anticipating 1 ppm accuracy for the $\mu$H $1S$ hfs experiment,  the Zemach radius will be determined with $5\times 10^{-3}$ relative uncertainty and $\Delta_\mathrm{pol}(\mu\text{H})$ with $40$ ppm absolute uncertainty. It will thus lead to the best empirical determination of the proton Zemach radius from spectroscopy, without the uncertainty
associated with the polarizability contribution. 
\begin{marginnote}[]
Leveraging radiative corrections allows to disentangle the Zemach radius from H and $\mu$H hfs.
\end{marginnote}

\subsection{Pinning down the $1S$  hyperfine splitting in $\mu$H} 
\label{Sec-scaling}

\begin{figure}[t]
\hsize38pc
\begin{minipage}{0.35\textwidth}
 \includegraphics[width=\textwidth]{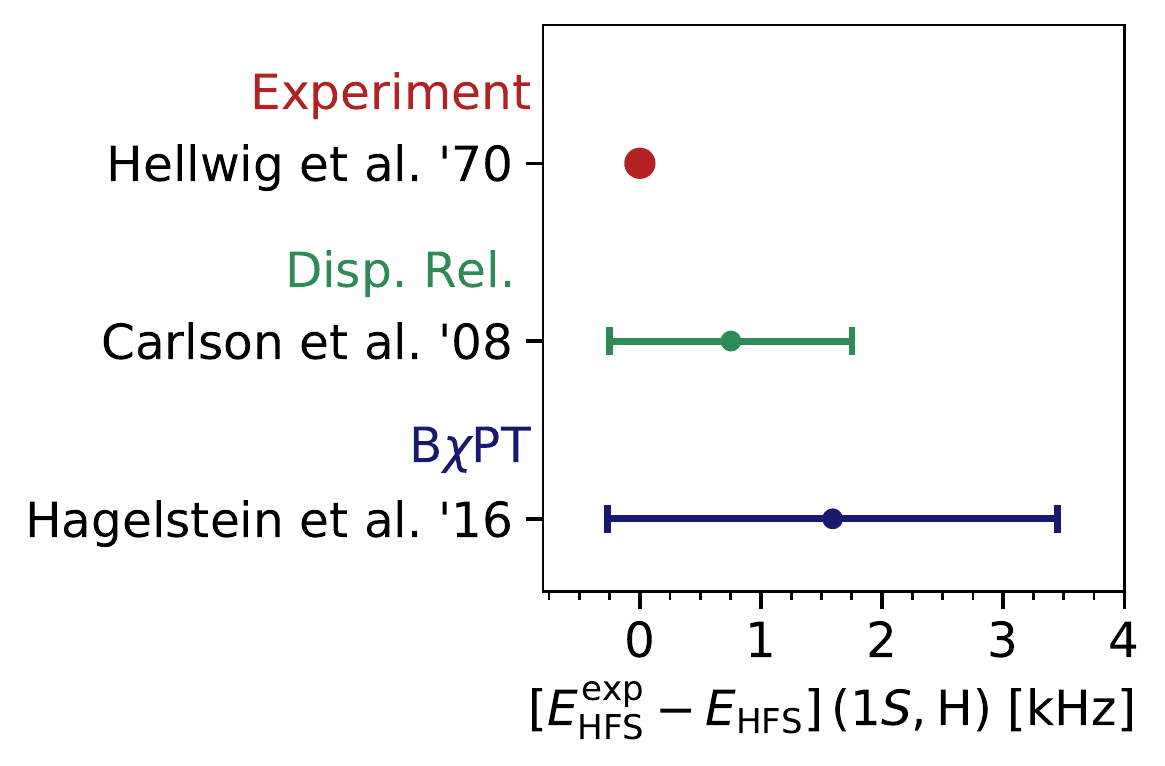}
    \end{minipage}\hfill
\begin{minipage}{0.43\textwidth}
\includegraphics[width=\textwidth]{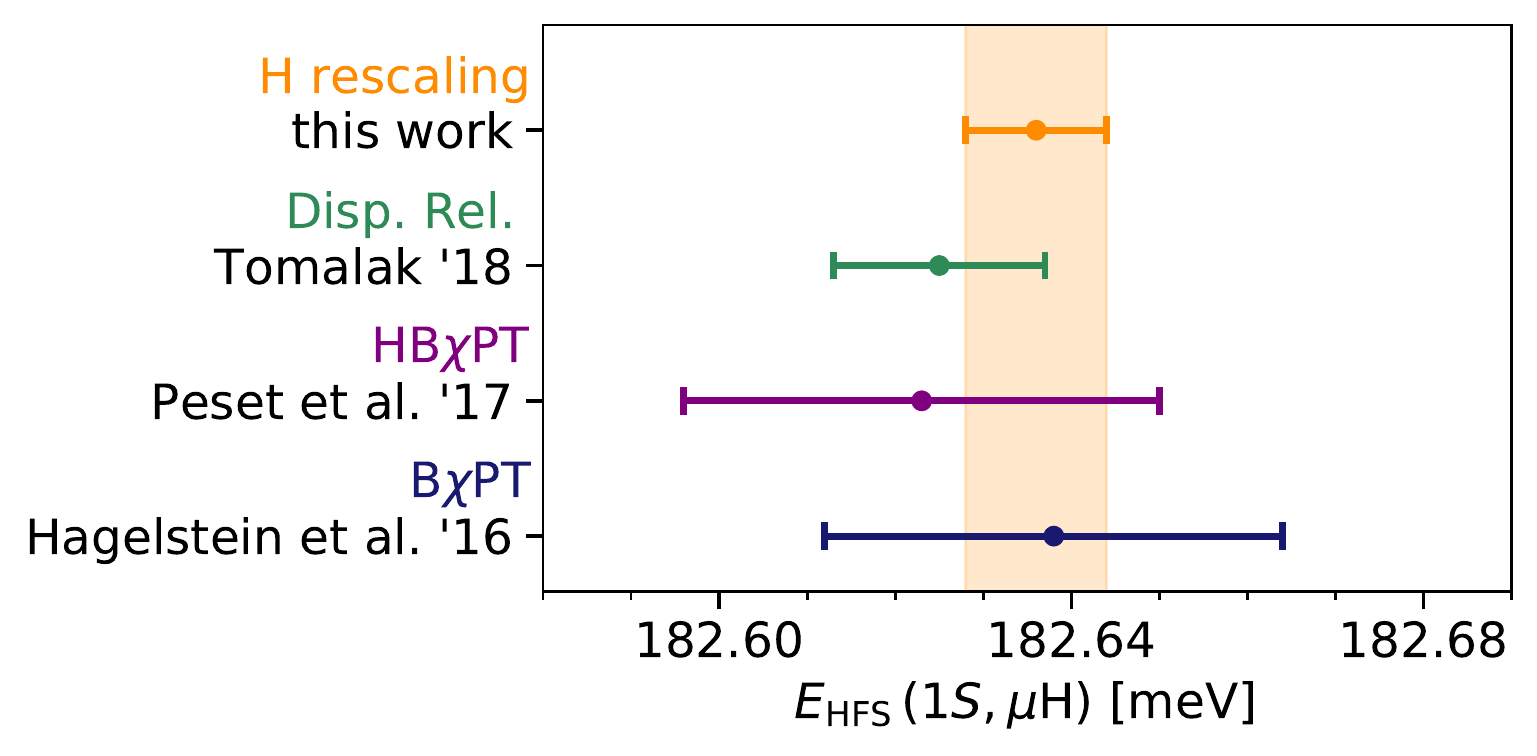}
\end{minipage}\hfill
\begin{minipage}{0.45\textwidth}
\includegraphics[width=\textwidth]{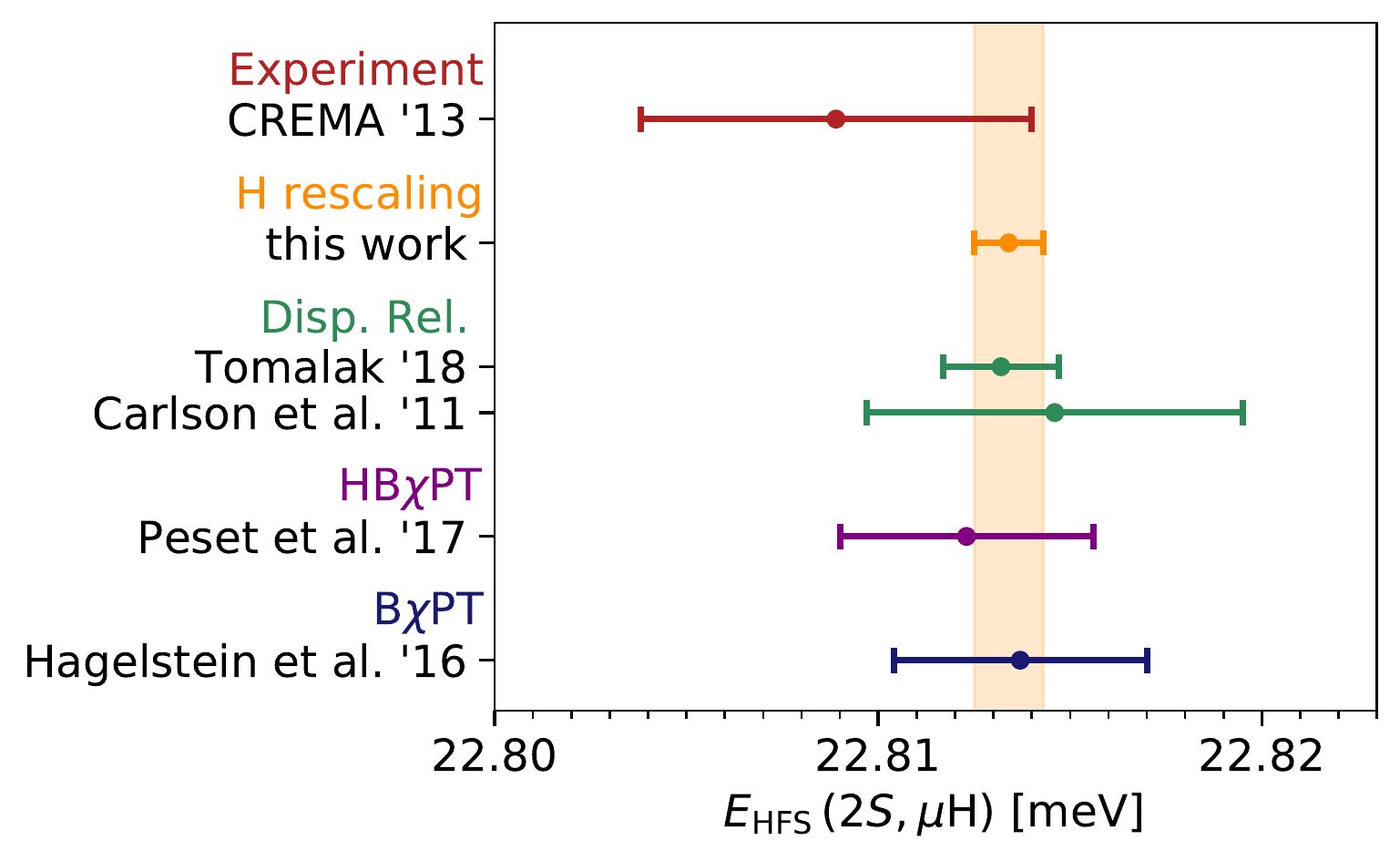}
\end{minipage}
\caption{Experimental values and theoretical predictions for the $1S$ and $2S$ hfs in H and $\mu$H \cite{Hellwig1970,Carlson:2008ke,Carlson:2011af,Hagelstein:2015lph,Antognini:1900ns,Tomalak:2018uhr,Peset:2016wjq}.}
        \label{fig:HFSSummary}
\end{figure}

The success of the $1S$ $\mu$H hfs experiments relies critically on the precision and accuracy of the theory prediction. The CREMA Collaboration is expecting 2 hours of data taking time per frequency point to observe an excess of events over background. The $1S$ hfs resonance would need to be searched in a more than $40$~GHz wide frequency range to be compared with a  linewidth of about $200$~MHz at FWHM resulting from Doppler broadening ($60$ MHz), laser bandwidth ($100$ MHz) and collisional effects.  We estimate the search range to cover a $\pm 3\sigma$ band over the present spread of \2PE-exchange theory predictions, cf.\ Fig.\ \ref{fig:HFSSummary}. Given the limited access to the PSI accelerator facility, it is important to further narrow it down as much as possible. 

\begin{marginnote}[]
\entry{Fractional uncertainty  of a quantity $X$}{$\delta =\sigma_X/X$, with $\sigma_X$ the absolute uncertainty.}
\end{marginnote}
The $1S$ hfs in H has already been measured with a fractional accuracy of $\delta=7\times10^{-13}$  \cite{Hellwig1970,Karshenboim:2000rg}:
\bea
E^{\,\text{exp.}}_{1S\text{-hfs}}(\mathrm{H})&=&1\,420.405\,751\,768(1)\,\text{MHz}.\eqlab{expHHFS}
\eea
The corresponding theory prediction is compiled in \Eqref{HFSHtheory}. 
Compared to a previous compilation by Volotka~\cite{Volotka:2004zu}, we have recalculated the $\mu$VP correction  which agrees with Ref.~\citenum{Karshenboim:1996ew}.  We have updated also the hVP, rescaling the recent result obtained for muonium \cite{Karshenboim:2021jsc}. These $\mu$VP and hVP results are considerably larger (roughly by a factor of 3 and 5, respectively) than quoted in \cite{Volotka:2004zu}.

\begin{textbox}[h]
\section{The hyperfine splitting of H (theory update):}
\bea
E_{1S\text{-hfs}}(\text{H})&=&  \Big[\underbrace{1\,418\,840.082(9)}_{E_\mathrm{F}}\;\underbrace{+1\,612.673(3)}_\text{QED+weak}\;\underbrace{+0.274}_{\mu\text{VP}}\;\underbrace{+0.077}_\text{hVP} \nn\\
&&\underbrace{-54.430(7)\,\left(\frac{r_{\mathrm{Z}p}}{\text{fm}}\right)+E_\mathrm{F}\,\Big(0.99807(13)\,\Delta_\mathrm{recoil}+1.00002\,\Delta_\mathrm{pol}\Big)}_\text{\2PE incl.\ radiative corr.}\Big]\,\text{kHz} \eqlab{HFSHtheory}
\eea
\end{textbox}

In Refs.~\citenum{Peset:2016wjq} and \citenum{Tomalak:2017lxo}, this high-precision hfs measurement was already exploited to constrain the
\2PE-exchange contribution and its effect in  the hfs of $\mu$H. 
Here we shall use a somewhat different procedure, where all the uncertainty of rescaling from H to $\mu$H is limited to radiative corrections. 
Combining the empirical and theoretical values for the $1S$ hfs in H, Eqs.~\eref{expHHFS} and \eref{HFSHtheory}, we deduce a subset of the \2PE-exchange contribution, containing the Zemach radius and polarizability corrections:
 \beq
E^\mathrm{\mathrm{Z}+\mathrm{pol} }_{1S\text{-hfs}}( \text{H})=E_\mathrm{F}(\text{H})\left[ b_{1S}(\text{H})\, \Delta_{\mathrm{Z}}(\text{H}) + c_{1S}(\text{H})\, \Delta_\mathrm{pol}(\text{H}) \right]
=-54.900(71)\,\mathrm{kHz},
\eeq
 where $b_{1S}$(H)$\simeq 1+2 \times 10^{-5}+0.01846-5\al/4\pi$ and $c_{1S}$(H)$\simeq 1+2 \times 10^{-5}$ are the radiative-correction factors shown explicitly in \Eqref{HFSHtheory}. The correction factors correspond to, respectively, the one-loop eVP correction to the wave function, see~\Eqref{WFcorrection},
 the one-loop eVP insertion in the elastic \2PE-exchange diagram, see Eqs.~43a of the Supplement, as well as self-energy and muon anomalous magnetic moment corrections to the Zemach-radius contribution, see Eq.~45 of the Supplement.
 We choose not to lump in here the recoil corrections to the Zemach term, because they are
 known rather precisely. We use \cite{Carlson:2008ke,Tomalak:2018uhr}: $\Delta_\mathrm{recoil}(\text{H})=5.33(5)$ ppm and $\Delta_\mathrm{recoil}(\mu\text{H})=846(6)$ ppm.

To go from H to $\mu$H, we assume that only the radiative factors scale non-trivially with the reduced mass, and that $\Delta_\mathrm{Z}$ and $\Delta_\mathrm{pol}$ scale  linearly: 
\beq 
\frac{\De_i (\text{H})}{m_r(\text{H})} =\frac{\De_i (\mu\text{H})}{m_r(\mu\text{H})}  , \quad \mbox{ $i = $ Z, pol.} 
\eeq 
This scaling is obvious for the Zemach contribution (cf.~Eqs.~\eref{Zcontr}), whereas 
for the polarizability contribution this has been verified numerically to better than $2\,\%$  \cite{Carlson:2008ke}.
Therefore, the sum of Zemach radius and polarizability corrections in $\mu$H, $E^\mathrm{\mathrm{Z}+\mathrm{pol} }_{nS\text{-hfs}}( \mu \text{H})$,  can be expressed via the one in the H $1S$ hfs, $E^\mathrm{\mathrm{Z}+\mathrm{pol} }_{1S\text{-hfs}}( \text{H})$, as follows: 
\bea
E^\mathrm{\mathrm{Z}+\mathrm{pol} }_{nS\text{-hfs}}( \mu \text{H}) & = &
\frac{E_\mathrm{F}(\mu\text{H})\,m_r(\mu\text{H})\,b_{nS}(\mu\text{H})}{n^3 E_\mathrm{F}(\text{H})\,m_r(\text{H})\,b_{1S}(\text{H})}E^\mathrm{\mathrm{Z}+\mathrm{pol} }_{1S\text{-hfs}}(\text{H})\nn\\
&-&\frac{E_\mathrm{F}(\mu\text{H})}{n^3}\,\Delta_\mathrm{pol}(\mu\text{H})\underbrace{\left[c_{1S}(\text{H}) \frac{b_{nS}(\mu\text{H})}{b_{1S}(\text{H})}-c_{nS}(\mu\text{H})\right]}_{\substack{=-6\times 10^{-5} \text{ for n=1}\\ =-5\times 10^{-5} \text{ for n=2}}}\eqlab{HFSpredformula}
\eea 
where $b_{1S}(\mu\text{H})\simeq 1+0.00402+0.01846-5\al/4\pi$, $b_{2S}(\mu\text{H})\simeq 1+0.00326+0.01846-5\al/4\pi$, $c_{1S}(\mu\text{H})\simeq 1+ 0.00402$, and $c_{2S}(\mu\text{H})\simeq 1+0.00326$  are the radiative-correction factors shown explicitly in \Eqref{HFStheory1S}. The second term in \Eqref{HFSpredformula} is 
negligible because the coefficient given by the square brackets is very small. We thus only evaluate the
first term and obtain:
\beq
E^\mathrm{\mathrm{Z}+\mathrm{pol} }_{1S\text{-hfs}}( \mu\text{H})=-1.318(2)\,\mathrm{meV},\quad 
E^\mathrm{\mathrm{Z}+\mathrm{pol} }_{2S\text{-hfs}}( \mu\text{H})=-0.1646(2)\,\mathrm{meV}.\eqlab{HFSpredformulaSet}
\eeq 

 The main source of uncertainty here is the \2PE recoil contribution $\Delta_\mathrm{recoil}(\text{H})$. 
Adding the \2PE recoil contribution $\Delta_\mathrm{recoil}(\mu\text{H})$ to \Eqref{HFSpredformulaSet}, we obtain a prediction for the full \2PE-exchange  contribution to the hfs in $\mu$H:
\beq
E^{\langle \mathrm{2\upgamma}\rangle}_{1S\text{-hfs}}(\mu\text{H})=-1.161(2)\,\mathrm{meV},\quad
E^{\langle \mathrm{2\upgamma}\rangle}_{2S\text{-hfs}}(\mu\text{H})=-0.1450(2)\,\mathrm{meV}.\eqlab{h2PEemp}
\eeq

With this, we arrive at a complete prediction of the hfs in $\mu$H:
\beq
E_{1S\text{-hfs}}(\mu\text{H})=182.636(8)\,\mathrm{meV}, \quad 
E_{2S\text{-hfs}}(\mu\text{H})=22.8134(9)\,\mathrm{meV},
\eqlab{HFSempPrediction}
\eeq 
where we have also included an uncertainty due to possible scaling violation of $\Delta_\mathrm{pol}$ at the level of
$2\,\%$ (assuming a very generous size for this contribution, $\Delta_\mathrm{pol}(\mu\text{H}) =400\,\text{ppm}$).
 Our result is shown in Fig.~\ref{fig:HFSSummary}, together with the existing $\mu$H $2S$ hfs measurement. The theory predictions based on the empirical hfs in H, \Eqref{HFSempPrediction}, are up to a factor 5 better than results that do not use the H hfs.


Note that all theory predictions shown in Fig.~\ref{fig:HFSSummary} are in agreement, even though the data-driven dispersive evaluations and the B$\chi$PT prediction disagree in the polarizability contribution  (cf.\ Fig.~\ref{fig:HFSpol}, Table \ref{Table:Summary2PEHFS}). This is because most works use the experimental H hfs to refine their prediction for the total \2PE-exchange effect. Hence the discrepancy in polarizability is compensated 
by slightly different Zemach radii. 

In future, reversing the above procedure to obtain a prediction of the \2PE-exchange contribution to the $1S$ hfs in H from a measurement of the $1S$ hfs in $\mu$H, might allow for a benchmark test of the H hfs theory. 
This, however, would also require further improvements for the recoil corrections from \2PE exchange, as well as for the uncertainty from missing contributions in the $\mu$H theory.  Note that a slightly better benchmark test ($\delta\sim 2\times 10^{-9}$) of bound-state
QED for a hyperfine transitions can be achieved for the muonium hfs, which the 
MuSEUM experiment~\cite{Kanda:2020mmc} aims to measure with
$\delta\sim2\times 10^{-9}$ relative accuracy. To test the muonium hfs on this level, the MuMass experiment~\cite{Ohayon:2021dec,Janka:2021xxr} has to determine the
$m_\mu/m_e$ ratio  to better than $\delta\sim1\times
10^{-9}$ from the $1S$-$2S$ transition in muonium.

\section{Bound-state QED tests of simple atomic and molecular systems}\label{Sec5}

The simplicity of two- and three-body atomic-molecular  systems combined with the precision of laser spectroscopy  permit unique confrontations between theory and experiments.
The predictive power of bound-state QED, however, depends on the
knowledge of fundamental constants such as the masses of the
involved particles, $\alpha$, $R_\infty$,
and nuclear properties such as the nuclear charge radii or magnetic
moments.

\begin{figure}[t]
              \includegraphics[width=\textwidth]{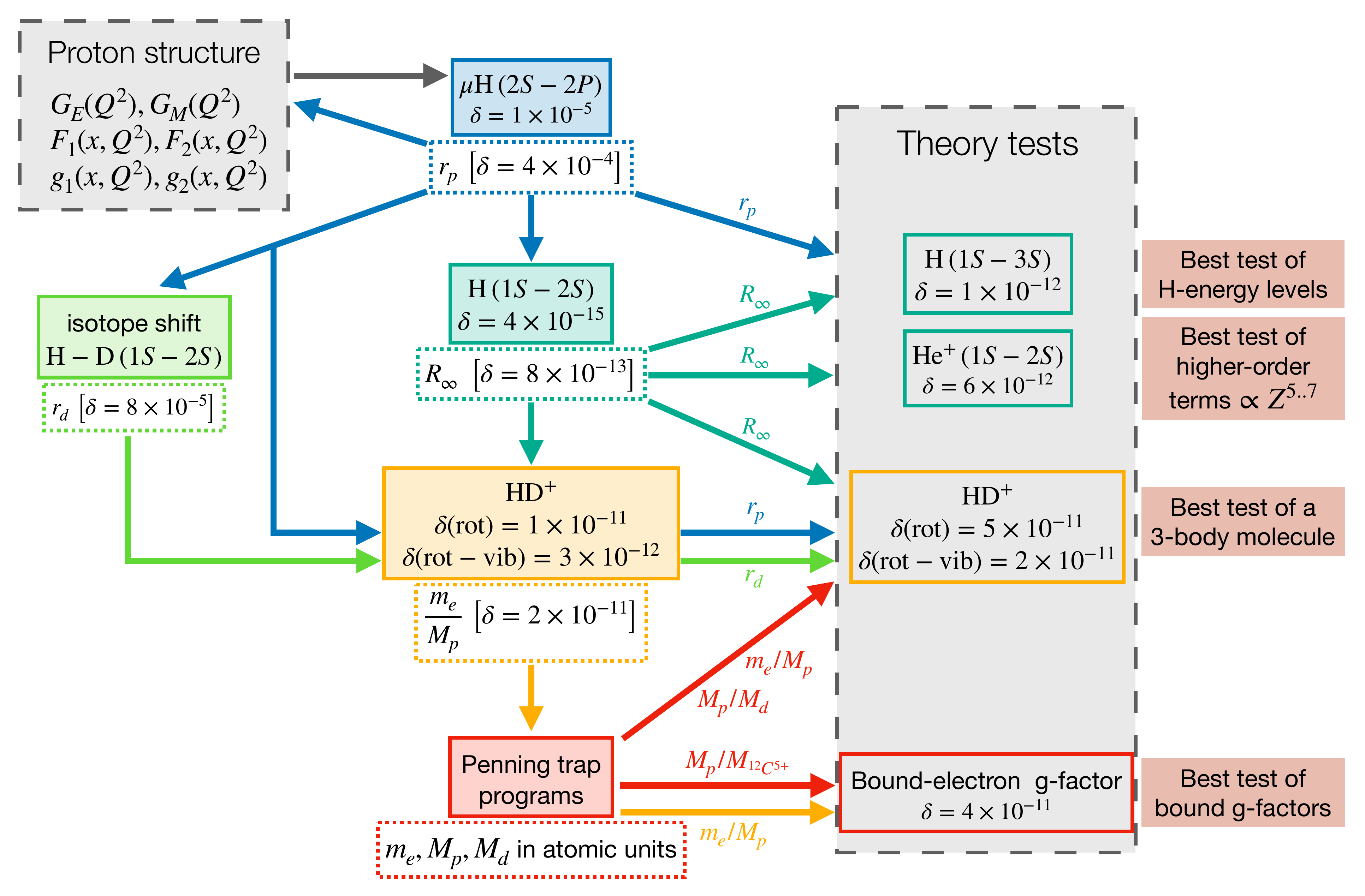}
\caption{Simplified scheme showing the impact of $r_p(\mu\text{H})$ on improving fundamental constants and bound-state QED tests.}
        \label{fig:Flow-chart}
\end{figure}

While the $\mu$H and $\mu$D measurements have been taken into
account in the CODATA-2018  adjustment of the 
fundamental constants  yielding   
$r_p=0.8414(19)\,\mathrm{fm}$~\cite{Tiesinga:2021myr}, 
its uncertainty is  $5$ times larger than the uncertainty from the muonic measurement alone \cite{Antognini:1900ns}, cf.\ Eq.~\eref{rpNEW}. Hence, the  $r_p$ value from CODATA-2018 does not completely reflect the potential of the $\mu\text{H}(2S-2P)$  measurements.
We thus sketch in the following the impact of $r_p(\mu\text{H})$  
by combining it with some selected measurements and corresponding theory predictions in simple systems   with distinctive precision and sensitivity. 
Figure~\ref{fig:Flow-chart} illustrates the impact of the $\mu$H spectroscopy and its connection to H, HD$^+$ and Penning trap measurements that leads to cutting edge tests of bound-state QED for H-like systems,  simple molecular systems, and  bound-electron  g-factors while improving on fundamental constant such as the $r_p$, $r_d$, $R_\infty$, $m_e$ and $M_p$. 
Throughout this section we use the SI units.

\subsection{$\mu$H to H: testing the H energy levels and extracting $R_\infty$ }

Even though the recent H($2S$-$8D$) measurement~\cite{Brandt:2021yor} is at some tension with the $\mu$H results, here we exploit the  agreement between the $r_p$ values from  H~\cite{Bezginov:2019mdi,Grinin:2020:Science_H1S3S, Beyer:2017:Science_2S4P} and $\mu$H to illustrate the potential of combining $\mu$H and H measurements  for testing the H energy levels and 
improving on $R_\infty$, the most precisely known fundamental constant and a major player in the adjustment of fundamental constants.
$R_\infty$ also sets the energy scale for atoms, ions and molecules, so that
precise predictions of transition frequencies in these systems  require its
precise value.
\begin{marginnote}[]
\entry{Rydberg constant in SI units}{$  R_\infty=\frac{\alpha^2 m_e c}{2h}$}
\end{marginnote}

In a simplified form, the H energy levels with principal quantum numbers
$n$ and angular momentum $l$ can be expressed as
\begin{equation}
    f_{nl}^\mathrm{th} \approx -\frac{R_\infty c}{n^2}\frac{1}{1+\frac{m_e}{M_p}}+ \frac{\mathrm{QED}_{nl}}{n^3} + \delta_{l0}\,\frac{\al\,c^4}{3\pi\,a_\mathrm{B}^3\,\hbar^3}\frac{r_p^2}{n^3} +\dots \; \: .
\end{equation}
The first term accounts for the Bohr structure 
corrected for the finite  proton mass $M_p$.
The second term QED$_{nl}$, scaling dominantly as $1/n^3$, accounts for radiative, relativistic and higher-order recoil effects while the third term is the finite-size effect.
The relevant unknowns in this equation are thus $R_\infty$ and $r_p$: in comparison the uncertainties of all the other constants involved can be neglected.

Hence, to determine both $R_\infty$ and $r_p$ two transition frequency measurements in  H are  needed  while QED$_{nl}$ are taken from theory.
When combining the two most precise measurements in H, the H($1S$-$2S$) transition with precision 
$\delta(1S-2S)=4.2\times 10^{-15}$~\cite{Parthey:2011lfa} and the H($1S$-$3S$) transition with precision
$\delta(1S-3S)=2.5\times 10^{-13}$~\cite{Fleurbaey:2018fih},  a Rydberg constant with a fractional precision of
$\delta(R_\infty)=3.5\times 10^{-12}$ can be obtained. 

A more precise $R_\infty$ value can be  determined by
combining the H($1S$-$2S$)  with the $\mu$H($2S$-$2P$) measurements.
Inserting $r_p(\mu\text{H})$ into the H($1S$-$2S$) theory prediction: 
\beq
f^\mathrm{th}_{2S-1S}(\text{H})=\left[0.74960091418756\,\frac{R_\infty c}{\mathrm{Hz}}-7\,126\,781\,916(1\,813)-1\,368\, 229\left(\frac{r_p}{\mathrm{fm}}\right)^2\right]\,\mathrm{Hz}\label{Htheory},
\eeq
and comparing it to the measured transition \cite{Parthey:2011lfa}:
\beq
f^\mathrm{exp}_{2S-1S}(\text{H})=2\,466\,061\,413\,187\,035(10)\,\mathrm{Hz},
\eeq
yields a Rydberg constants of
\beq
R_\infty c=3.289\,841\,960\,2509(11)_{r_p}(24)_\mathrm{H-theory}\times 10^{15}\,\mathrm{Hz},
\eqlab{Rydberg-muH-H}
\eeq
with a total uncertainty of 2.7~kHz corresponding to  $\delta(R_\infty)=8\times
10^{-13}$. 
Even though Eq.~\ref{Htheory} accounts for several recent updates -- hVP~\cite{Karshenboim:2021jsc}, two-loop and three-loop QED contributions \cite{Karshenboim:2019iuq} (e.g., the previously neglected light-by-light contribution $B_{61}^\text{LbL}(nS)$ at order $\al^2(Z\al)^6m \ln Z \al$), and inelastic three-photon ($3\upgamma$) exchange~\cite{Pachucki:2018yxe} -- (see Ref.\ \citenum{Yerokhin2018} for a recent review)
the obtained value is in perfect agreement with Ref.\ \cite[Eq.~22]{Pohl:2016glp}. 
Notice from \Eqref{Rydberg-muH-H} that the $R_\infty$ accuracy  is limited by the  uncertainty of the H theory (2.4 kHz) while the uncertainty from $r_p(\mu\text{H})$ is only of
1.1~kHz.
\begin{marginnote}[]
\entry{$R_\infty$ fractional uncertainty}{$\delta(R_\infty)=8\times 10^{-13}$}
\end{marginnote}

A test of the H energy levels 
requires combining theory and measurement of three transitions in H: two of them to determine $R_\infty$ and $r_p$, the third to check for consistency.
This test is presently  limited  by the uncertainty of the third best measurement in H (the $1S$ hfs excluded) and by the correlations
of the various contributions to the energy splittings.
Hence, a more sensitive  way to test the H energy levels is to
use of the precise  $r_p(\mu\text{H})$ value and to
combine it with two most precise  measurements in
H: the H($1S$-$2S$) and  H($1S$-$3S$) transitions.
Agreement between theory and experiment  has been verified on the $1\times 10^{-12}$ level, limited by theory.

\subsection{ $\mu^4$He$^+$ and He$^+$: testing higher-order QED and nuclear models}
\label{Sec-He}

An interesting test of bound-state QED can be obtained when the ongoing efforts to measure the $1S$-$2S$ transition in the hydrogen-like He$^+$ ion
in LaserLaB, Amsterdam~\cite{Krauth:2019kei} and MPQ, Garching \cite{PhysRevA.79.052505} will be accomplished.
To understand the interplay between measurements in  He$^+$, $\mu^4$He$^+$, H and $\mu$H we express the He$^+$($1S$-$2S$)  with explicit $Z$-dependence:
\begin{subequations}
\label{isoHE}
\begin{align}
f_{2S-1S}^{\mathrm{th}}(\mathrm{He^+}) & \approx &  \frac{3 Z^2 cR_\infty}{4}\frac{1}{1+\frac{m_e}{M_\alpha}} &  & +\mathrm{QED}_{\text{He}^+}\left(Z^{3.7},Z^{5  \dots 7}\right) &  & -\frac{7 (Z\al)c^4}{24\pi\,a_\mathrm{B}^3\,\hbar^3}\,r_\alpha^2, \eqlab{1S2SHe} \\ 
   (1\,\mathrm{kHz})                           &         &     (9\,\mathrm{kHz})                 &  &                                   (40\,\mathrm{kHz})        &   & (61\,\mathrm{kHz}) \eqlab{1S2SHe-unc}
\end{align} 
\end{subequations}
with $M_\alpha$ being the alpha-particle mass.
The Bohr structure scales only with $Z^2$, the finite size with $Z^4$, the
one-loop QED contributions scale approximately as $Z^{3.7}$, while
the challenging higher-order contributions (e.g., the two-loop $B_{60}$ term at order $\al^2 (Z\al)^6m$, the three-loop $C_{50}$ term at order $\al^3(Z\al)^5 m$) scaling as $Z^{5..7}$ are strongly enhanced in He$^+$.
\Eqref{1S2SHe-unc} illustrates the uncertainties: 1 kHz uncertainty is
expected from the  LaserLaB  experiment in the first phase~\cite{Krauth:2019kei}, while
an analysis of typical systematic effects of the MPQ experiment promises uncertainties far below that level, on the order Hz level~\cite{PhysRevA.79.052505}.
The 9~kHz is from the uncertainty 
of $R_\infty(\mu \text{H} +\text{H})$ (\Eqref{Rydberg-muH-H}), the 40~kHz represents the present uncertainty of the QED theory~\cite{Yerokhin2018, Karshenboim:2019iuq}, and the 61 kHz is the uncertainty resulting from the alpha-particle charge radius, $r_\alpha=1.67824(13)_\mathrm{exp}(82)_\mathrm{th}$~fm, from $\mu^4$He$^+$\cite{Krauth:2021foz}
spectroscopy limited by the uncertainty of the \2PE-exchange  contribution in $\mu^4$He$^+$~\cite{Diepold:2016cxv, Ji:2018ozm}.

By considering these uncertainties, it is clear that the $1S$-$2S$ transition in He$^+$ can be tested after completion of the measurement in He$^+$ down to an accuracy of $\sim$ 60~kHz limited by $r_\alpha$ from $\mu^4$He$^+$. This correspond to a test at the $6\times 10^{-12}$ level.
Even though the energy levels  in H are tested on the $1\times 10^{-12}$ level,
He$^+$ has  a superior sensitivity to higher-order QED contributions that scale with $Z^5=32$ and $Z^6=64$.

To push further the QED test in He$^+$ requires  reducing the uncertainty of $r_\alpha$,  achievable by progressing the \2PE- and $3\upgamma$-exchange contributions in $\mu^4$He$^+$: $E^{\langle 2\upgamma \rangle A+N}_{2P-2S}=9.34(20)_\mathrm{N}(11)_\mathrm{A}\,\mathrm{meV}$ \cite{Diepold:2016cxv}, with $A$ and $N$ the nuclear and nucleon contributions, and $E^{\langle 3\upgamma \rangle}_{2P-2S}=- 0.150(150)\,\mathrm{meV}$ \cite{Krauth:2021foz}.
In order of importance, the  \2PE-exchange theory  can be advanced by  improving on the nucleon-polarizability contribution (primarily the neutron), on the nuclear-polarizability contribution  (whose precision is presently limited by the spread from various parametrizations of the nuclear potential), and  on the electric form factor needed to compute the elastic part~\cite{Diepold:2016cxv, Ji:2018ozm}.

\begin{marginnote}[]
\entry{Neutron \2PE-contribution}{will limit the test of He$^+$ energy levels.}
\end{marginnote}

Conversely, the comparison theory-measurement  via \Eqref{1S2SHe} can be used to extract  $r_\alpha$ to a better precision than from $\mu^4$He$^+$.
With $r_\alpha$ from He$^+$, the comparison theory-measurement in $\mu^4$He$^+$ can be used to determine the \2PE-exchange contribution in $\mu^4$He$^+$ providing  a precise  benchmark for nuclear theories to guide future advances.

\subsection{HD$^+$, H$_2^+$ and H$_2$: from $r_p$ to $m_e$ and the bound-electron g-factor}
\label{Sec-HD}

An impressive improvement has been witnessed in recent years in HD$^+$ theory~\cite{Karr:2020jbg} and experiments~\cite{Alighanbari2020, Patra2020, Kortunov:2021rfe}, so that the  precision reached has become sensitive to the proton radius puzzle. 
The transition frequencies in HD$^+$ can be expressed as:
\begin{equation}
  f=cR_\infty\left[K_\mathrm{NR}\left(\frac{m_e}{M_p},\; \frac{M_p}{M_d}\right)
    +\alpha^2 K_\mathrm{QED}\left(\alpha, \frac{m_e}{M_p},\; \frac{M_p}{M_d}\right)+k_p\left(r_p^2\right)+k_d\left(r_d^2\right) \right]\, ,
\end{equation}
where the first term corresponds to the non-relativistic energy, the
second to QED and relativistic corrections, and the last two terms to
finite-size corrections. 
HD$^+$ provides thus an independent access to $R_\infty$,
$m_e/M_p$, $M_p/M_d$, $r_p$ and $r_d$, where $m_e/M_p$ and $M_p/M_d$ are electron-to-proton  and proton-to-deuteron  mass ratios, respectively.

Agreement between theory and experiment in HD$^+$ has been demonstrated for various transitions~\cite{Alighanbari2020, Patra2020, Kortunov:2021rfe} down to the $10^{-11}$ level, representing  the best tests of quantum-three-body predictions.
For the rotational transitions presented in Ref.~\citenum{Alighanbari2020} the comparison theory-experiment  is limited to the  $5\times 10^{-11}$ level by the $m_e/M_p$ uncertainty
while the uncertainties of $R_\infty$, $r_p$ and $r_d$ play a minor role.
Nonetheless, the reached  precision is  sensitive to the proton radius puzzle: the $r_p$ value from $\mu$H is favoured as it yields to a better agreement between theory and experiment~\cite{Alighanbari2020}.

Conversely, we can equate theory and experiment in HD$^+$  to extract $m_e/M_p$ with a fractional precision of $\delta(m_e/M_p)=2\times10^{-11}$\cite{Alighanbari2020, Patra2020}, i.e. a factor of 2 better
than achievable combining  the electron mass of $\delta(m_e/M_{^{12}\text{C}^{5+}})=3\times 10^{-11}$ \cite{Sturm:2014bla} with the proton mass of $\delta(M_p/M_{^{12}\text{C}^{5+}})=3\times 10^{-11}$ \cite{Heisse:2019xnz} as obtained from Penning traps. 
The $m_e/M_p$ ratio from HD$^+$ can then be combined with $M_p$ \cite{Heisse:2019xnz} from Penning traps to improve on $m_e$. This allows an extraction of the electron bound g-factor from the measurement of Ref.~\citenum{Sturm:2014bla} to be confronted with corresponding theoretical predictions~\cite{Zatorski:2017vro}: agreement on the $4\times10^{-11}$ level is observed, making this the best test of any bound-electron g-factor.
\begin{marginnote}[]
\entry{HD$^+$ impact}{HD$^+$ links spectroscopy of simple atoms to Penning trap programs}
\end{marginnote}

The precision recently reached in HD$^+$ has established a link between $r_p$ and the electron, proton, and deuteron masses.
The rapid progresses observed in recent years in HD$^+$ theory and experimental techniques  promises a fruitful exploitation of this link that connects two very active precision fields: laser spectroscopy of simple atoms and  Penning traps~\cite{Zatorski:2017vro}. 
The potential of HD$^+$ roots in the   Hz to kHz line widths given by the tens of milliseconds lifetimes of its ro-vibrational states. This has to be compared with the MHz linewidths of the recently measured $2S$-$4P$ and $1S$-$3S$ transitions in H.
Even higher precision is expected in H$_2^+$  given the day-long lifetimes of its states. Novel quantum-logic schemes and state preparation methods are being developed for this purpose~\cite{PhysRevApplied.14.024053}. Also spectroscopy of H$_2$, D$_2$ and HD --cornerstones of quantum chemistry-- will need soon precise values of $r_p$ and $r_d$, expanding the impact of the  $\mu$H and $\mu$D measurements to chemical bonds and four-body QED~\cite{Czachorowski:2018xvk, PhysRevLett.122.103002}.

\subsection{New Physics searches}\label{sec:BSM}

Precision spectroscopy of atoms and molecules could sense energy shifts caused by physics beyond the standard model (BSM) involving a low-mass and weakly coupled sector that escapes detection in high-energy colliders \cite{Karshenboim:2014tka,Liu:2018qgl,Safronova:2017xyt}.
This searches typically involve a comparison between  theoretical predictions and experiments  that eventually will be limited by hadronic effects.
In our context, the explicit way to searches for BSM physics is to look for deviations between $r_p$  values as extracted from the various systems: $ep$ scattering,  H,  $\mu$H and  molecules. Any deviation  might reveal an inconsistency of the theoretical framework  pointing to the existence of  BSM physics.
Presently, these searches are limited by the uncertainty of the $r_p$ as determined from measurements other than $\mu$H. 

Along these lines of investigation, Ref.~\citenum{Brandt:2021yor} highlights  that $R_\infty$ extracted from  H tends to decrease as the $n$ of either the upper or lower state increases. Such a trend could be explained by a fifth-force expressed as a Yukawa-like potential  with a large length scale~\cite{Jones:2019qny}  mitigating the new tension between $\mu$H and recent H measurements~\cite{Brandt:2021yor}. 

\begin{marginnote}[]
\entry{$\mu$H sensitivity to BSM}{weakly-coupled light dark sector}
\end{marginnote}
A recent  study \cite{Frugiuele:2021bic} highlighted the peculiar sensitivity of $\mu$H, $\mu$D  and H($1S$-$2S$) to a dark sector with masses in the keV to GeV range. 
The sensitivity presented in this study is greatly enhanced when accounting for the upcoming measurement of the $1S$ hfs in $\mu$H, and  improved determinations of $r_p$.

The implicit way to exploit $r_p$ for BSM searches, is simply by using its accuracy to improve other fundamental constants increasing the predictive power of our theories. Advancing the \2PE- and $3\upgamma$-exchange contributions is essential.

\section{Future prospects}\label{sec:Conclusions}

\begin{issues}[Experimental prospects]
\begin{enumerate}

\item \textit{Precise measurement of the $1S$ hfs  in $\mu$}H: Three collaborations (CREMA~\cite{Amaro:2021goz}, FAMU~\cite{Pizzolotto:2021dai, Pizzolotto:2020fue} and J-PARC/Riken~\cite{Sato:2014uza}) are aiming at a measurement with up to 1 ppm relative precision to extract the \2PE-exchange contribution. The narrow  line width (relative to the $2S$-$2P$ splitting) promises  improvements in a second phase.

\item \textit{Improved $2S$-$2P$ measurements  in $\mu$}H:
An upgraded  CREMA-2010 setup~\cite{Pohl:2010zza, Antognini:1900ns} holds the potential
of improving the $2S$-$2P$ measurements by at least a  factor of 5.
This can be  obtained principally by increasing the data taking time and using a laser technology capable of delivering mJ-scaled pulses at 6~$\upmu$m with bandwidths smaller than 100 MHz under development  for the hfs experiment.
Improving the $2S$-$2P$ measurements by 5, would pave the way for $r_p(\mu\text{H})$ and $R_\infty$ determinations down to  $\delta (r_p)\lesssim 1\times 10^{-4}$ and  
$\delta (R_\infty)\lesssim 1\times 10^{-13}$, respectively.

\item He\textit{$^+$ $1S$-$2S$ measurements (Sec.~\ref{Sec-He}):} Two groups~\cite{Krauth:2019kei, PhysRevA.79.052505} are addressing this transition using novel frequency comb and trap technologies. Completion of their experiments will contribute to the proton radius solution and enable testing the higher-order QED contributions (scaling as $Z^{5..7}$) in the He$^+$ ion when assuming $r_\alpha$ from $\mu^4$He$^+$. Conversely, comparing theory to experiment in He$^+$ yields to an alternative and improved determination of $r_\alpha$ that can be used to extract the \2PE-exchange contribution in $\mu^4$He$^+$ benchmarking nuclear and nucleon models.

\item \textit{Ultra precision-spectroscopy in simple system}: Spectroscopy of H, HD$^+$, H$_2^+$, H$_2$, He, 
has the potential to not only resolve the proton radius puzzle but to improve  fundamental constants and theory tests to unprecedented levels of accuracy.

\item \textit{Proton radius from scattering experiments}: The upgraded PRad experiment (PRad-II) will reduce the experimental uncertainties by a factor of $3.8$ and reach down to the $Q^2$ range of $10^{-5}\, \text{GeV}^2$. The  $\mu p $ scattering experiments by the MUSE (PSI)  and AMBER (CERN) Collaborations 
are underway. The PRES Collaboration is building a new experiments in the A2 Hall of MAMI to measure $ep$ scattering
using, for the first time, over-determined kinematics, i.e., detecting both the scattered electron and the recoil proton.
\item \textit{Spin structure functions}: Results from the g2p experiment at JLab Hall A \cite{JeffersonLabHallAg2p:2022qap}
will improve evaluations of the polarizability contribution in the H and $\mu$H hfs.

\end{enumerate}
\end{issues}

\begin{issues}[Theory prospects]
\begin{enumerate}
\item \textit{Lattice QCD calculations}: Direct calculations of the nucleon  radii using lattice QCD will soon reach the precision comparable to the
$ep$ scattering experiments. Also highly anticipated are the lattice calculations of the polarizability effects in the $\mu$H Lamb shift.

\item \textit{Next-to-leading order $\chi$PT calculations}: The present NLO $\chi$PT calculations, which agree with the wealth of low-energy Compton scattering data,
can be extended to muonic atoms, to improve the predictions of the polarizability effect. 

\item \textit{Theory prediction of $\mu$}H \textit{hfs with ppm accuracy}: The accuracy of the 
present empirical constraint on the $\mu$H hfs, see Sec.~\ref{Sec-scaling}, is limited by missing higher-order QED contributions and recoil effects.
An improvement in these directions is desirable for finding this transition experimentally and the interpretation of results. 

\item \textit{QED for}  H \textit{$S$-levels}: The two most precise measurements in H are the $S$-level transitions. They are measured with $\delta(1S-2S)=4.2\times 10^{-15}$~\cite{Parthey:2011lfa} and $\delta(1S-3S)=2.5\times 10^{-13}$~\cite{Fleurbaey:2018fih} relative precision, and thus have a smaller uncertainty compared to the uncertainty of the theoretical predictions. An improvement on the theory side will allow for a better extraction of the Rydberg constant and, in turn, better tests of the H energy levels. 

\item \textit{Nucleon \2PE-exchange contribution}: The
biggest uncertainty in the $\mu^4$He$^+$ is presently given by the nucleon \2PE-exchange contribution. Improving the latter, and in particular the neutron \2PE-exchange contribution, will allow for an improved extraction of  $r_\alpha$, which can then be used for QED tests of He$^+$.
\end{enumerate}
\end{issues}

\section*{DISCLOSURE STATEMENT}
The authors are not aware of any affiliations, memberships, funding, or financial holdings that
might be perceived as affecting the objectivity of this review. 

\section*{ACKNOWLEDGMENTS}
We  gratefully acknowledge useful discussions with Carl Carlson, Misha Gorchtein, Vadim Lensky, and Marc Vanderhaeghen.  
V.P.\ thanks his co-authors for the warm (yet unacceptable) hospitality at the Paul Scherrer Institute, where this work had began. 
This work is supported by the Swiss National Science Foundation (SNSF) through the project 
$200020\_197052$ and the Ambizione Grant PZ00P2\_193383, the European Research Council (ERC) through CoG. \# 725039,
the Deutsche Forschungsgemeinschaft (DFG) 
through the Emmy Noether Programme under the grant 449369623 and through the project 204404729-SFB1044.

%

\bibliography{lowQ}

\begin{thebibliography}{145}
\expandafter\ifx\csname natexlab\endcsname\relax\def\natexlab#1{#1}\fi

\bibitem{Pohl:2010zza}
Pohl R, et~al.
\newblock \textit{Nature} 466:213--216 (2010)

\bibitem{Antognini:1900ns}
Antognini A, Nez F, Schuhmann K, Amaro FD, et~al.
\newblock \textit{Science} 339:417--420 (2013)

\bibitem{Mohr:2012aa}
Mohr PJ, Taylor BN, Newell DB.
\newblock \textit{Rev. Mod. Phys.} 84(4):1527--1605 (2012)

\bibitem{Carlson:2015jba}
Carlson CE.
\newblock \textit{Prog. Part. Nucl. Phys.} 82:59--77 (2015)

\bibitem{Karr:2020wgh}
Karr JP, Marchand D, Voutier E.
\newblock \textit{Nature Rev. Phys.} 2(11):601--614 (2020)

\bibitem{Gao:2021sml}
Gao H, Vanderhaeghen M.
\newblock \textit{Rev. Mod. Phys.} 94(1):015002 (2022)

\bibitem{Peset:2021iul}
Peset C, Pineda A, Tomalak O.
\newblock \textit{Prog. Part. Nucl. Phys.} 121:103901 (2021)

\bibitem{CREMA:2016idx}
Pohl R, et~al.
\newblock \textit{Science} 353(6300):669--673 (2016)

\bibitem{Jentschura:2011is}
Jentschura U, Matveev A, Parthey C, Alnis J, Pohl R, et~al.
\newblock \textit{Phys. Rev. A} 83(4):042505 (2011)

\bibitem{Krauth:2021foz}
Krauth JJ, et~al.
\newblock \textit{Nature} 589(7843):527--531 (2021)

\bibitem{Lensky:2022tue}
Lensky V, Hagelstein F, Blin AH, Pascalutsa V. 2022.
\newblock In \textit{{10th International workshop on Chiral Dynamics}}.
\newblock 2203.13030

\bibitem{Lensky:2022bid}
Lensky V, Hagelstein F, Pascalutsa V  (2022), nucl-th/2206.14756

\bibitem{Lensky:2022fif}
Lensky V, Hagelstein F, Pascalutsa V  (2022), nucl-th/2206.14066

\bibitem{Kalinowski:2018rmf}
Kalinowski M.
\newblock \textit{Phys. Rev. A} 99(3):030501 (2019)

\bibitem{Tiesinga:2021myr}
Tiesinga E, Mohr PJ, Newell DB, Taylor BN.
\newblock \textit{Rev. Mod. Phys.} 93(2):025010 (2021)

\bibitem{Mohr:2015ccw}
Mohr PJ, Newell DB, Taylor BN.
\newblock \textit{Rev. Mod. Phys.} 88(3):035009 (2016)

\bibitem{Brandt:2021yor}
Brandt AD, Cooper SF, Rasor C, Burkley Z, Yost DC, Matveev A.
\newblock \textit{Phys. Rev. Lett.} 128(2):023001 (2022)

\bibitem{Grinin:2020:Science_H1S3S}
Grinin A, Matveev A, Yost D, Maisenbacher L, Wirthl V, et~al.
\newblock \textit{Science} 370(6520):1061--1066 (2020)

\bibitem{Bezginov:2019mdi}
Bezginov N, Valdez T, Horbatsch M, Marsman A, Vutha AC, Hessels EA.
\newblock \textit{Science} 365(6457):1007--1012 (2019)

\bibitem{Fleurbaey:2018fih}
Fleurbaey H.
\newblock \textit{Phys. Rev. Lett.} 120(18):183001 (2018)

\bibitem{Beyer:2017:Science_2S4P}
Beyer A, Maisenbacher L, Matveev A, Pohl R, Khabarova K, et~al.
\newblock \textit{Science} 358:79 (2017)

\bibitem{Xiong:2019umf}
Xiong W, et~al.
\newblock \textit{Nature} 575(7781):147--150 (2019)

\bibitem{Horbatsch:2016ilr}
Horbatsch M, Hessels EA, Pineda A.
\newblock \textit{Phys. Rev. C} 95(3):035203 (2017)

\bibitem{Higinbotham:2015rja}
Higinbotham DW, Kabir AA, Lin V, Meekins D, et~al.
\newblock \textit{Phys. Rev.} C 93(5):055207 (2016)

\bibitem{Lee:2015jqa}
Lee G, Arrington JR, Hill RJ.
\newblock \textit{Phys. Rev.} D 92(1):013013 (2015)

\bibitem{Sick:2012zz}
Sick I.
\newblock \textit{Prog. Part. Nucl. Phys.} 67:473--478 (2012)

\bibitem{Bernauer:2010wm}
Bernauer JC, et~al.
\newblock \textit{Phys. Rev. Lett.} 105:242001 (2010)

\bibitem{Lin:2021xrc}
Lin YH, Hammer HW, Mei\ss{}ner UG.
\newblock \textit{Phys. Rev. Lett.} 128(5):052002 (2022)

\bibitem{Alarcon:2018zbz}
Alarc\'on JM, Higinbotham DW, Weiss C, Ye Z.
\newblock \textit{Phys. Rev. C} 99(4):044303 (2019)

\bibitem{Lorenz:2014yda}
Lorenz I, Mei{\ss}ner UG, Hammer HW, Dong YB.
\newblock \textit{Phys. Rev.} D 91(1):014023 (2015)

\bibitem{Belushkin:2006qa}
Belushkin MA, Hammer HW, Meissner UG.
\newblock \textit{Phys. Rev. C} 75:035202 (2007)

\bibitem{Mihovilovic:2019jiz}
Mihovilovi\v{c} M, et~al.
\newblock \textit{Eur. Phys. J. A} 57(3):107 (2021)

\bibitem{Strauch:2018ros}
Strauch S.
\newblock \textit{PoS} NuFACT2018:136 (2018)

\bibitem{COMPASSAMBERworkinggroup:2019amp}
Dreisbach C, et~al.
\newblock \textit{PoS} DIS2019:222 (2019)

\bibitem{PRad:2020oor}
Gasparian A, et~al.  (2020), 2009.10510

\bibitem{Scheidegger_Merkt_2020}
Scheidegger S, Merkt F.
\newblock \textit{CHIMIA} 74(4):285 (2020)

\bibitem{Krauth:2019kei}
Krauth JJ, Dreissen LS, Roth C, Gr\"undeman EL, Collombon M, et~al.
\newblock \textit{PoS} FFK2019:049 (2020)

\bibitem{PhysRevA.79.052505}
Herrmann M, Haas M, Jentschura UD, Kottmann F, Leibfried D, et~al.
\newblock \textit{Phys. Rev. A} 79(5):052505 (2009)

\bibitem{Alighanbari2020}
Alighanbari S, Giri GS, Constantin FL, Korobov VI, Schiller S.
\newblock \textit{Nature} 581(7807):152--158 (2020)

\bibitem{Pospelov:2017kep}
Pospelov M, Tsai YD.
\newblock \textit{Phys. Lett. B} 785:288--295 (2018)

\bibitem{Karshenboim:2014tka}
Karshenboim SG, McKeen D, Pospelov M.
\newblock \textit{Phys. Rev. D} 90(7):073004 (2014), [Addendum: Phys.Rev.D 90,
  079905 (2014)]

\bibitem{Liu:2018qgl}
Liu YS, Clo\"et IC, Miller GA.
\newblock \textit{Nucl. Phys.} B:114638 (2019)

\bibitem{Carlson:2015poa}
Carlson CE, Freid M.
\newblock \textit{Phys. Rev. D} 92(9):095024 (2015)

\bibitem{Sick:1998cvq}
Sick I, Trautmann D.
\newblock \textit{Nucl. Phys.} A 637:559--575 (1998)

\bibitem{Pohl:2016glp}
Pohl R, et~al.
\newblock \textit{Metrologia} 54(2):L1 (2017)

\bibitem{Sato:2014uza}
Sato M, et~al. 2014.
\newblock In \textit{{Proceedings, 20th International Conference on Particles
  and Nuclei (PANIC 14), Hamburg, Germany, August 24-29, 2014}}

\bibitem{Adamczak:2001uvg}
Adamczak A.
\newblock \textit{Hyperfine Interact.} 138(1-4):343--350 (2001)

\bibitem{Dupays:2003zz}
Dupays A, Beswick A, Lepetit B, Rizzo C, Bakalov D.
\newblock \textit{Phys. Rev.} A 68:052503 (2003)

\bibitem{Bakalov:2015xya}
Bakalov D, Adamczak A, Stoilov M, Vacchi A.
\newblock \textit{Hyperfine Interact.} 233(1-3):97--101 (2015)

\bibitem{Pizzolotto:2021dai}
Pizzolotto C, et~al.
\newblock \textit{Phys. Lett. A} 403:127401 (2021)

\bibitem{Pizzolotto:2020fue}
Pizzolotto C, et~al.
\newblock \textit{Eur. Phys. J. A} 56(7):185 (2020)

\bibitem{Pachucki:1996zza}
Pachucki K.
\newblock \textit{Phys. Rev.} A 53:2092--2100 (1996)

\bibitem{Eides:2000xc}
Eides MI, Grotch H, Shelyuto VA.
\newblock \textit{Phys. Rept.} 342:63--261 (2001)

\bibitem{Borie:2012zz}
Borie E.
\newblock \textit{Annals Phys.} 327:733--763 (2012)

\bibitem{Indelicato:2012pfa}
Indelicato P.
\newblock \textit{Phys. Rev.} A 87(2):022501 (2013)

\bibitem{Karshenboim:2015}
Karshenboim SG, Korzinin EY, Shelyuto VA, Ivanov VG.
\newblock \textit{Journal of Physical and Chemical Reference Data} 44(3):031202
  (2015)

\bibitem{BetheSalpeter}
Bethe HA, Salpeter EE.
\newblock Quantum mechanics of one- and two-electron atoms (1957)

\bibitem{LandauLifshitz4}
Berestetskii VB, Lifshitz EM, Pitaevskii LP.
\newblock Course of theoretical physics, vol. 4. quantum electrodynamics (1982)

\bibitem{Friar:1978wv}
Friar JL.
\newblock \textit{Annals Phys.} 122:151 (1979)

\bibitem{DeRujula:2010dp}
De~Rujula A.
\newblock \textit{Phys. Lett.} B 693:555--558 (2010)

\bibitem{Distler:2010zq}
Distler MO, Bernauer JC, Walcher T.
\newblock \textit{Phys. Lett.} B 696:343--347 (2011)

\bibitem{Hagelstein:2015aa}
Hagelstein F, Pascalutsa V.
\newblock \textit{Phys. Rev. A} 91:040502 (R) (2015)

\bibitem{Hagelstein:2017ldk}
Hagelstein F. 2017.
\newblock {Exciting nucleon in Compton scattering and hydrogen-like atoms}.
\newblock Ph.D. thesis, Mainz U.
\newblock 1710.00874

\bibitem{Hagelstein:2015lph}
Hagelstein F, Pascalutsa V.
\newblock \textit{PoS} CD15:077 (2016)

\bibitem{Huong:2015naj}
Huong NT, Kou E, Moussallam B.
\newblock \textit{Phys. Rev.} D 93(11):114005 (2016)

\bibitem{Zhou:2015bea}
Zhou HQ, Pang HR.
\newblock \textit{Phys. Rev.} A 92(3):032512 (2015), [Erratum: Phys. Rev. A 93
  (2016) 069903]

\bibitem{Dorokhov:2017gst}
Dorokhov AE, Kochelev NI, Martynenko AP, Martynenko FA, Faustov RN.
\newblock \textit{Phys. Part. Nucl. Lett.} 14(6):857--864 (2017)

\bibitem{Karshenboim:2021jsc}
Karshenboim SG, Shelyuto VA.
\newblock \textit{Eur. Phys. J. D} 75(2):49 (2021)

\bibitem{Iddings:1965zz}
Iddings CK.
\newblock \textit{Phys. Rev. B} 138:446--458 (1965)

\bibitem{Drell:1966kk}
Drell S, Sullivan JD.
\newblock \textit{Phys. Rev.} 154:1477--1498 (1967)

\bibitem{Hagelstein:2015egb}
Hagelstein F, Miskimen R, Pascalutsa V.
\newblock \textit{Prog. Part. Nucl. Phys.} 88:29--97 (2016)

\bibitem{Pasquini:2018wbl}
Pasquini B, Vanderhaeghen M.
\newblock \textit{Ann. Rev. Nucl. Part. Sci.} 68:75--103 (2018)

\bibitem{GellMann:1954db}
Gell-Mann M, Goldberger M, Thirring WE.
\newblock \textit{Phys. Rev.} 95:1612--1627 (1954)

\bibitem{Pascalutsa:2018ced}
Pascalutsa V.
\newblock {Causality Rules}: {A light treatise on dispersion relations and sum
  rules} (2018)

\bibitem{Pachucki:1999zza}
Pachucki K.
\newblock \textit{Phys. Rev.} A 60:3593--3598 (1999)

\bibitem{Martynenko:2005rc}
Martynenko A.
\newblock \textit{Phys. Atom. Nucl.} 69:1309--1316 (2006)

\bibitem{Carlson:2011zd}
Carlson CE, Vanderhaeghen M.
\newblock \textit{Phys. Rev.} A 84:020102 (2011)

\bibitem{Birse:2012eb}
Birse MC, McGovern JA.
\newblock \textit{Eur. Phys. J.} A 48:120 (2012)

\bibitem{Gorchtein:2013yga}
Gorchtein M, Llanes-Estrada FJ, Szczepaniak AP.
\newblock \textit{Phys. Rev.} A 87(5):052501 (2013)

\bibitem{Hill:2016bjv}
Hill RJ, Paz G.
\newblock \textit{Phys. Rev. D} 95(9):094017 (2017)

\bibitem{Tomalak:2018uhr}
Tomalak O.
\newblock \textit{Eur. Phys. J. A} 55(5):64 (2019)

\bibitem{Alarcon:2013cba}
Alarc{\'o}n JM, Lensky V, Pascalutsa V.
\newblock \textit{Eur. Phys. J.} C 74(4):2852 (2014)

\bibitem{Lensky:2017bwi}
Lensky V, Hagelstein F, Pascalutsa V, Vanderhaeghen M.
\newblock \textit{Phys. Rev. D} 97(7):074012 (2018)

\bibitem{Fu:2022fgh}
Fu Y, Feng X, Jin LC, Lu CF.
\newblock \textit{Phys. Rev. Lett.} 128(17):172002 (2022)

\bibitem{Hagelstein:2020awq}
Hagelstein F, Pascalutsa V.
\newblock \textit{Nucl. Phys. A} 1016:122323 (2021)

\bibitem{Miller:2012ne}
Miller GA.
\newblock \textit{Phys. Lett. B} 718:1078--1082 (2013)

\bibitem{Pauk:2020gjv}
Pauk V, Carlson CE, Vanderhaeghen M.
\newblock \textit{Phys. Rev. C} 102(3):035201 (2020)

\bibitem{Can:2020}
Can K, Hannaford-Gunn A, Horsley R, Nakamura Y, Perlt H, et~al.
\newblock \textit{Phys.~Rev.~D} 102(11) (2020)

\bibitem{Hannaford-Gunn:2020pvu}
Hannaford-Gunn A, Horsley R, Nakamura Y, Perlt H, Rakow P, et~al.
\newblock \textit{PoS} LATTICE2019:278 (2020)

\bibitem{Chambers:2017dov}
Chambers A, Horsley R, Nakamura Y, Perlt H, Rakow P, et~al.
\newblock \textit{Phys. Rev. Lett.} 118(24):242001 (2017)

\bibitem{Fu:2021dja}
Fu Y. 2021.
\newblock In \textit{{38th International Symposium on Lattice Field Theory}}.
\newblock 2112.14913

\bibitem{Can:2022rgi}
Can KU, et~al. 2022.
\newblock In \textit{{33rd International (ONLINE) Workshop on High Energy
  Physics}: {Hard Problems of Hadron Physics: Non-Perturbative QCD \& Related
  Quests}}.
\newblock 2201.08367

\bibitem{Borah:2020gte}
Borah K, Hill RJ, Lee G, Tomalak O.
\newblock \textit{Phys. Rev. D} 102(7):074012 (2020)

\bibitem{Volotka:2004zu}
Volotka A, Shabaev V, Plunien G, Soff G.
\newblock \textit{Eur. Phys. J.} D 33:23--27 (2005)

\bibitem{Carlson:2011af}
Carlson CE, Nazaryan V, Griffioen K.
\newblock \textit{Phys. Rev.} A 83:042509 (2011)

\bibitem{Kelly:2004hm}
Kelly JJ.
\newblock \textit{Phys. Rev.} C 70:068202 (2004)

\bibitem{Bradford:2006yz}
Bradford R, Bodek A, Budd HS, Arrington J.
\newblock \textit{Nucl. Phys. Proc. Suppl.} 159:127--132 (2006)

\bibitem{Arrington:2007ux}
Arrington J, Melnitchouk W, Tjon JA.
\newblock \textit{Phys. Rev.} C 76:035205 (2007)

\bibitem{Arrington:2006hm}
Arrington J, Sick I.
\newblock \textit{Phys. Rev.} C 76:035201 (2007)

\bibitem{Drell:1966jv}
Drell S, Hearn AC.
\newblock \textit{Phys. Rev. Lett.} 16:908--911 (1966)

\bibitem{Gerasimov:1965et}
Gerasimov S.
\newblock \textit{Sov. J. Nucl. Phys.} 2:430--433 (1966)

\bibitem{Pascalutsa:2014zna}
Pascalutsa V, Vanderhaeghen M.
\newblock \textit{Phys. Rev.} D 91:051503 (R) (2015)

\bibitem{Lensky:2017dlc}
Lensky V, Pascalutsa V, Vanderhaeghen M, Kao C.
\newblock \textit{Phys. Rev. D} 95(7):074001 (2017)

\bibitem{Faustov:2006ve}
Faustov R, Gorbacheva I, Martynenko A.
\newblock \textit{Proc. SPIE Int. Soc. Opt. Eng.} 6165:0M (2006)

\bibitem{Carlson:2008ke}
Carlson CE, Nazaryan V, Griffioen K.
\newblock \textit{Phys. Rev.} A 78:022517 (2008)

\bibitem{Zielinski:2017gwp}
Zielinski R. 2017.
\newblock {The g2p Experiment: A Measurement of the Proton's Spin Structure
  Functions}.
\newblock Ph.D. thesis, New Hampshire U., nucl-ex/1708.08297

\bibitem{Hagelstein:2018bdi}
Hagelstein F.
\newblock \textit{Few Body Syst.} 59(5):93 (2018)

\bibitem{CLAS:2017qga}
Fersch R, et~al.
\newblock \textit{Phys. Rev. C} 96(6):065208 (2017)

\bibitem{Slifer:2009ik}
Slifer K.
\newblock \textit{AIP Conf. Proc.} 1155 (2009)

\bibitem{Zielinski:2016khh}
Zielinski R.
\newblock \textit{PoS} CD15:090 (2016)

\bibitem{JeffersonLabHallAg2p:2022qap}
Ruth D, et~al.  (2022), nucl-ex/2204.10224

\bibitem{Lensky:2009uv}
Lensky V, Pascalutsa V.
\newblock \textit{Eur. Phys. J. C} 65:195--209 (2010)

\bibitem{Lensky:2015awa}
Lensky V, McGovern J, Pascalutsa V.
\newblock \textit{Eur. Phys. J.} C 75(12):604 (2015)

\bibitem{Korzinin:2013uia}
Korzinin EY, Ivanov VG, Karshenboim SG.
\newblock \textit{Phys. Rev. D} 88(12):125019 (2013)

\bibitem{Karshenboim:2018iyl}
Karshenboim SG, Korzinin EY, Shelyuto VA, Ivanov VG.
\newblock \textit{Phys. Rev. A} 98(6):062512 (2018)

\bibitem{Amaro:2021goz}
Amaro P, et~al.
\newblock \textit{SciPost Phys.} 13:020 (2022)

\bibitem{Faustov:1997rc}
Faustov RN, Martynenko AP.
\newblock \textit{Phys. Atom. Nucl.} 61:471--475 (1998)

\bibitem{Karshenboim2009}
Karshenboim SG, Korzinin EY, Ivanov VG.
\newblock \textit{JETP Letters} 89(4):216--216 (2009)

\bibitem{Brodsky:1966vn}
Brodsky SJ, Erickson GW.
\newblock \textit{Phys. Rev.} 148:26--46 (1966)

\bibitem{Peset:2016wjq}
Peset C, Pineda A.
\newblock \textit{JHEP} 04:060 (2017)

\bibitem{Hellwig1970}
Hellwig H, Vessot RFC, Levine MW, Zitzewitz PW, Allan DW, Glaze DJ.
\newblock \textit{IEEE Transactions on Instrumentation and Measurement}
  19(4):200--209 (1970)

\bibitem{Karshenboim:2000rg}
Karshenboim SG.
\newblock \textit{Can. J. Phys.} 78:639--678 (2000)

\bibitem{Karshenboim:1996ew}
Karshenboim SG.
\newblock \textit{Phys. Lett. A} 225:97 (1997)

\bibitem{Tomalak:2017lxo}
Tomalak O.
\newblock \textit{Eur. Phys. J. A} 54(1):3 (2018)

\bibitem{Kanda:2020mmc}
Kanda S, et~al.
\newblock \textit{Phys. Lett. B} 815:136154 (2021)

\bibitem{Ohayon:2021dec}
Ohayon B, Burkley Z, Crivelli P.
\newblock \textit{SciPost Phys. Proc.} 5:029 (2021)

\bibitem{Janka:2021xxr}
Janka G, Ohayon B, Crivelli P.
\newblock \textit{EPJ Web Conf.} 262:01001 (2022)

\bibitem{Parthey:2011lfa}
Parthey CG, et~al.
\newblock \textit{Phys. Rev. Lett.} 107:203001 (2011)

\bibitem{Karshenboim:2019iuq}
Karshenboim SG, Ozawa A, Shelyuto VA, Szafron R, Ivanov VG.
\newblock \textit{Phys. Lett. B} 795:432--437 (2019)

\bibitem{Pachucki:2018yxe}
Pachucki K, Patk\'o\v{s} V, Yerokhin VA.
\newblock \textit{Phys. Rev. A} 97(6):062511 (2018)

\bibitem{Yerokhin2018}
Yerokhin VA, Pachucki K, Patkóš V.
\newblock \textit{Annalen der Physik} 531(5):1800324 (2019)

\bibitem{Diepold:2016cxv}
Diepold M, Franke B, Krauth JJ, Antognini A, Kottmann F, Pohl R.
\newblock \textit{Annals Phys.} 396:220--244 (2018)

\bibitem{Ji:2018ozm}
Ji C, Bacca S, Barnea N, Hernandez OJ, Nevo-Dinur N.
\newblock \textit{J. Phys. G} 45(9):093002 (2018)

\bibitem{Karr:2020jbg}
Karr JP, Haidar M, Hilico L, Korobov VI.
\newblock \textit{Springer Proc. Phys.} 238:75--81 (2020)

\bibitem{Patra2020}
Patra S, Germann M, Karr JP, Haidar M, Hilico L, et~al.
\newblock \textit{Science} 369(6508):1238--1241 (2020)

\bibitem{Kortunov:2021rfe}
Kortunov IV, Alighanbari S, Hansen MG, Giri GS, Korobov VI, Schiller S.
\newblock \textit{Nature Phys.} 17(5):569 (2021)

\bibitem{Sturm:2014bla}
Sturm S, K\"ohler F, Zatorski J, Wagner A, Harman Z, et~al.
\newblock \textit{Nature} 506(7489):467--470 (2014)

\bibitem{Heisse:2019xnz}
Hei\ss{}e F, Rau S, K\"ohler-Langes F, Quint W, Werth G, et~al.
\newblock \textit{Phys. Rev. A} 100(2):022518 (2019)

\bibitem{Zatorski:2017vro}
Zatorski J, Sikora B, Karshenboim SG, Sturm S, K\"ohler-Langes F, et~al.
\newblock \textit{Phys. Rev. A} 96(1):012502 (2017)

\bibitem{PhysRevApplied.14.024053}
Schmidt J, Louvradoux T, Heinrich J, Sillitoe N, Simpson M, et~al.
\newblock \textit{Phys. Rev. Applied} 14(2):024053 (2020)

\bibitem{Czachorowski:2018xvk}
Czachorowski P, Puchalski M, Komasa J, Pachucki K.
\newblock \textit{Phys. Rev. A} 98(5):052506 (2018)

\bibitem{PhysRevLett.122.103002}
H\"olsch N, Beyer M, Salumbides EJ, Eikema KSE, Ubachs W, et~al.
\newblock \textit{Phys. Rev. Lett.} 122(10):103002 (2019)

\bibitem{Safronova:2017xyt}
Safronova MS, Budker D, DeMille D, Kimball DFJ, Derevianko A, Clark CW.
\newblock \textit{Rev. Mod. Phys.} 90(2):025008 (2018)

\bibitem{Jones:2019qny}
Jones MPA, Potvliege RM, Spannowsky M.
\newblock \textit{Phys. Rev. Res.} 2(1):013244 (2020)

\bibitem{Frugiuele:2021bic}
Frugiuele C, Peset C.
\newblock \textit{JHEP} 05:002 (2022)

\end{thebibliography}
\bibliographystyle{ar-style5}



\section*{Supplementary material}

# Supplement to "The proton structure in and out of muonic hydrogen"

Aldo Antognini,[1,2] Franziska Hagelstein,[1,3] and Vladimir Pascalutsa[3]

[1]Laboratory for Particle Physics, Paul Scherrer Institute, 5232 Villigen-PSI, Switzerland
[2]Institute for Particle Physics and Astrophysics, ETH, 8093 Zurich, Switzerland
[3]Institut für Kernphysik, Johannes Gutenberg Universität Mainz, 55099 Mainz, Germany
email: aldo.antognini@psi.ch, hagelste@uni-mainz.de, pascalut@uni-mainz.de

## Abstract

In this Supplement, we give a few more details on the theoretical description of nuclear effects in hydrogen-like atoms.

**Contents**

## 1. Quantum-mechanical Coulomb problem

To set the stage, we recall that the hydrogen energy spectrum:

$$E_n = -(Z\alpha)^2 m_r/2n^2, \qquad 1.$$

with $m_r = mM/(M+m)$, where $m, M$ are the masses of the lepton and the nucleus, as well as the corresponding wave functions $\phi_{nlm}(\boldsymbol{r})$, are obtained from the Schrödinger equation with the Coulomb potential:

$$\left[-\frac{\nabla^2}{2m_r} + V_\mathrm{C}(r) - E_n\right]\phi_{nlm}(\boldsymbol{r}) = 0, \qquad V_\mathrm{C}(r) = -\frac{Z\alpha}{r}, \qquad 2.$$

or, in momentum space, from the homogeneous Lippmann-Schwinger equation:

$$\left(\frac{\boldsymbol{p}^2}{2m_r} - E_n\right)\varphi_{nlm}(\boldsymbol{p}) = \int\frac{\mathrm{d}\boldsymbol{p}'}{(2\pi)^3} V_\mathrm{C}(|\boldsymbol{p}-\boldsymbol{p}'|)\,\varphi_{nlm}(\boldsymbol{p}'), \qquad V_\mathrm{C}(|\boldsymbol{q}|) = -\frac{4\pi Z\alpha}{\boldsymbol{q}^2}. \qquad 3.$$

The coordinate and momentum representations are related via the 3-dimensional Fourier transform. To compute perturbative effects due to a small correction $V_\epsilon(\boldsymbol{p}'-\boldsymbol{p};\boldsymbol{p}',\boldsymbol{p})$, the following matrix elements are required:

$$\langle n'l'm'|V_\epsilon|nlm\rangle = \int\frac{\mathrm{d}\boldsymbol{p}\,\mathrm{d}\boldsymbol{p}'}{(2\pi)^3}\,V_\epsilon(\boldsymbol{p}'-\boldsymbol{p};\boldsymbol{p}',\boldsymbol{p})\,\varphi^*_{n'l'm'}(\boldsymbol{p}')\,\varphi_{nlm}(\boldsymbol{p}). \qquad 4.$$

For a central potential, $V_\epsilon = V_\epsilon(|\boldsymbol{p}'-\boldsymbol{p}|)$,

$$\langle n'l'm'|V_\epsilon|nlm\rangle = \delta_{ll'}\delta_{mm'}\int_0^\infty\frac{\mathrm{d}|\boldsymbol{q}|\,\boldsymbol{q}^2}{2\pi^2}\,V_\epsilon(|\boldsymbol{q}|)\,w_{nl}(\boldsymbol{q}^2), \qquad 5.$$

$$w_{nl}(\boldsymbol{q}^2)\,Y^*_{lm}(\Omega_q) \equiv \int\mathrm{d}\boldsymbol{p}\,\varphi^*_{nlm}(\boldsymbol{p}+\boldsymbol{q})\,\varphi_{nlm}(\boldsymbol{p}). \qquad 6.$$

For a zero-range correction, $V_\epsilon$ is constant and we are left with

$$\frac{1}{(2\pi)^3}\int\mathrm{d}\boldsymbol{p}\,\varphi_{nlm}(\boldsymbol{p})\int\mathrm{d}\boldsymbol{p}'\,\varphi^*_{n'l'm'}(\boldsymbol{p}') = \phi_{nlm}(0)\,\phi_{n'l'm'}(0) = \frac{(Z\alpha m_r)^3}{\pi(nn')^{3/2}}\delta_{l'0}\delta_{m'0}\delta_{l0}\delta_{m0}. \qquad 7.$$

## 2. One-photon exchange in dispersive representation

In Sec. 2 of the main Review, we introduced finite-size, polarizability and radiative corrections in hydrogen-like atoms using dispersive representations of the nuclear form factors, the two-photon-exchange potential and the scalar part of the vacuum polarization. In the following subsections, we, firstly, expand on the finite-size effects and express the spherical charge and magnetization distributions through the absorptive parts of the electromagnetic form factors, and secondly, discuss the Uehling contribution.

### 2.1. Finite-size effects

The spherically-symmetric charge and magnetization distributions, $\rho_E(r)$ and $\rho_M(r)$, are Lorentz invariant and related to the form factors and their absorptive parts by, respectively, the Bessel and Laplace transforms. We can see this from the Fourier transforms of the electromagnetic form factors in the Breit frame:

$$\rho(r) = \int \frac{d\boldsymbol{q}}{(2\pi)^3} G(Q^2) e^{-i\boldsymbol{q}\cdot\boldsymbol{r}} \qquad 8a.$$

$$= \frac{1}{2\pi^2} \int_0^\infty dQ\, Q^2 j_0(Qr)\, G(Q^2), \qquad 8b.$$

where $G(Q^2)$ stands for the electric and magnetic Sachs form factors normalized to unity, $G_E(Q^2)$ and $\frac{G_M(Q^2)}{1+\kappa_N}$, and $j_0(Qr) = \frac{\sin Qr}{Qr}$ is the spherical Bessel function. With the dispersive representation of the form factors:

$$G(Q^2) = \frac{1}{\pi} \int_{t_0}^\infty dt\, \frac{\text{Im}\, G(t)}{t + Q^2 - i0^+}, \qquad 9.$$

and $t_0$ the lowest particle-production threshold, it follows in turn that:

$$\rho_E(r) = \frac{1}{4\pi^2 r} \int_{t_0}^\infty dt\, \text{Im}\, G_E(t)\, e^{-r\sqrt{t}}, \qquad 10a.$$

$$\rho_M(r) = \frac{1}{4\pi^2 r} \int_{t_0}^\infty dt\, \frac{\text{Im}\, G_M(t)}{1+\kappa_N}\, e^{-r\sqrt{t}}. \qquad 10b.$$

The finite-size potentials in coordinate space, corresponding to the momentum-space potentials defined in Eqs. 1 and 11 of the main Review, are of Yukawa type:

$$V_{\text{eFF}}(r) = \frac{Z\alpha}{\pi} \int_{t_0}^\infty \frac{dt}{t}\, \text{Im}\, G_E(t)\, \frac{1}{r} e^{-r\sqrt{t}}, \qquad 11a.$$

$$V_{\text{mFF}}^F(r) = \frac{Z\alpha}{3\pi m M}\left[F(F+1) - \tfrac{3}{2}\right] \int_{t_0}^\infty dt\, \text{Im}\, G_M(t)\, \frac{1}{r} e^{-r\sqrt{t}}. \qquad 11b.$$

By comparing to Eq. 10b, one can see that $V_{\text{mFF}}^F(r)$ is directly proportional to $\rho_M(r)$. To compute their effect in perturbation theory, as done in Sec. 2 of the main Review, we only need the matrix elements of the Yukawa potential; e.g., at $1^{\text{st}}$ order:

$$\left\langle nlm \left| \frac{1}{r} e^{-r\sqrt{t}} \right| nlm \right\rangle = \frac{Z\alpha m_r x_n^{l+1}}{n^2 (1+\sqrt{x_n})^{2n}} \frac{(n+l)!}{(n-l-1)!(2l+1)!}\, _2F_1(1+l-n, 1+l-n, 2+2l; x_n), \qquad 12.$$

with $x_n = (2Z\alpha m_r)^2/(n^2 t)$, and $_2F_1(a,b,c;x)$ the hypergeometric function of the second kind. For the effect of the $V_{\text{eFF}}(r)$ potential on the classic $(2S-2P)$ Lamb shift, this leads to:

$$E_{2S-2P}^{\langle\text{eFF}\rangle} \equiv \langle 2S|V_{\text{eFF}}|2S\rangle - \langle 2P|V_{\text{eFF}}|2P\rangle = \frac{(Z\alpha)^4 m_r^3}{2\pi} \int_{t_0}^\infty dt\, \frac{\text{Im}\, G_E(t)}{(\sqrt{t} + Z\alpha m_r)^4}. \qquad 13.$$

Expanding for $Z\alpha m_r \ll t_0$, one obtains the moments of the charge distribution $\rho_E(r)$, starting with the squared charge radius $\langle r_E^2 \rangle$, cf. Eq. 5 of the main Review. The correction to the HFS from the magnetic form factor, $V_{\text{mFF}}$, is calculated analogously. One obtains the Fermi energy and the magnetic radius as shown in Eq. 12 of the main Review.

## 2.2. Vacuum polarization (Uehling) contribution

The dispersive representation applies to many other effects, such as the vacuum-polarization and self-energy contributions. The former, shown in Fig. 3(a) of the main Review, is obtained simply by replacing $\text{Im}\,G_E(t)$ with $\text{Im}\,\Pi(t)$ in the above Eq. 13. For instance, substituting the one-loop eVP,

$$\text{Im}\,\Pi^{(1)}(t) = -\tfrac{1}{3}\alpha \left(1 + 2m_e^2/t\right)\sqrt{1 - 4m_e^2/t}, \qquad 14.$$

one obtains (note the convention of the Lamb shift for muonic atoms: $2P$-$2S$),

$$E_{2P-2S}^{(\text{eVP})(1)} = -\frac{(Z\alpha)^4 m_r^3}{2\pi} \int_{4m_e^2}^{\infty} dt \frac{\text{Im}\,\Pi^{(1)}(t)}{(\sqrt{t} + Z\alpha m_r)^4} \qquad 15a.$$

$$= \frac{\alpha(Z\alpha)^2 m_r}{3\kappa^3}\left(1 + \frac{\kappa\left(2\kappa^6 - 13\kappa^4 + 44\kappa^2 - 24\right)}{12\pi(1-\kappa^2)^2} - \frac{15\kappa^4 - 20\kappa^2 + 8}{4\pi(1-\kappa^2)^{5/2}}\arccos\kappa\right), \qquad 15b.$$

with $\kappa = \frac{Z\alpha m_r}{2m_e}$. For $\mu$H this gives $205.0074\,\text{meV}$, the well-known Uehling correction. In this example, the integrand cannot be expanded in $Z\alpha$, since $\kappa$ is of order 1, i.e., the Bohr radius is comparable with the Compton wavelength of the virtual $e^+e^-$ pair. For this reason, the Uehling correction is effectively of order $\alpha(Z\alpha)^2$, rather than $\alpha(Z\alpha)^4$ as suggested by naive counting.

## 3. Proton self-energy and the charge-radius definition

Let us briefly consider the correction resulting from the difference between the self-energy of the bound and free proton, see Fig. 3(e) in the main Review. This contribution has the same topology as the finite-size effect, Fig. 3(b) in the main Review, hence, could in principle be absorbed into the proton form factors. The problem is that it differs slightly between H and $\mu$H and thus requires extra care. This difference is exactly the same for the electric and magnetic form factor (affects only the Dirac form factor), and is proportional to the logarithm of the reduced-mass ratio for the two hydrogens, i.e.,

$$\text{Im}\,G_E^{\text{H}-\mu\text{H}}(t) = \text{Im}\,G_M^{\text{H}-\mu\text{H}}(t) = Z^2\alpha\frac{t - 2M^2}{\sqrt{t(t-4M^2)}}\log\frac{m_r^{(\mu\text{H})}}{m_r^{(\text{H})}}. \qquad 16.$$

Using this expression, one can match all of the aforementioned finite-size contributions. For instance, for the charge radius this correction would be:

$$r_p^2(\text{H}) - r_p^2(\mu\text{H}) = \frac{6}{\pi}\int_{4M^2}^{\infty} dt \frac{\text{Im}\,G_E^{\text{H}-\mu\text{H}}(t)}{t^2} = \frac{2Z^2\alpha}{\pi M^2}\log\frac{m_r^{(\mu\text{H})}}{m_r^{(\text{H})}} = 0.0010737\,\text{fm}^2, \qquad 17.$$

which is at the level of the $\mu$H precision. Presently, this difference is already accounted for (in the charge radius only) by shifting these logarithms, accompanied by the Bethe logarithms

on the proton line, into the QED contribution, cf. (1, 2). This practice essentially establishes a unique definition of the charge radius of a bound proton.

The correction shown in Fig. 3(e) is infrared divergent for the free proton and hence the relation to the charge radius extracted from $ep$ scattering depends on the treatment of radiative corrections in these experiments. As these radiative corrections on the proton line are known to be negligibly small (3), at least at the present level of precision of scattering experiments, a possible mismatch with the atomic definition can be ignored.

## 4. Forward two-photon exchange

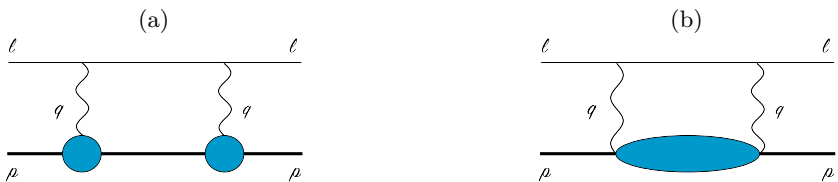

**Figure 1**

$2\gamma$-exchange diagrams in forward kinematics: the horizontal lines correspond to the lepton and the nucleus (bold). (a) Elastic contribution to the $2\gamma$-exchange diagram. (b) Inelastic contribution to the $2\gamma$-exchange diagram, where the "blob" represents all possible excitations. The crossed diagrams are not drawn.

Section 3 of the main Review is dedicated to "Evaluations of the forward two-photon exchange". In this section, we will introduce the finite-size and polarizability contributions to the Lamb shift and hfs stemming from forward $2\gamma$-exchange, including the lepton mass $m$, and describe their derivation a bit more detailed. Note that we will limit the discussion to the spin-1/2 case. We will further try to shed light on the often confusing terminology, and present evaluations for the forward $2\gamma$-exchange contributions to the $\mu$H hfs, including the Zemach radius and recoil contributions omitted in the review.

### 4.1. Relation to Compton scattering and electroproduction data

Looking at the forward $2\gamma$-exchange diagrams in Fig. 1, we see that they are related to the process of forward doubly-virtual Compton scattering (VVCS) off a nucleus. Therefore, the forward $2\gamma$-exchange contributions to the Lamb shift and hfs can be written as integrals over the spin-independent and spin-dependent scalar VVCS amplitudes, $T_i$ and $S_i$, respectively:

$$E_{nS}^{\langle 2\gamma \rangle} = 8\pi\alpha m\, \phi_{nS}^2(0) \int \frac{\mathrm{d}^4 q}{i(2\pi)^4} \frac{(Q^2 - 2\nu^2) T_1(\nu, Q^2) - (Q^2 + \nu^2) T_2(\nu, Q^2)}{Q^4(Q^4 - 4m^2\nu^2)}, \quad \text{18a.}$$

$$E_{nS\text{-hfs}}^{\langle 2\gamma \rangle} = \frac{32\pi\alpha}{3M} \phi_{nS}^2(0) \int \frac{\mathrm{d}^4 q}{i(2\pi)^4} \frac{1}{Q^4 - 4m^2\nu^2} \left\{ \frac{(2Q^2 - \nu^2)}{Q^2} S_1(\nu, Q^2) + \frac{3\nu}{M} S_2(\nu, Q^2) \right\}, \quad \text{18b.}$$

where $M$ is the nuclear mass, $\nu$ is the photon energy in the lab frame, $q^2 = -Q^2$ is the photon virtuality, and $\phi_{nS}^2(0) = 1/(\pi a_B^3 n^3)$ is the wave function at the origin of the atomic $nS$-level, with $a_B = 1/(Z\alpha m_r)$ the Bohr radius and $m_r$ the reduced mass of the atomic bound state. Using a Wick rotation, $q_0 \to iQ_0$, and substituting hyperspherical coordinates, the integrals

simplify to:

$$E_{nS}^{\langle 2\gamma \rangle} = \frac{\alpha}{2\pi^2 m} \phi_{nS}^2(0) \int_0^\infty \frac{\mathrm{d}Q}{Q} \int_0^\pi \mathrm{d}\chi\, \sin^2\chi \times \qquad \text{19a.}$$
$$\times \frac{\left(1 + 2\cos^2\chi\right) T_1(iQ\cos\chi, Q^2) - \sin^2\chi\, T_2(iQ\cos\chi, Q^2)}{\tau_l + \cos^2\chi},$$

$$E_{nS\text{-hfs}}^{\langle 2\gamma \rangle} = \frac{2\alpha}{3\pi^2 m^2 M} \phi_{nS}^2(0) \int_0^\infty \mathrm{d}Q\, Q \int_0^\pi \mathrm{d}\chi\, \sin^2\chi \times \qquad \text{19b.}$$
$$\times \frac{(2 + \cos^2\chi)\, S_1(iQ\cos\chi, Q^2) + \frac{3iQ\cos\chi}{M} S_2(iQ\cos\chi, Q^2)}{\tau_l + \cos^2\chi}.$$

The VVCS amplitudes can be related to empirical data by applying the general principles of analyticity and causality. In other words, combining the optical theorem:[1]

$$\mathrm{Im}\, T_1(\nu, Q^2) = \frac{4\pi^2 Z^2 \alpha}{M} F_1(x, Q^2) = \nu\, \sigma_T(\nu, Q^2), \qquad \text{20a.}$$

$$\mathrm{Im}\, T_2(\nu, Q^2) = \frac{4\pi^2 Z^2 \alpha}{\nu} F_2(x, Q^2) = \frac{Q^2 \nu}{\nu^2 + Q^2} \left[\sigma_T + \sigma_L\right](\nu, Q^2), \qquad \text{20b.}$$

$$\mathrm{Im}\, S_1(\nu, Q^2) = \frac{4\pi^2 Z^2 \alpha}{\nu} g_1(x, Q^2) = \frac{M\nu^2}{\nu^2 + Q^2} \left[\frac{Q}{\nu}\sigma_{LT} + \sigma_{TT}\right](\nu, Q^2), \qquad \text{20c.}$$

$$\mathrm{Im}\, S_2(\nu, Q^2) = \frac{4\pi^2 Z^2 \alpha M}{\nu^2} g_2(x, Q^2) = \frac{M^2 \nu}{\nu^2 + Q^2} \left[\frac{\nu}{Q}\sigma_{LT} - \sigma_{TT}\right](\nu, Q^2), \qquad \text{20d.}$$

and dispersion relations, one arrives at:

$$T_1(\nu, Q^2) - T_1(0, Q^2) = \frac{32\pi Z^2 \alpha M \nu^2}{Q^4} \int_0^1 \frac{\mathrm{d}x\, x}{1 - x^2(\nu/\nu_{\mathrm{el}})^2 - i0^+} F_1(x, Q^2), \qquad \text{21a.}$$

$$\begin{Bmatrix} T_2(\nu, Q^2) \\ S_1(\nu, Q^2) \\ \frac{\nu}{M} S_2(\nu, Q^2) \end{Bmatrix} = \frac{16\pi Z^2 \alpha M}{Q^2} \int_0^1 \frac{\mathrm{d}x}{1 - x^2(\nu/\nu_{\mathrm{el}})^2 - i0^+} \begin{Bmatrix} F_2(x, Q^2) \\ g_1(x, Q^2) \\ g_2(x, Q^2) \end{Bmatrix}, \qquad \text{21b.}$$

with $x = Q^2/2M\nu$ the Bjorken variable, and $Z$ the nuclear charge ($Z = 1$ for the proton). Here, $F_i(x, Q^2)$ and $g_i(x, Q^2)$ are the unpolarized and spin structure functions of the nucleus that are directly related to photoabsorption cross sections measurable in lepton-scattering experiments. The cross sections in Eq. 20 are the usual combinations of helicity cross sections: $\sigma_T = 1/2\,(\sigma_{1/2} + \sigma_{3/2})$ and $\sigma_{TT} = 1/2\,(\sigma_{1/2} - \sigma_{3/2})$ for transversely polarized photons, and $\sigma_L = 1/2\,(\sigma_{1/2} + \sigma_{-1/2})$ for longitudinal photons, where the subscript on the right-hand-side denotes the total helicity of the $\gamma^* N$ state. The cross section $\sigma_{LT}$ describes a simultaneous helicity change of the photon (from longitudinal to transverse) and a nucleon spin-flip.

While the forward 2γ-exchange contribution to the hfs is fully constrained by electroproduction data, the contribution to the Lamb shift is not. The high-energy asymptotics of $F_1(x, Q^2)$, unfortunately, prevent the converge of an unsubtracted dispersion relation. Therefore, one has to rely on a once-subtracted dispersion relation for $T_1(\nu, Q^2)$, see Eq. 21a, where the subtraction function $T_1(0, Q^2)$ is not constrained by data. This introduces an unavoidable model dependence in the data-driven dispersive evaluation of the polarizability contribution to the Lamb shift. In chiral perturbation theory ($\chi$PT) calculations this problem does not arise. See Sec. 6 for further details.

---

[1] For discussion of the photon flux factor see Ref. 4. Here we chosen $K = \nu$.

## 4.2. Finite-size and polarizability contributions

There are two different criteria that can be used to split the VVCS amplitudes into distinctive contributions. We either consider the simplest tree-level diagrams (aka, the Born diagrams) and everything else (i.e., the non-Born diagrams). Or, we distinguish contributions from the simplest tree-level diagrams which have a pole for the nucleon mass, $s = M^2$, (aka, the pole part), and everything else (i.e., the non-pole part). In other words:

$$T_i(\nu, Q^2) = T_i^{(\text{Born})}(\nu, Q^2) + \overline{T}_i(\nu, Q^2) = T_i^{(\text{pole})}(\nu, Q^2) + T_i^{(\text{non-pole})}(\nu, Q^2), \quad 22\text{a}.$$

$$S_i(\nu, Q^2) = S_i^{(\text{Born})}(\nu, Q^2) + \overline{S}_i(\nu, Q^2) = S_i^{(\text{pole})}(\nu, Q^2) + S_i^{(\text{non-pole})}(\nu, Q^2), \quad 22\text{b}.$$

where the non-Born parts of the VVCS amplitudes are denoted as $\overline{T}_i$ and $\overline{S}_i$. The Born and pole contributions are not necessarily the same. For the spin-1/2 case considered here, one finds $T_2^{(\text{pole})}(\nu, Q^2) = T_2^{(\text{Born})}(\nu, Q^2)$ and $S_2^{(\text{pole})}(\nu, Q^2) = S_2^{(\text{Born})}(\nu, Q^2)$, but also:

$$T_1^{(\text{pole})}(\nu, Q^2) = T_1^{(\text{Born})}(\nu, Q^2) + \frac{4\pi Z^2 \alpha}{M} F_1^2(Q^2), \quad 23\text{a}.$$

$$S_1^{(\text{pole})}(\nu, Q^2) = S_1^{(\text{Born})}(\nu, Q^2) + \frac{2\pi Z^2 \alpha}{M} F_2^2(Q^2). \quad 23\text{b}.$$

Here, $F_1(Q^2)$ and $F_2(Q^2)$ are the Dirac and Pauli form factors, related to the electromagnetic Sachs from factors: $G_E(Q^2) = F_1(Q^2) - \tau F_2(Q^2)$ and $G_M(Q^2) = F_1(Q^2) + F_2(Q^2)$ with $\tau = Q^2/4M^2$. Therefore, one ought to be careful in adding the different contributions to the VVCS amplitudes and $2\gamma$ exchange consistently.

Both notations have their advantageous. Speaking in terms of pole and non-pole contributions is convenient if the VVCS amplitudes are expressed as integrals over experimentally observable cross sections by means of dispersion relations, as is done in Eq. 21. In this case, the pole contributions are related to the simple tree-level cross sections of $\gamma N \to N$. These 1-particle production cross sections have fixed center-of-mass energy $s = M^2$, or equivalently $x = 1$. The non-pole contributions are related to inelastic proton structure functions: $\gamma N \to$ anything. These are functions of $x$ and $Q^2$, with $0 < x < x_0$ and $x_0$ the inelastic threshold, e.g., the threshold for pion-production off the proton: $x_0 = Q^2/(2M_p m_\pi + m_\pi^2 + Q^2)$.

Speaking in terms of Born and non-Born contributions, on the other hand, allows to conveniently split into finite-size and polarizability contributions. As one can see from Fig. 1, there are contributions to the $2\gamma$ exchange, where (a) the nucleus in the intermediate state is intact, and (b) the intermediate state is excited, or a nucleus has been broken up into its constituents. A main focus in the review has been on the polarizability effect, Fig. 1(b), which is defined through the non-Born part of the VVCS amplitudes. Figure 1 a) is usually referred to as the elastic $2\gamma$-exchange contribution. It is described through the Born part of the VVCS amplitudes, and attributed to the finite-size effects at order $(Z\alpha)^5$.

### 4.3. Lamb shift formalism

As explained in Sec. 3.1 of the main Review, the $2\gamma$-exchange effect in the Lamb shift is conventionally split into the contribution from elastic structure functions:

$$E_{nS}^{(\text{el})} = 8(Z\alpha)^2 \phi_{nS}^2(0) \int_0^\infty \frac{\mathrm{d}Q}{Q^2} \left[ 4m_r G'_E(0) + \frac{m}{M}\frac{1}{Q}\left(-\frac{v_l+2}{(1+v_l)^2}\left(F_1^2(Q^2)-1\right)\right.\right. \quad 24.$$

$$+\frac{1}{(1+\tau)(v_l+v)}\left\{\left[\tau + \frac{3+2\tau}{(1+v_l)(1+v)}\right](G_M^2(Q^2)-1)\right.$$

$$\left.\left.\left.-\frac{1}{\tau}\left[1-\frac{1}{(1+v_l)(1+v)}\right](G_E^2(Q^2)-1)\right\}\right)\right],$$

with $G'_E(0) = -\langle r_E^2\rangle/6$, and a polarizability contribution $E_{nS}^{(\text{pol})}$. The latter is the sum of the contributions from inelastic structure functions:

$$E_{nS}^{(\text{inel})} = -32(Z\alpha)^2 Mm\,\phi_{nS}^2(0) \int_0^\infty \frac{\mathrm{d}Q}{Q^5} \int_0^{x_0} \mathrm{d}x \frac{1}{(1+v_l)(1+v_x)} \times \quad 25.$$

$$\times \left\{\left[1+\frac{v_l v_x}{v_l+v_x}\right]F_2(x,Q^2) + \frac{2x}{(1+v_l)(1+v_x)}\left[2+\frac{3+v_l v_x}{v_l+v_x}\right]F_1(x,Q^2)\right\},$$

and the non-Born subtraction function:

$$E_{nS}^{(\text{subt})} = \frac{2\alpha m}{\pi} \phi_{nS}^2(0) \int_0^\infty \frac{\mathrm{d}Q}{Q^3} \frac{2+v_l}{(1+v_l)^2} \overline{T}_1(0,Q^2). \quad 26.$$

Here, we introduced the following definitions:

$$v_x = \sqrt{1+x^2\tau^{-1}}, \qquad v = \sqrt{1+\tau^{-1}}, \qquad v_l = \sqrt{1+\tau_l^{-1}}, \qquad \tau_l = \frac{Q^2}{4m^2}. \quad 27.$$

Let us sketch how the above formulas are derived. Plugging the dispersion relations from Eq. 21 into Eq. 19a, the $nS$-level shift generated by the forward $2\gamma$-exchange can be described by:

$$E_{nS}^{[F_1(x,Q^2)]} = -64(Z\alpha)^2 Mm\,\phi_{nS}^2(0) \int_0^\infty \frac{\mathrm{d}Q}{Q^5} \int_0^1 \frac{\mathrm{d}x\,x\,F_1(x,Q^2)}{(1+v_l)^2(1+v_x)^2}\left[2+\frac{3+v_l v_x}{v_l+v_x}\right], \quad 28\text{a}.$$

$$E_{nS}^{[F_2(x,Q^2)]} = -32(Z\alpha)^2 Mm\,\phi_{nS}^2(0) \int_0^\infty \frac{\mathrm{d}Q}{Q^5} \int_0^1 \frac{\mathrm{d}x\,F_2(x,Q^2)}{(1+v_l)(1+v_x)}\left[1+\frac{v_l v_x}{v_l+v_x}\right], \quad 28\text{b}.$$

$$E_{nS}^{[T_1(0,Q^2)]} = \frac{2\alpha m}{\pi} \phi_{nS}^2(0) \int_0^\infty \frac{\mathrm{d}Q}{Q^3} \frac{2+v_l}{(1+v_l)^2} T_1(0,Q^2). \quad 28\text{c}.$$

Now, we need to identify the individual contributions: $E_{nS}^{(\text{el})}$, $E_{nS}^{(\text{inel})}$ and $E_{nS}^{(\text{subt})}$.

To obtain $E_{nS}^{(\text{inel})}$ in Eq. 25, one has to plug the inelastic structure functions into Eqs. 28a and 28b, where the $x$ integration goes from 0 to the inelastic threshold $x_0$.

To obtain $\mathrm{E}_{nS}^{(\text{subt})}$ in Eq. 26, one has to take into account one subtlety. As one can see from Eq. 26, the so-called subtraction-function contribution conventionally only contains the non-Born part $\overline{T}_1(0,Q^2)$ and not the Born part $T_1^{(\text{Born})}(0,Q^2)$. The latter is treated separately because it is known in terms of elastic form factors:

$$T_1^{(\text{Born})}(0,Q^2) = \frac{4\pi Z^2\alpha}{M}\left[G_M^2(Q^2) - F_1^2(Q^2)\right]. \quad 29.$$

It is therefore included in $E_{nS}^{(\mathrm{el})}$. Note that Eq. 29 also contains the important conversion factor between pole and Born VVCS amplitudes shown in Eq. 23a.

To obtain $E_{nS}^{(\mathrm{el})}$ in Eq. 24, one then needs Eq. 28c evaluated with Eq. 29, as well as Eqs. 28a and 28b evaluated with the elastic structure functions:

$$F_1^{(\mathrm{el})}(x, Q^2) = \frac{1}{2} G_M^2(Q^2)\, \delta(1-x), \qquad \text{30a.}$$

$$F_2^{(\mathrm{el})}(x, Q^2) = \frac{1}{1+\tau}\left[G_E^2(Q^2) + \tau G_M^2(Q^2)\right]\delta(1-x). \qquad \text{30b.}$$

In this way, one will obtain a few contributions that are taken into account already through the one-photon exchange potential. These contributions need to be subtracted in order to avoid double-counting. For one thing, we subtract the contribution of a static, structureless nucleus by replacing $G_i^2(Q^2)$ with $G_i^2(Q^2) - 1$. Furthermore, we subtract the charge-radius term, which is effectively of order-$(Z\alpha)^4$, but disguised as:

$$\frac{16}{3}(Z\alpha)^2 m_r\, \phi_{nS}^2(0) \int_0^\infty \frac{\mathrm{d}Q}{Q^2} \langle r_E^2 \rangle. \qquad \text{31.}$$

This leads to $E_{nS}^{(\mathrm{el})}$ as given in Eq. 24. See Refs. 5 and 6 for more details.

In Ref. (7), it has been suggested to chose the Euclidean subtraction $T_1(iQ, Q^2)$ instead of the conventional subtraction $T_1(0, Q^2)$ used in Eq. 21a. In this way, one obtains an approximate formula for the proton-polarizability effect in the Lamb shift:

$$E_{nS}^{\prime\,(\mathrm{subt})} = \frac{2Z\alpha m}{\pi} \phi_{nS}^2(0) \int_0^{x_0} \frac{\mathrm{d}Q}{Q^3} \frac{2+v_l}{(1+v_l)^2} \overline{T}_1(iQ, Q^2). \qquad \text{32.}$$

The remaining contribution of the inelastic structure functions is negligible, due to current conservation (Callan-Gross relation). The single integral in Eq. 32 might hold advantages for future effective field-theory as well as lattice-QCD calculations. See Sec. 6 for further details.

### 4.4. Hyperfine splitting formalism

The dispersive formalism for the hfs is derived analogously. Here, we only give the final formulas. The leading order in $\alpha$ ground-state hfs is given by the Fermi energy:

$$E_\mathrm{F} = \frac{8(Z\alpha)^4 m_r^3 (1+\kappa_N)}{3mM}, \qquad \text{33.}$$

with $\kappa_N$ the anomalous magnetic moment of the nucleus. The $2\gamma$-exchange effects are proportional to $E_\mathrm{F}$, and usually split into three terms:

$$E_{nS\text{-hfs}}^{(2\gamma)} = \frac{E_\mathrm{F}}{n^3} \left(\Delta_\mathrm{Z} + \Delta_\mathrm{recoil} + \Delta_\mathrm{pol}\right), \qquad \text{34.}$$

referred to as the Zemach-radius, recoil, and polarizability contributions. The former two originate from the so-called elastic $2\gamma$-exchange diagram in Fig. 1 a). They are expressed through the elastic form factors as follows (8):

$$\Delta_\mathrm{Z} = \frac{8Z\alpha m_r}{\pi} \int_0^\infty \frac{\mathrm{d}Q}{Q^2}\left[\frac{G_E(Q^2)G_M(Q^2)}{1+\kappa_N} - 1\right] \equiv -2Z\alpha m_r r_\mathrm{Z}, \qquad \text{35.}$$

$$\Delta_\mathrm{recoil} = \frac{Z\alpha}{\pi(1+\kappa_N)} \int_0^\infty \frac{\mathrm{d}Q}{Q} \left\{ \frac{G_M(Q^2)}{Q^2}\frac{8mM}{v_l+v}\left(2F_1(Q^2) + \frac{F_1(Q^2) + 3F_2(Q^2)}{(v_l+1)(v+1)}\right) \right. \qquad \text{36.}$$
$$\left. - \frac{8m_r\, G_M(Q^2) G_E(Q^2)}{Q} - \frac{m F_2^2(Q^2)}{M}\frac{5+4v_l}{(1+v_l)^2} \right\}.$$

The polarizability contribution originates from the so-called inelastic $2\gamma$-exchange diagram in Fig. 1 b). It can be expressed through the inelastic structure functions $g_i(x, Q^2)$ and the Pauli form factor $F_2(Q^2)$:

$$\Delta_{\rm pol} = \Delta_1 + \Delta_2 = \frac{Z\alpha m}{2\pi(1+\kappa)M}\left[\delta_1 + \delta_2\right], \qquad \text{37a.}$$

with:

$$\delta_1 = 2\int_0^\infty \frac{{\rm d}Q}{Q}\left(\frac{5+4v_l}{(v_l+1)^2}\left[4I_1(Q^2)/Z^2 + F_2^2(Q^2)\right] - \frac{32M^4}{Q^4}\int_0^{x_0}{\rm d}x\, x^2 g_1(x,Q^2)\right. \qquad \text{37b.}$$
$$\left.\left\{\frac{1}{(v_l+v_x)(1+v_x)(1+v_l)}\left[4 + \frac{1}{1+v_x} + \frac{1}{v_l+1}\right]\right\}\right),$$

$$\delta_2 = 96M^2\int_0^\infty \frac{{\rm d}Q}{Q^3}\int_0^{x_0}{\rm d}x\, g_2(x,Q^2)\frac{1-v_x}{(1+v_l)(v_l+v_x)}. \qquad \text{37c.}$$

Here, we introduced $I_1(Q^2)$ as the first moment of the $g_1$ structure function:

$$I_1(Q^2) \equiv \frac{2M^2}{Q^2}\int_0^{x_0}{\rm d}x\, g_1(x,Q^2), \qquad \text{38.}$$

whose polarizability part reads:

$$I_1^{(\rm pol)}(Q^2) = I_1(Q^2) + \tfrac{1}{4}F_2^2(Q^2). \qquad \text{39.}$$

Note that the $F_2^2(Q^2)$ term is the important conversion factor between pole and Born VVCS amplitudes shown in Eq. 23b. The $m=0$ limit of $\Delta_{\rm pol}$ is presented in Sec. 3.2.2 of the main Review, where the polarizability contribution is discussed in details.

In Table 1, we summarize results for the $2\gamma$-exchange contribution to the $\mu$H hfs. While $\Delta_{\rm recoil}$ is known with the best accuracy, it is a limiting factor when narrowing down the search range for the $1S$ hfs transition in $\mu$H with the help of the precisely measured $1S$ hfs transition in H, as done in Sec. 4.3 of the main Review.

### 4.5. Off-forward two-photon exchange

As explained in Sec. 2.2 of the main Review, the leading order-$(Z\alpha)^5$ $2\gamma$-exchange corrections originate from the $2\gamma$-exchange diagram in forward kinematics, cf. Fig. 1, while off-forward kinematics ($t \neq 0$, i.e., a non-vanishing momentum transfer between initial and final state) are further suppressed in $Z\alpha$. The forward $2\gamma$-exchange is best described by the dispersive approach introduced above. It can be evaluated based on empirical input for the elastic form factors and inelastic structure functions, and models for the subtraction function, or predicted from $\chi$PT.

When considering different contributions of $2\gamma$-exchange type, it is important to distinguish forward and off-forward kinematics and avoid double counting. Take for instance the $2\gamma$-exchange diagrams with $t$-channel meson exchanges discussed in Refs. 21–26 for $\mu$H. Their forward kinematics are already accounted for in the parametrizations of proton structure functions used in the data-driven dispersive approach. Therefore, when combining the recent results for scalar (21, 22), axial-vector (23–25) and tensor (26) mesons exchanges with data-driven evaluations of $E_{nS}^{(2\gamma)}$, it is important to remove the order-$(Z\alpha)^5$ effect stemming from the forward limit. To do so, one can apply the once-subtracted coordinate-space

**Table 1** Forward 2γ-exchange contribution to the HFS in $\mu$H.

| Reference | $\Delta_{\mathrm{Z}}$ [ppm] | $\Delta_{\mathrm{recoil}}$ [ppm] | $\Delta_{\mathrm{pol}}$ [ppm] | $\Delta_1$ [ppm] | $\Delta_2$ [ppm] | $E_{1S\text{-hfs}}^{\langle 2\gamma\rangle}$ [meV] |
|---|---|---|---|---|---|---|
| DATA-DRIVEN | | | | | | |
| Pachucki '96 (1) | −8025 | 1666 | 0(658) | | | −1.160 |
| Faustov et al. '01 (9)[a] | −7180 | | 410(80) | 468 | −58 | |
| Faustov et al. '06 (10)[b] | | | 470(104) | 518 | −48 | |
| Carlson et al. '11 (11)[c] | −7703 | 931 | 351(114) | 370(112) | −19(19) | −1.171(39) |
| Tomalak '18 (12)[d] | −7333(48) | 846(6) | 364(89) | 429(84) | −65(20) | −1.117(19) |
| HEAVY-BARYON $\chi$PT | | | | | | |
| Peset et al. '17 (13) | | | | | | −1.161(20) |
| LEADING-ORDER B$\chi$PT | | | | | | |
| Hagelstein et al. '16 (14) | | | 37(95) | 29(90) | 9(29) | |
| Hagelstein et al. '18 (15)[e] | | | −13 | 84 | −97 | |

[a]Adjusted values: $\Delta_{\mathrm{pol}}$ and $\Delta_1$ corrected by −46 ppm as described in Ref. 16.
[b]Different convention was used to calculate the Pauli form factor contribution to $\Delta_1$, which is equivalent to the approximate formula in the limit of $m = 0$ used for H in Ref. 11.
[c]Elastic form factors from Ref. 17 and updated error analysis from Ref. 16. Note that this result already includes radiative corrections for the Zemach-radius contribution, $(1+\delta_{\mathrm{Z}}^{\mathrm{rad}})\Delta_{\mathrm{Z}}$ with $\delta_{\mathrm{Z}}^{\mathrm{rad}} \sim 0.0153$ (18, 19), as well as higher-order recoil corrections with the proton anomalous magnetic moment, cf. (11, Eq. 22) and (18).
[d]Uses $r_p$ from $\mu$H (20) as input.
[e]Partially includes the $\Delta(1232)$-isobar contribution.

2γ-exchange potential:

$$V_{2\gamma}(\boldsymbol{r}) = \delta(\boldsymbol{r})\, V_{2\gamma}(t=0) - \frac{1}{\pi}\int_0^\infty \mathrm{d}t\; \mathrm{Im}\, V_{2\gamma}(t) \left[\frac{\delta(\boldsymbol{r})}{t} - \frac{e^{-r\sqrt{t}}}{4\pi r}\right], \qquad 40.$$

where the $\delta(\boldsymbol{r})$-function potential, related to the forward limit, is subtracted from the Yukawa potential in the dispersive integral describing the off-forward 2γ exchange. Note that the pseudoscalar-meson exchanges, on the other hand, are purely off-forward. Thus, their numerically small order-$(Z\alpha)^6$ effects can be included as they are (14, 27–29).

## 5. Radiative corrections to spin-dependent two-photon exchange

It is important to include also radiative corrections to the 2γ exchange. Recently, the initial tension between deuteron charge radius extractions from the $\mu$D Lamb shift, on the one hand, and the $1S-2S$ H-D isotopic shift paired with the $\mu$H Lamb shift, on the other hand, was resolved by amending the $\mu$D theory (30) to include subleading $O(\alpha^6)$ eVP effects (31), as well as inelastic three-photon exchange (3γ exchange) (32). Different radiative corrections due to vacuum polarization were discussed in Sec. 2.3 of the main Review. In the following, we will evaluate the eVP insertion in the elastic 2γ-exchange diagram, shown in Fig. 2, for $\mu$H based on empirical proton form factors.

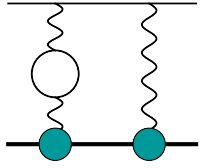

**Figure 2**

Elastic 2γ exchange with vacuum-polarization insertion at $O(\alpha^6)$.

To a good approximation, the different radiative corrections to the elastic 2γ exchange considered here can be expressed through numerical scaling factors:

$$E_{nS\text{-hfs}}^{(\text{el})} = \frac{E_{\text{F}}}{n^3}\left[\left(1 + \frac{\alpha\, C_1(nS)}{\pi} + \sum_i \delta_{\text{Z}}^{\text{rad},i}\right)\Delta_{\text{Z}} + \left(1 + \frac{\alpha\, C_1(nS)}{\pi} + \sum_i \delta_{\text{recoil}}^{\text{rad},i}\right)\Delta_{\text{recoil}}\right], \quad 41.$$

where $C_1(nS)$ is the eVP correction to the wave function from $2^{\text{nd}}$-order perturbation theory, see Eq. 20 in the main Review, and $\delta^{\text{rad},i}$ are other radiative corrections. We will denote the correction factors corresponding to the diagram in Fig. 2 at $1^{\text{st}}$ order in perturbation theory by $\delta_{\text{Z}}^{\text{eVP}}$ and $\delta_{\text{recoil}}^{\text{eVP}}$, respectively.

To account for the insertions of one-loop eVP into the elastic 2γ exchange, we multiply the integrands in Eqs. 35 and 36 with $\left[1-\overline{\Pi}_1(Q^2)\right]^{-2}$, where $\overline{\Pi}_1(Q^2)$ is the one-loop eVP:

$$\overline{\Pi}_1(Q^2) = \Pi_1(Q^2) - \Pi_1(0) = \frac{\alpha}{3\pi}\left[2\left(1-\frac{1}{2\tau_e}\right)\left(\sqrt{1+\frac{1}{\tau_e}}\,\text{arccoth}\sqrt{1+\frac{1}{\tau_e}} - 1\right) + \frac{1}{3}\right], \quad 42.$$

with $\tau_e = Q^2/4m_e^2$ and $m_e$ the electron mass. Several modern proton form factor parametrizations are considered besides the basic dipole form factor: Kelly (33), Bradford et al. (34), Arrington et al. (17), and Borah et al. (35). We find:

$$\delta_{\text{Z}}^{\text{eVP}} = 0.01846(13), \quad 43a.$$
$$\delta_{\text{recoil}}^{\text{eVP}} = 0.01254(4), \quad 43b.$$

where the errors cover all form factor parametrizations, including the dipole. The largest corrections are found for the most recent proton form factor parametrization (35), which uses $r_p(\mu\text{H})$ as a constraint. Our result for $\delta_{\text{Z}}^{\text{eVP}}$ applies to $\mu$H and H. It is in good agreement with Ref. 19:

$$\delta_{\text{Z}}^{\text{eVP}} = \frac{\alpha}{3\pi}\left[2\ln\frac{\Lambda^2}{m_e^2} - \frac{634}{315}\right] \sim 0.0182, \quad 44.$$

with the standard value for the dipole form factor of the proton $\Lambda^2 = 0.71\,\text{GeV}^2$. Due to the size of the Zemach-radius contribution, this is an important correction. The $\alpha(Z\alpha)^5$ effect related to $\delta_{\text{recoil}}^{\text{eVP}}$ is relevant as well, since it is slightly larger than the present uncertainty of the pure $(Z\alpha)^5$ 2γ-exchange recoil contribution. It is interesting to note that the eVP recoil correction in H has a different sign, i.e., reduces the recoil contribution: $\delta_{\text{recoil}}^{\text{eVP}}(\text{H}) = -0.00195(13)$. For the polarizability contribution from inelastic 2γ exchange, we expect the eVP correction to be smaller than the present uncertainty. However, it should be included in future calculations, as we aim for an improved prediction of the polarizability contribution.

For the self-energy and muon anomalous magnetic moment corrections to the Zemach-radius contribution, we use (19, Eq. 6):

$$\delta_Z^{\mu-\text{line}} = -\frac{5\alpha}{4\pi}. \qquad 45.$$

The radiative corrections to the muon line found in Ref. (36, 37) for the $2\gamma$ finite-nuclear-size contribution are slightly larger in magnitude. As these corrections will become more relevant in the future, an independent cross-check would be desirable. Particular care has to be paid to avoid double counting of contributions that are effectively of lower order.

## 6. Model-independent predictions of the Lamb shift polarizability contribution

The subtraction function contribution, Eq. 28c, whose $Q^2$-dependence is not constrained by data, is the bottleneck of the dispersive approach in the calculation of the Lamb shift polarizability contribution. A systematically improvable calculations is necessary in order to refine the theoretical predictions of the $2\gamma$ exchange. In the following, we briefly discuss the prediction from baryon $\chi$PT (B$\chi$PT), focusing in particular on an updated error estimate for the LO plus $\Delta(1232)$ isobar prediction (38) (presented in Table 1 of the main Review), as well as future prospects for lattice QCD.

### 6.1. Chiral perturbation theory

Low-energy effective field-theories, such as B$\chi$PT, are expected to be best applicable to atomic systems, where the energies are naturally very small. B$\chi$PT is a low-energy effective field-theory of QCD at energies well below 1 GeV. It is formulated in terms of hadronic degrees of freedom; here: pions, nucleons and the lowest nucleon-excitation $\Delta(1232)$. It has predictive power for the nucleon polarizabilities up to and including NLO. That means all relevant low-energy constants can be matched to other observables. Thus, B$\chi$PT is used to make predictions for the non-Born VVCS amplitudes, $\overline{T}_i$ and $\overline{S}_i$, and in turn, the proton-polarizability contribution to the Lamb shift and HFS. The main purpose of the predictions from B$\chi$PT, discussed below, as well as from heavy-baryon $\chi$PT (39, 40), is to provide a consistency check for the dispersive evaluations presented in Sec. 4 and remove model dependence.

In Table 1 of the main Review, we show the LO and LO plus $\Delta(1232)$ isobar predictions from B$\chi$PT (38, 41). A few remarks are in order when it comes to the $\Delta$-exchange contribution from B$\chi$PT. It is customary to include a dipole form factor on the dominant magnetic coupling of the $N \to \gamma\Delta$ transition to model a vector-meson type of dependence. It has been shown that in this way B$\chi$PT is able to reproduce pion electroproduction data. In addition, we can see from Fig. 4 in Ref. 38 that the B$\chi$PT prediction of $\overline{T}_1(0, Q^2)/Q^2$ with inclusion of the $\Delta$ form factor is in better agreement with the empirical super-convergence relation estimate from Ref. 42. Both display a sign change in the region of $Q^2$ between 0.03 and 0.28 GeV$^2$. Furthermore, we need to ensure that the contributions to the $2\gamma$-exchange integral from beyond the scale at which B$\chi$PT as an effective field-theory is safely applicable do not exceed the expected uncertainty of such calculation. To regularize the behaviour, one can go a step further away from the pure B$\chi$PT framework and relate the $\gamma N\Delta$ couplings to Jones-Scadron form factors (43), which in turn can be related to electromagnetic nucleon form factors through the use of large-$N_c$ relations (44). This approach has been used in Refs. 15, 38.

In the following, we present an improved error estimate for the LO+$\Delta$ prediction described above. The incomplete NLO calculation (missing $\pi\Delta$ loops), as well as the departure from pure B$\chi$PT by the use of large-$N_c$ relations and form factors, complicate the error estimate of the theory prediction (38). Our improved error estimate is motivated by the low-energy expansion of the Lamb shift polarizability contribution in terms of the electric and magnetic dipole polarizabilities.

The 2$\gamma$-exchange contribution to the Lamb shift is dominated by longitudinal photons and, in turn, the electric dipole polarizability and electric Sachs form factor. Let us start from a low-energy expansion of the spin-independent 2$\gamma$ exchange in Eq. 18a, see (41):

$$E_{nS}^{\langle 2\gamma\rangle} \simeq \frac{\alpha}{\pi}\phi_n^2(0) \int_0^\infty \frac{\mathrm{d}Q}{Q^2} \left(\sqrt{\tau_l} - \sqrt{1+\tau_l}\right) T_L(0, Q^2), \qquad 46.$$

with the purely longitudinal amplitude:

$$T_L(\nu, Q^2) = -T_1(\nu, Q^2) + (1 + \nu^2/Q^2)\, T_2(\nu, Q^2). \qquad 47.$$

That the latter is longitudinal follows from Eqs. 20a and 20b. In a next step we consider the low-energy expansion of the non-Born part of the VVCS amplitudes:

$$\overline{T}_1(0, Q^2) \simeq 4\pi\beta_{M1}Q^2 + O(Q^4), \qquad 48\text{a.}$$
$$\overline{T}_2(0, Q^2) \simeq 4\pi\left(\alpha_{E1} + \beta_{M1}\right)Q^2 + O(Q^4), \qquad 48\text{b.}$$

and an analogous expansion of the Born part:

$$T_L^{(\text{Born})}(0, Q^2) = \frac{16\pi\alpha m^2}{MQ^2}\left[G_E^2(Q^2) + O(Q^2)\right]. \qquad 49.$$

Combining Eqs. 48a and 48b into $\overline{T}_L(0, Q^2)$, we can see that the polarizability contribution to the Lamb shift is indeed dominated by the electric dipole polarizability $\alpha_{E1}$. Looking at the low-$Q$ part of Eq. 49, confirms that the elastic contribution to the Lamb shift is dominated by the electric Sachs form factor $G_E(Q^2)$. Equation 48a also shows that the subtraction function is, on the contrary, dominated by transverse photons and the magnetic dipole polarizability $\beta_{M1}$. Thus, the dominant polarizability effect must be contained in the inelastic contribution. We can think of the latter as approximated by $\overline{T}_2(0, Q^2)$, see Ref. (41, Eq. 17), and the sum of dipole polarizabilities.

A leading-order (LO: $\pi$N loop) plus next-to-leading-order (NLO: $\Delta$ exchange + $\pi\Delta$ loop) B$\chi$PT prediction of the static proton polarizabilities reads (in units of $10^{-4}$ fm$^3$) (45):

$$\alpha_{E1}^{(p)} = 6.8\,(\pi\text{N loop}) - 0.1\,(\Delta\text{ exchange}) + 4.5\,(\pi\Delta\text{ loop}) = 11.2, \qquad 50\text{a.}$$
$$\beta_{M1}^{(p)} = -1.8\,(\pi\text{N loop}) + 7.1\,(\Delta\text{ exchange}) - 1.4\,(\pi\Delta\text{ loop}) = 3.9. \qquad 50\text{b.}$$

The large contribution of the $\Delta$ exchange to $\beta_{M1}^{(p)}$ is expected, since the nucleon-to-Delta transition is dominantly of magnetic-dipole type. The subleading $\Delta$ exchange, thus, has a negligible effect on the polarizability contribution, but will dominate the subtraction-function contribution. We can see this from Table 1 of the main Review, where Alarcon et al. (41) is the LO B$\chi$PT prediction and Lensky et al. (38) is the LO+$\Delta$ prediction.

Since polarizability, subtraction-function and inelastic contributions are dominated by $\alpha_{E1}$, $\beta_{M1}$ and their sum, respectively, we can check how well the LO+$\Delta$ calculation agrees

with empirical determinations of the static polarizabilities and use this as a criterion for our error estimate. The sum of dipole polarizabilities is constrained by the well known Baldin sum rule (46). It is precisely evaluated for the proton based on empirical total photoabsorption cross sections: $\alpha_{E1} + \beta_{M1} = 14.0(2) \times 10^{-4}\,\text{fm}^3$ (47). The PDG recommended values for the individual proton dipole polarizabilities are: $\alpha_{E1} = 11.2(4) \times 10^{-4}\,\text{fm}^3$ and $\beta_{M1} = 2.5(4) \times 10^{-4}\,\text{fm}^3$ (48). Comparing these empirical values for the static polarizabilities to the LO+$\Delta$ predictions, we deduce a 53 % uncertainty for the subtraction-function contribution towards magnitude decrease, and a 62 % uncertainty for the polarizability contribution towards magnitude increase, see Table 1 in the main Review. This is much larger than the usual uncertainty estimate for a LO B$\chi$PT calculation, that would be 15 % ($\sim m_\pi/\text{GeV}$).

In summary, we can say that the B$\chi$PT and HB$\chi$PT predictions for $E_{nS}^{(\text{subt})}$ support the model-dependent results used in the dispersive approach. In view of a possible refined measurement of the Lamb shift in $\mu$H, the uncertainty of the theoretical 2$\gamma$-exchange predictions has to improve further. For a full NLO B$\chi$PT prediction, the contribution of the potentially important $\pi\Delta$ loops has to be evaluated. In the next subsection, we will discuss the Euclidean subtraction function $\overline{T}_1(iQ, Q^2)$ as a new ansatz to calculate the Lamb shift polarizability contribution with prospects for lattice QCD and the NLO $\pi\Delta$-loops in B$\chi$PT.

### 6.2. Prospects for lattice QCD

An ab-initio calculation of the 2$\gamma$-exchange effect would be most desirable. First lattice-QCD results to describe a small part of the 2$\gamma$-exchange contribution to the $\mu$H Lamb shift were recently published (49, 50). In the following, we present a different approach that would allow to access most of the 2$\gamma$-exchange contribution with a direct lattice-QCD calculation (7). Instead of the conventional subtraction function $T_1(0, Q^2)$, we suggest to use the Euclidean subtraction function $T_1(iQ, Q^2)$. It isolates the purely longitudinal amplitude:

$$\overline{T}_L(iQ, Q^2) = -\overline{T}_1(iQ, Q^2) = 4\pi\alpha_{E1}Q^2 + O(Q^4), \qquad 51.$$

thereby, providing an approximate formula for the proton-polarizability effect in the Lamb shift:

$$E_{nS}^{\prime(\text{subt})} = \frac{2Z\alpha m}{\pi}\phi_{nS}^2(0)\int_0^{x_0}\frac{dQ}{Q^3}\frac{2+v_l}{(1+v_l)^2}\overline{T}_1(iQ, Q^2). \qquad 52.$$

Here, the overline denotes the non-Born or polarizability part of the VVCS amplitudes. The remaining contribution of the inelastic structure functions, as we will show, is negligible. With the subtraction point at $\nu = iQ$, the $T_1$ dispersion relation reads as:

$$T_1(\nu, Q^2) = T_1(iQ, Q^2) + \frac{32\pi\alpha M}{Q^4}\left(\nu^2 + Q^2\right)\int_0^1 dx\, x\,\frac{F_1(x, Q^2)}{\left[1 - x^2(\nu/\nu_{\text{el}})^2\right]\left(1 + x^2\tau^{-1}\right)}. \qquad 53.$$

Obviously, the dispersive integral over the $F_1(x, Q^2)$ structure function differs from Eq. 21a, thus, also the $F_1(x, Q^2)$ contribution to the Lamb shift will change. A successive small-$x$ and low-$Q$ expansion of the inelastic contribution to the $nS$-level shift shows that:

$$E_{nS}^{\prime(\text{inel})} \sim -16\alpha^2 M\phi_{nS}^2(0)\int_0^\infty \frac{dQ}{Q^4}\int_0^{x_0} dx\, F_L(x, Q^2), \qquad 54.$$

where we identified the longitudinal structure function: $F_L(x, Q^2) = F_2(x, Q^2) - 2xF_1(x, Q^2)$. According to current conservation (or, Callan-Gross relation), the later is

vanishing for asymptotically low (or large) $Q$, and its moments go as:

$$\lim_{Q^2 \to 0} Q^{-4-2n} \int \mathrm{d}x\, x^{2n} F_L(x,Q^2) = 0. \qquad 55.$$

With the Bosted-Christy parametrization (51) of the structure functions, one can make the rough estimate (7):

$$E_{2S}'^{(\mathrm{inel})} \simeq 1.6\,\mu\mathrm{eV}. \qquad 56.$$

This shows that the inelastic contribution is indeed negligible at the current level of experimental precision ($\sim 2\,\mu\mathrm{eV}$), thus, Eq. 52 is a good approximation for the polarizability contribution to the Lamb shift.

The single integral in Eq. 52 holds advantages for effective field-theory as well as lattice-QCD calculations. For one thing, it can be used to calculate the $\pi\Delta$-loop contribution and study the size of contributions from beyond the scale at which B$\chi$PT is safely applicable. For another thing, one could use lattice QCD to calculate $T_1(iQ, Q^2)$. First lattice-QCD calculations of the nucleon VVCS amplitude $T_1(\nu, Q^2)$ in the unphysical region appeared in (52–55).

## LITERATURE CITED

1. Pachucki K. *Phys. Rev. A* 53:2092–2100 (1996)
2. Karshenboim SG, Korzinin EY, Shelyuto VA, Ivanov VG. *Journal of Physical and Chemical Reference Data* 44(3):031202 (2015)
3. Maximon LC, Tjon JA. *Phys. Rev. C* 62:054320 (2000)
4. Drechsel D, Pasquini B, Vanderhaeghen M. *Phys. Rept.* 378:99–205 (2003)
5. Carlson CE, Vanderhaeghen M. *Phys. Rev. A* 84:020102 (2011)
6. Hagelstein F, Miskimen R, Pascalutsa V. *Prog. Part. Nucl. Phys.* 88:29–97 (2016)
7. Hagelstein F, Pascalutsa V. *Nucl. Phys. A* 1016:122323 (2021)
8. Zemach AC. *Phys. Rev.* 104(6):1771–1781 (1956)
9. Faustov R, Cherednikova E, Martynenko A. *Nucl. Phys. A* 703:365–377 (2002)
10. Faustov R, Gorbacheva I, Martynenko A. *Proc. SPIE Int. Soc. Opt. Eng.* 6165:0M (2006)
11. Carlson CE, Nazaryan V, Griffioen K. *Phys. Rev. A* 78:022517 (2008)
12. Tomalak O. *Eur. Phys. J. C* 77(12):858 (2017)
13. Peset C, Pineda A. *JHEP* 04:060 (2017)
14. Hagelstein F, Pascalutsa V. *PoS* CD15:077 (2016)
15. Hagelstein F. *Few Body Syst.* 59(5):93 (2018)
16. Carlson CE, Nazaryan V, Griffioen K. *Phys. Rev. A* 83:042509 (2011)
17. Arrington J, Melnitchouk W, Tjon JA. *Phys. Rev. C* 76:035205 (2007)
18. Bodwin GT, Yennie D. *Phys. Rev. D* 37:498 (1988)
19. Karshenboim SG. *Phys. Lett. A* 225:97 (1997)
20. Antognini A, Nez F, Schuhmann K, Amaro FD, et al. *Science* 339:417–420 (2013)
21. Borisyuk D. *Phys. Rev. C* 96(5):055201 (2017)
22. Dorokhov AE, Martynenko AP, Martynenko FA, Radzhabov AE. *Phys. Part. Nucl. Lett.* 16(5):520–523 (2019)
23. Dorokhov AE, Kochelev NI, Martynenko AP, Martynenko FA, Radzhabov AE. *Phys. Lett. B* 776:105–110 (2018)
24. Dorokhov AE, Kochelev NI, Martynenko AP, Martynenko FA, Radzhabov AE. *Eur. Phys. J. A* 54(8):131 (2018)
25. Miranda A, Roig P, Sanchez-Puertas P. *Phys. Rev. D* 105(1):016017 (2022)
26. Dorokhov AE, Martynenko AP, Martynenko FA, Radzhabov AE. *J. Phys. Conf. Ser.* 1435(1):012004 (2020)

27. Huong NT, Kou E, Moussallam B. *Phys. Rev.* D 93(11):114005 (2016)
28. Zhou HQ, Pang HR. *Phys. Rev.* A 92(3):032512 (2015), [Erratum: Phys. Rev. A 93 (2016) 069903]
29. Dorokhov AE, Kochelev NI, Martynenko AP, Martynenko FA, Faustov RN. *Phys. Part. Nucl. Lett.* 14(6):857–864 (2017)
30. Krauth JJ, Diepold M, Franke B, Antognini A, et al. *Annals Phys.* 366:168–196 (2016)
31. Kalinowski M. *Phys. Rev. A* 99(3):030501 (2019)
32. Pachucki K, Patkóš V, Yerokhin VA. *Phys. Rev. A* 97(6):062511 (2018)
33. Kelly JJ. *Phys. Rev.* C 70:068202 (2004)
34. Bradford R, Bodek A, Budd HS, Arrington J. *Nucl. Phys. Proc. Suppl.* 159:127–132 (2006)
35. Borah K, Hill RJ, Lee G, Tomalak O. *Phys. Rev. D* 102(7):074012 (2020)
36. Faustov RN, Martynenko AP, Martynenko GA, Sorokin VV. *Phys. Lett. B* 733:354–358 (2014)
37. Faustov RN, Martynenko AP, Martynenko FA, Sorokin VV. *Phys. Part. Nucl.* 48(5):819–821 (2017)
38. Lensky V, Hagelstein F, Pascalutsa V, Vanderhaeghen M. *Phys. Rev. D* 97(7):074012 (2018)
39. Peset C, Pineda A. *Eur. Phys. J.* A 51(3):32 (2015)
40. Peset C, Pineda A. *Nucl. Phys.* B 887:69–111 (2014)
41. Alarcón JM, Lensky V, Pascalutsa V. *Eur. Phys. J.* C 74(4):2852 (2014)
42. Tomalak O, Vanderhaeghen M. *Eur. Phys. J.* C 76(3):125 (2016)
43. Jones HF, Scadron MD. *Annals Phys.* 81:1–14 (1973)
44. Pascalutsa V, Vanderhaeghen M. *Phys. Rev.* D 76:111501 (2007)
45. Alarcón JM, Hagelstein F, Lensky V, Pascalutsa V. *Phys. Rev. D* 102(1):014006 (2020)
46. Baldin AM. *Nucl. Phys.* 18:310–317 (1960)
47. Gryniuk O, Hagelstein F, Pascalutsa V. *Phys. Rev. D* 92:074031 (2015)
48. Zyla PA, et al. *PTEP* 2020(8):083C01 (2020)
49. Fu Y. 2021. In *38th International Symposium on Lattice Field Theory*. 2112.14913
50. Fu Y, Feng X, Jin LC, Lu CF. *Phys. Rev. Lett.* 128(17):172002 (2022)
51. Christy ME, Bosted PE. *Phys. Rev. C* 81(5):055213 (2010)
52. Can K, Hannaford-Gunn A, Horsley R, Nakamura Y, Perlt H, et al. *Phys. Rev. D* 102(11) (2020)
53. Hannaford-Gunn A, Horsley R, Nakamura Y, Perlt H, Rakow P, et al. *PoS* LATTICE2019:278 (2020)
54. Chambers A, Horsley R, Nakamura Y, Perlt H, Rakow P, et al. *Phys. Rev. Lett.* 118(24):242001 (2017)
55. Hannaford-Gunn A, et al. *PoS* LATTICE2021:028 (2022)



\end{document}